\documentclass{lmcs} 
\pdfoutput=1

\usepackage{lastpage}
\lmcsdoi{18}{2}{16}
\lmcsheading{}{\pageref{LastPage}}{}{}%
{Oct.~29,~2020}{Jun.~10,~2022}{}

\pdfoutput=1

\usepackage[utf8]{inputenc}
\usetikzlibrary{positioning}
\usepackage{hyperref}
\usepackage[numbers]{natbib}
\usepackage{amssymb}
\usepackage{amsmath}
\usepackage{multirow}
\usepackage{mathpartir}
\usepackage{packages/iris}
\usepackage{packages/actris}
\usepackage{packages/heaplang}
\usepackage{cleveref}
\usepackage{listings}
\usepackage{graphicx}
\usepackage{xspace}
\usepackage{pftools}
\usepackage{ifthen}
\usepackage{comment}
\newcommand{\safe}[1]{\mathsf{safe}\ #1}

\newcommand{\postvalid}[2]{\mathsf{post\_valid}\ (#1,#2)}

\newcommand{\Comment}[1]{\texttt{\color{gray}(*\ #1\ *)}}

\newcommand{\cmpspecname}{\logdefemph{\cmpword\_spec}}

\newcommand{\listsortprefix}{sort}
\newcommand{\listsortservicename}{\defemph{\listsortprefix\_service}}
\newcommand{\listsortclientname}{\defemph{\listsortprefix\_client}}
\newcommand{\listsortprotname}{\logdefemph{\listsortprefix\_prot}}

\newcommand{\listsortcmpservicename}{\listsortservicename_{\mathit{func}}}
\newcommand{\listsortcmpclientname}{\listsortclientname_{\mathit{func}}}
\newcommand{\listsortcmpprotname}{\listsortprotname_{\mathit{func}}}

\newcommand{\listsortloopclientname}{\listsortclientname_{\mathit{rec}}}
\newcommand{\listsortloopservicename}{\listsortservicename_{\mathit{rec}}}
\newcommand{\listsortloopprotname}{\listsortprotname_{\mathit{rec}}}

\newcommand{\listsortdelclientname}{\listsortclientname_{\mathit{del}}}
\newcommand{\listsortdelservicename}{\listsortservicename_{\mathit{del}}}
\newcommand{\listsortdelprotname}{\listsortprotname_{\mathit{del}}}

\newcommand{\listsortelemservicesplitname}{\defemph{split}_{\mathit{fg}}}
\newcommand{\listsortelemservicemergename}{\defemph{merge}_{\mathit{fg}}}
\newcommand{\listsortelemservicemergerecname}{\defemph{merge}_{\mathit{fg}}^{\mathit{aux}}}
\newcommand{\listsortelemservicetransfername}{\defemph{transfer}}
\newcommand{\listsortelemservicename}{\listsortservicename_{\mathit{fg}}}
\newcommand{\listsortelemclientname}{\listsortclientname_{\mathit{fg}}}

\newcommand{\listsortelemprottailname}{\listsortprotname_{\mathit{fg}}^{\mathit{tail}}}
\newcommand{\listsortelemprotheadname}{\listsortprotname_{\mathit{fg}}^{\mathit{head}}}
\newcommand{\listsortelemprotname}{\listsortprotname_{\mathit{fg}}}

\newcommand{\sendallname}{\defemph{send\_all}}
\newcommand{\recvallname}{\defemph{recv\_all}}

\newcommand{\cmpword}{cmp}
\newcommand{\cmpvar}{\logemph{\cmpword}}

\newcommand{\servername}{auth}
\newcommand{\clientname}{contrib}
\newcommand{\servervar}{\logdefemph{\servername}}
\newcommand{\clientvar}{\logdefemph{\clientname}}

\newcommand{\serverpredT}[2]{\servervar_{#1}\ \! #2}
\newcommand{\clientpredT}[1]{\clientvar_{#1}}

\newcommand{\simpleexampleprefix}{prog\_lock}

\newcommand{\simpleexamplename}{\defemph{\simpleexampleprefix}}
\newcommand{\lockprotname}{\defemph{lock\_prot}}

\newcommand{\serverpred}[3]{\serverpredT{#1}{#2}\ #3}
\newcommand{\clientpred}[2]{\clientpredT{#1}\ #2}
\newcommand{\mapperword}{mapper}
\newcommand{\mapperclientname}{\defemph{\mapperword\_client}}
\newcommand{\mapperservicename}{\defemph{\mapperword\_service}}
\newcommand{\mapperprotname}{\defemph{\mapperword\_prot}}
\newcommand{\parmapperword}{par\_\mapperword}
\newcommand{\parmapperworkername}{\defemph{\parmapperword\_worker}}
\newcommand{\parmapperclientname}{\defemph{\parmapperword\_client}}
\newcommand{\valuefy}[1]{#1_{\mathsf v}}
\newcommand{\mapword}{f}
\newcommand{\mapvar}{\logemph{\mapword}}
\newcommand{\vmapvar}{\valuefy{\mapvar}}
\newcommand{\parmapperprotname}{\logdefemph{\parmapperword\_prot}}

\newcommand{\mapspecname}{\logdefemph{\mapword\_spec}}
\newcommand{\flatmapname}{\defemph{flatMap}}

\newcommand{\ite}[3]{\langkw{if}\ #1\ \langkw{then}\ #2 \ \langkw{else}\ #3}

\newcommand{\lineref}[1]{\hyperref[#1]{line~\ref*{#1}}}

\newcommand{\fvar}{\mathit{f}}

\newcommand{\goname}{\mathit{go}}

\definecolor{darkred}{rgb}{0.7,0,0}
\definecolor{darkgreen}{rgb}{0,0.5,0}
\Crefname{part}{\S\!}{\S}
\Crefname{chapter}{\S\!}{\S}
\Crefname{section}{\S\!}{\S}

\crefname{thm}{Theorem}{Theorems}

\renewcommand{\cite}{\citep}
\setcitestyle{%
  alphaurl,%
  open={[},close={]},citesep={;},%
  aysep={},yysep={,},%
  notesep={, }}

\begin{document}

\title[Actris 2.0]{Actris 2.0: Asynchronous Session-Type Based Reasoning in Separation Logic}

\author{Jonas Kastberg Hinrichsen\rsuper{a}}	
\address{IT University of Copenhagen and Aarhus University, Denmark}	
\email{hinrichsen@cs.au.dk}

\author{Jesper Bengtson\rsuper{b}}	
\address{IT University of Copenhagen, Denmark}	
\email{bengtson@itu.dk}

\author{Robbert Krebbers\rsuper{c}}	
\address{Radboud University and Delft University of Technology, The Netherlands}	
\email{mail@robbertkrebbers.nl}




\begin{abstract}
  Message passing is a useful abstraction for implementing concurrent
programs.
For real-world systems, however, it is often combined with other programming
and concurrency paradigms, such as
higher-order functions, mutable state, shared-memory concurrency, and locks.
We present \textbf{\lname}: a logic for proving functional correctness of
programs that use a combination of the aforementioned features.
\lname combines the power of modern concurrent separation logics with a
first-class protocol mechanism---based on session types---for
reasoning about message passing in the presence of other concurrency paradigms.
We show that \lname provides a suitable level of abstraction by proving
functional correctness of a variety of examples, including
a channel-based merge sort, a channel-based load-balancing mapper, and a variant
of the map-reduce model, using concise specifications.

While \lname was already presented in a conference paper (POPL'20), this paper
expands the prior presentation significantly.
Moreover, it extends \lname to \textbf{\lname 2.0} with a notion of
\emph{subprotocols}---based on session-type subtyping---that permits additional flexibility when
composing channel endpoints, and that takes full advantage of the asynchronous
semantics of message passing in Actris.
Soundness of \lname 2.0 is proven using a model of its protocol mechanism in the
Iris framework.
We have mechanised the theory of \lname, together with custom tactics, as well as
all examples in the paper, in the Coq proof assistant.

\end{abstract}

\maketitle

\section{Introduction}
\label{sec:intro}

Message-passing programs are ubiquitous in modern computer systems,
emphasising the importance of their functional correctness.
Programming languages, like Erlang, Elixir, and Go, have built-in
primitives that handle spawning of processes and intra-process
communication, while other mainstream languages, such as Java,
Scala, F\#, and C\#, have introduced an Actor model~\cite{hewitt_IJCAI1973} to achieve similar
functionality. In both cases the goal remains the same---help design
reliable systems, often with close to constant up-time, using
lightweight processes that can be spawned by the hundreds of
thousands and that communicate via asynchronous message passing.

While message passing is a useful abstraction, it is not
a silver bullet of concurrent programming. In a qualitative study of
larger Scala projects Tasharofi \etal~\cite{tasharofi-ECOOP2013} write:

\begin{quote}
  We studied 15 large, mature, and actively maintained actor programs
  written in Scala and found that 80\% of them mix the actor model
  with another concurrency model.
\end{quote}

\noindent In this study, 12 out of 15 projects did not entirely stick to the Actor
model, hinting that even for projects that embrace message passing,
low-level concurrency primitives like locks (\ie mutexes) still have
their place.
Tu \etal~\cite{tu-ASPLOS2019} came to a similar conclusion when studying 6 large and
popular Go programs.
A suitable solution for reasoning about message-passing
programs should thus integrate with other programming and concurrency paradigms.

In this paper we introduce \textbf{\lname}---a concurrent separation logic for
proving functional correctness of programs that combine message passing with
other programming and concurrency paradigms.
\lname can be used to reason about programs written in a language that mimics the
important features found in aforementioned languages such as higher-order functions,
higher-order references, fork-based concurrency, locks, and primitives for
asynchronous message passing over channels.
The channels of our language are first-class and can be
sent as arguments to functions, be sent over other channels
(often referred to as delegation), and be stored in references.

Program specifications in \lname are written in an impredicative higher-order
concurrent separation logic built on top of the Iris framework
\cite{jung-POPL2015,krebbers-ESOP2017,jung-ICFP06,jung-JFP2018}.
In addition to the usual features of Iris, \lname
provides a notion of \emph{\pname} to reason about message passing over
channels, inspired by 
binary session types~\cite{honda-ESOP1998}.
We show that \pname integrate seamlessly with other concurrency
paradigms, allow delegation of resources,
support channel sharing over multiple concurrent threads using locks, and more.

\subsection{Message passing in concurrent separation logic}
\label{sec:intro_sep_logic}

Over the last decade, there has been much work on
extensions of concurrent separation logic with
reasoning principles for message passing~\cite{francalanza-LMCS2011,villard-ICE2012,craciun-ICECCS2015,oortwijn-PLACES2016}.
These logics typically include some form of mechanism for writing
protocol specifications in a high-level manner,
to elegantly reason about message passing in some specific context.

In a different line of work, researchers have developed more expressive
extensions of concurrent separation logic that support proving strong
specifications of programs involving features such as
higher-order functions, fine-grained shared-memory concurrency, and locks.
Examples of such logics are TaDA \cite{pinto-ECOOP2014}, iCAP~\cite{svendsen-ESOP2014},
Iris~\cite{jung-POPL2015}, FCSL~\cite{nanevski-ESOP2014}, and
VST~\cite{appel2014vst}.
However, only a few variants and extensions of these logics provide a
high-level reasoning mechanism specific to message-passing concurrency.

First off, there has been work on the use of Iris-like separation logic to reason about programs that communicate via message passing over a network.
The reasoning principles in such logics are geared towards different programming patterns than the ones used in high-level languages like Erlang, Elixir, Go, and Scala.
Namely, on networks all data must be serialised, and packets can be lost or delivered out of order.
In high-level languages messages cannot get lost, are ensured to be delivered in order, and are allowed to contain many types of data, including functions, references, and even channel endpoints.
Two examples of network logics are Disel by Sergey \etal~\cite{sergey-POPL2018} and
Aneris by Krogh-Jespersen \etal~\cite{krogh-jespersen}.
Additionally, there has been work on the use of separation logic to prove compiler correctness of high-level message-passing languages.
Tassarotti \etal~\cite{tassarotti-ESOP2017} verified a small compiler of a session-typed language into a language where channel buffers are modelled on the heap.

The primary reasoning principle to model the interaction between processes in
the aforementioned
logics is the notion of a State Transition
System (STS).
As a simple example, consider the following program, which is
borrowed from Tassarotti \etal~\cite{tassarotti-ESOP2017}:
\[
{\mathit{prog}}_1\langdef\Let {(\chan, \chan')} = \newchan in \Fork{\send {\chan'} {42}}; \recv \chan
\]
\noindent This program creates two channel endpoints $c$ and $c'$, forks off a
new thread, and sends the number $42$ over the channel $c'$, which is then
received by the initiating thread.
Modelling the behaviour of this program in an STS typically requires three states:
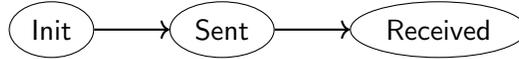
\begin{equation*}
\begin{tikzpicture}
\node[ellipse,draw=black] (init) {$\mathsf{Init}$};
\node[right=of init,ellipse,draw=black] (send) {$\mathsf{Sent}$};
\node[right=of send,ellipse,draw=black] (received) {$\mathsf{Received}$};
\draw[thick,->] (init) -- (send);
\draw[thick,->] (send) -- (received);
\end{tikzpicture}
\end{equation*}
\noindent The three states model that no message has been
sent ($\mathsf{Init}$), that a message has been sent but not received
($\mathsf{Sent}$),
and finally that the message has been sent and received ($\mathsf{Received}$).
Exactly what this STS represents is made precise by the underlying logic, which
determines what constitutes a state and a transition, and how these are related
to the channel buffers.

While STSs appear like a flexible and intuitive abstraction to reason about
message-passing concurrency, they have their problems:

\begin{itemize}
\item Coming up with a good STS that makes the appropriate abstractions is
  difficult because
  the STS has to keep track of all possible states that the channel buffers can
  be in, including all possible interleavings of messages in transit.
\item While STSs used for the verification of different modules
  can be composed at the level of the logic, there is no canonical way of
  composing them due to their unrestrained structure.
\item Finally, STSs are first-order meaning that their states and
  transitions cannot be indexed by propositions of the underlying logic, which
  limits what they can express when sending messages containing functions or other channels.
\end{itemize}

\subsection{\lname 1.0: \Pname}
\label{sec:intro_protocols}

\lname extends separation logic with a notion called \emph{\pname}.
This notion is inspired by the session type community, pioneered by
Honda \etal~\cite{honda-ESOP1998}, where channel endpoints are given types that describe
the expected exchanges.
Using session types, the channels $\chan$ and $\chan'$ in the program
${\mathit{prog}}_1$ in \Cref{sec:intro_sep_logic}
would have the types $\chan :\ \strecv {\integer} \stend$ and
$\chan':\ \stsend {\integer} \stend$,
where $\SEND T$ and $\RECV T$ denotes that a value of type $T$ is sent or
received, respectively.
Moreover, the types of the channels $\chan$ and $\chan'$ are \emph{duals}---when one does a send the other
does a receive, and \viceversa.

While session types provide a compact way of specifying the behaviour of channels,
they can only be used to talk
about the type of data that is being passed around---not their payloads.
In this paper, we build on prior work by Bocchi \etal~\cite{bocchi-CONCUR2010} and
Craciun \etal~\cite{craciun-ICECCS2015} to attach logical predicates to session types to say more about the payloads, thus vastly extending the expressivity.
Concretely, we port session types into separation logic in the
form of a construct $\interp \chan \prot$, which denotes ownership of a channel
$\chan$ with \pnameSingular $\prot$.
\Pname $\prot$ are streams of
$\sendprot {\xdots} \val \iprop \prot$ and
$\recvprot {\xdots} \val \iprop \prot$ constructors that are either infinite or finite,
where finite streams are ultimately terminated by an $\protend$ constructor.
Here, $\val$ is the value that is
being sent or received, $\iprop$ is a separation logic proposition denoting
the ownership of the resources being transferred as part of the message, and the variables $\xdots$ bind into $\val$, $\iprop$, and $\prot$.
The \pname for the above example are:
\[
\interp{\chan}{\recvprot {} {42} \TRUE \protend}
\quad\mathtt{and}\quad
\interp{\chan'}{\sendprot {} {42} \TRUE \protend}
\]

\noindent These protocols state that the endpoint $\chan$ expects the number $42$ to
be sent along it, and that the endpoint $\chan'$ expects to send the number
$42$.
Using this protocol, we can prove that ${\mathit prog}_1$
has the specification
$\hoare \TRUE {\mathit {prog}_1} {\Ret \val. \val = 42}$, where $\val$ is its
resulting value.

\Pname $\sendprot \xdots \val \iprop \prot$ and
$\recvprot \xdots \val \iprop \prot$ are \emph{dependent}, meaning that
the tail $\prot$ can be defined in terms of the previously bound variables $\xdots$.
A sample program showing the use of such dependency is:
\[
{\mathit {prog}_2}\langdef
\begin{array}[t]{@{} l}
\Let {(\chan, \chan')} = \newchan in \\
\Fork{ \Let {\var} = \recv {\chan'} in {\send {\chan'} (\var + 2 )} }; \\
\send \chan 40;\ \recv \chan
\end{array}
\]
In this program, the main thread sends the number $40$ to the forked-off thread,
which then adds two to it, and sends it back.
This program has the same specification as ${\mathit prog}_1$, while we change the
\pnameSingular as follows (we omit the \pnameSingular for the dual endpoint $\chan'$):
\[
\interp{\chan}{\sendprot{(\var:\integer)}{\var}{\TRUE}{\recvprot{}{\var+2}{\TRUE} \protend}}
\]
This protocol states that the second exchanged value is exactly the first with two
added to it.
To do so, it makes use of a dependency on the variable $\var$, which is
used to describe the contents of the first message, which the second message then
depends on.
This variable is bound in the protocol and it is instantiated only when a message is sent.
This is different from the logic by Craciun \etal~\cite{craciun-ICECCS2015}, which does not support dependent protocols.
Their logic is limited to protocols analogous to $\sendprot{}{\var}{\TRUE} \recvprot{}{\var+2}{\TRUE} \protend$ where $\var$ is free, which means the value of $\var$ must be known when the protocol is created.

While the prior examples could have been type-checked and verified using
the formalisms of Bocchi \etal~\cite{bocchi-CONCUR2010} and Craciun \etal~\cite{craciun-ICECCS2015},
the following stateful example cannot:
\[
{\mathit {prog}_3}\langdef
\begin{array}[t]{@{} l}
  \Let {(\chan,\chan')} = \newchan in \\
  \Fork{\Let{\locvar} = \recv{\chan'} in \locvar \gets (\deref{\locvar} + 2);\
  \send {\chan'} {\TT}}; \\
  \Let \locvar = \newref{40} in \send{\chan}{\locvar};\ \recv {\chan};\ \deref \locvar
\end{array}
\]
Here, the main thread stores the value $40$ on the heap, and sends a
reference $\locvar$ over the channel $\chan$ to the forked-off thread.
The main thread then awaits a signal $\TT$, notifying that the
reference has been updated to $42$ by the forked-off thread.
This program has the same specification as ${\mathit prog}_1$ and ${\mathit prog}_2$, but the \pnameSingular is updated:
\[
\interp{\chan}{\sendprot{(\loc: \Loc)\ (\var:\integer)}{\loc}{\loc \mapsto \var}
               \recvprot{}{\TT}{\loc \mapsto (\var+2)}
               \protend}
\]

\noindent
This protocol denotes that the endpoints first exchange a reference $\loc$, as well as a
\emph{points-to} connective $\loc\mapsto x$ that describes the ownership and
value of the reference $\loc$.
To perform the exchange $\chan$ has to give up ownership of the
location, while $\chan'$ acquires it---which is why it can then safely
update the received location to $42$ before sending the ownership back along with the
notification $\TT$.

The type system by Bocchi \etal~\cite{bocchi-CONCUR2010} cannot verify this program because it does not support mutable state, while Actris can verify the program because it is a separation logic.
The logic by Craciun \etal~\cite{craciun-ICECCS2015} cannot verify this program because it does not support dependent protocols, which are crucial here as they make it possible to
delay picking the location $\loc$ used in the protocol until the send operation is performed.

Dependent protocols are also useful to define recursive protocols to reason about programs that use a channel in a loop.
Consider the following variant of ${\mathit prog}_1$:
\[
{\mathit {prog}_4}\langdef
\begin{array}[t]{@{} l}
\Let {(\chan,\chan')} = \newchan in \\
\Fork{\Let {\mathit{go}\ \TT} =
  {\big(\send {\chan'} ({\recv {\chan'} + 2});\ \mathit{go}\ \TT\big)} in \mathit{go}\ \TT}; \\
\send \chan 18;\ \Let \var = {\recv \chan} in \\
\send \chan 20;\ \Let \varB = {\recv \chan} in \var + \varB
\end{array}
\]
The forked-off thread will repeatedly interleave receiving values with
sending those values back incremented by two.
The program ${\mathit prog}_4$ has the same specification as before, but now we use the following recursive \pnameSingular:
\[
\interp {\chan} {\MU \recvar. \sendprot{(\var:\integer)}{\var}{\TRUE}
                            \recvprot{}{\var+2}{\TRUE} \recvar}
\]
This protocol expresses that it is possible to make repeated exchanges with the forked-off thread to increment a number by two.
The fact that the variable $\var$ is bound in the protocol is once again crucial---it allows the use of different numbers for each exchange.

Furthermore, \lname inherently includes some features of conventional
session types.
One such example is the \emph{delegation} of channels as seen in the following program:
\[
{\mathit prog}_5
\langdef{}
\begin{array}[t]{@{} l}
  \Let {(\chan_1, \chan_1')} = \newchan in \\
  \Fork{ \Let \chan = \recv {\chan_1'} in
    \Let \varB = \recv {\chan_1'} in
    \send \chan \varB;\ \send {\chan_1'} \TT};\\
  \Let {(\chan_2, \chan_2')} = \newchan in \\
  \Fork{ \Let {\var} = \recv {\chan_2'} in {\send {\chan_2'} (\var + 2 )} }; \\
  \send {\chan_1} {\chan_2};\ \send {\chan_1} {40};\ \recv {\chan_1};\ \recv {\chan_2}
\end{array}
\]
This program uses the channel pair $\chan_2,\chan_2'$ to exchange the number $40$ with the second forked-off thread, which adds $2$ to it, and sends it back.
Contrary to the programs we have seen before, it uses the
additional channel pair $\chan_1,\chan_1'$ to delegate the endpoint
$\chan_2$ to the first forked-off thread, which then sends the number over $\chan_2$.
While this program is intricate, the following \pname describe the communication concisely:
\begin{align*}
\interp{\chan_1{}&}{
  \begin{array}[t]{@{} l @{}}
  \sendprot {(\chan:\Val)} \chan {\textcolor{darkred}{
    \interp \chan {
      \sendprot {(\var:\integer)} \var \TRUE {
      \recvprot {} {\var+2} \TRUE \protend}
    }}} \\
  \sendprot {(\varB:\integer)} \varB {\TRUE}
  \recvprot {} \TT {\textcolor{darkred}{
    \interp \chan {\recvprot {} {\varB+2} \TRUE \protend}
  }} \protend
  \end{array}} \\
\interp{\chan_2{}&}{
  \sendprot {(\var:\integer)} \var \TRUE
  \recvprot {} {\var + 2} \TRUE \protend}
\end{align*}

\noindent
The first protocol states that the initial value sent must be a channel
endpoint $\chan$ with the protocol used in ${\mathit prog}_1$.
This means that the main thread must give up ownership of the channel endpoint
$\chan_2$, thereby delegating it.
The first protocol then expects a value $\varB$ to be sent,
and finally to receive a notification $\TT$, along with ownership of the channel
$\chan_2$, which has since taken one step by sending $\varB$.
Note that $\chan$ is of the type $\Val$ of programming language values
due to the programming language being untyped.

Lastly, the dependencies in \pname are not limited to first-order data, but can also be used
in combination with functions.
For example:
\[
{\mathit prog}_6\langdef
\begin{array}[t]{@{} l}
  \Let {(\chan,\chan')} = \newchan in \\
  \Fork{\Let{f} = \recv{\chan'} in \send {\chan'} {(\Lam \_. f\TT+2)}}; \\
  \Let \locvar = \newref{40} in \send{\chan}{(\Lam \_. \deref \locvar)};\ \recv {\chan}\ \TT
\end{array}
\]

\noindent
This program exchanges a value to which $2$ is added,
but postpones the evaluation by wrapping the computation in a closure.
The following protocol is used to verify this program:
\begin{align*}
\interp{\chan}{
  \begin{array}[t]{@{} l}
  \sendprot{(\iprop\,\ipropB : \iProp)\,(f:\Val)}{f}
    {\textcolor{darkred}{\hoare \iprop {f\;\TT} {\Ret \val.\val \in \integer * \ipropB(\val)}}}
  \\
  \recvprot{(g:\Val)}{g}
    {\textcolor{darkred}{\hoare \iprop {g\;\TT} {\Ret \val. \Exists \valB. (\val = \valB + 2) * \ipropB(\valB)}}}
    \protend
  \end{array}}
\end{align*}

\noindent
The send constructor ($\SEND$) does not just bind the function value
$f$, but also the precondition $\iprop$ and postcondition $\ipropB$ of its
Hoare triple.
In the second message, a Hoare triple is returned that maintains the original
pre- and postconditions, but returns an integer of two higher.
To send the function, the main thread would let $\iprop \eqdef \loc \mapsto 40$ and $Q(\val) \eqdef (\val = 40)$, and prove
$\hoare { \iprop } {(\Lam \_. \deref \loc)\ \TT} { \ipropB }$.
This example demonstrates that the state space of \pname can be
higher-order---it is indexed by the precondition $\iprop$ and postcondition
$\ipropB$ of $f$---which means that they do not have to be agreed upon
when creating the protocol, masking the internals of the function from the forked-off thread.

It is worth noting that using dependent recursive protocols it is possible to
keep track of a history of what actions have been performed, which, as is shown in
\Cref{sec:integration},
is especially useful when combining channels with locks.

\subsection{\lname 2.0: Subprotocols}
\label{sec:subprotocol_intro}
While \lname 1.0's notion of \pname is expressive enough to specify advanced
exchanges, as indicated by the examples in the previous section,
they can only reason about interactions that are strictly dual.
In particular, the dual nature of \lname 1.0 requires that:
\begin{itemize}
\item Sends ($\sendprotHead {\xdots} \val \iprop$)
  are matched up with receives ($\recvprotHead {\xdots} \val \iprop$), and \viceversa,
\item The \binders $\xdots$ of matched sends and receives are the same, and,
\item The propositions $\iprop$ of matched send and receives are the same.
\end{itemize}
Reasoning about programs with a more relaxed duality principle
has been studied in the session type community,
namely in the context of \emph{asynchronous session subtyping}
\cite{mostrous-ESOP2009,mostrous-IaC2015}.
A subtyping relation $\ksubtype{\stype_1 \!}{\stype_2}$ captures that the session type $\stype_2$ can be used in place of $\stype_1$
when type checking a program.
Channel endpoints are then allocated with strictly dual session types, after which
either side can be weakened based on the subtyping relation.
For one, the subtyping relation captures that sends can be swapped \emph{ahead of} receives
$\ksubtype{\strecv{\tvar}\stsend{\tvarB}\stype}
{\stsend{\tvarB}\strecv{\tvar}\stype}$.
Swapping sends ahead of receives is safe to do, as the messages are simply enqueued
into the corresponding channel buffer earlier than necessary.
The following program illustrates such a non-dual yet safe interaction:
\[
{\mathit {prog}_7}\langdef
\begin{array}[t]{@{} l}
\Let {(\chan,\chan')} = \newchan in \\
\Fork{\send{\chan'}{20};\
      \send {\chan'} ({\recv {\chan'} + 2})}; \\
\send \chan 20; \\
\Let \var = {\recv \chan} in\\
\Let \varB = {\recv \chan} in \var + \varB
\end{array}
\]
Here, both threads first send the value 20, which is enqueued into both of the
channel buffers, after which they receive the value of the other thread.
After this, they follow a dual behaviour, where the forked-off thread sends a value,
which the main thread receives.

In this paper, we show that \pname are compatible with the idea of asynchronous
session subtyping.
This gives rise to \textbf{\lname 2.0}, which supports
so-called \emph{subprotocols}.
Subprotocols are formalised by a preorder $\subprot{\prot_1}{\prot_2}$,
which captures (among others) a notion of swapping sends ahead of receives
(provided that the send does not depend on the \binders of the receive).
We can prove that $\mathit{prog}_7$ results in $42$ by picking the
following \pname:
\[
\begin{array}{@{} r @{} l l @{}}
\interp{\chan &{}}{
\sendprot{(\var:\integer)}{\var}{\TRUE}
\recvprot{}{20}{\TRUE}
\recvprot{}{\var+2}{\TRUE}
\protend} & \mathtt{and}\\
\interp{\chan' &{}}{
\recvprot{(\var:\integer)}{\var}{\TRUE}
\sendprot{}{20}{\TRUE}
\sendprot{}{\var+2}{\TRUE}
\protend}
\end{array}
\]
While the main thread satisfies the protocol of $\chan$ immediately, the
forked-off thread does not satisfy the protocol of $\chan'$, as it sends the
first
value before receiving.
However, it is possible to weaken the protocol of $\chan'$ using \lname 2.0's
notion of subprotocols:
\[\begin{array}{@{} l @{\ } l}
\subprot
{&
\color{darkgreen}\recvprot{(\var:\integer)}{\var}{\TRUE}
\color{darkred}\sendprot{}{20}{\TRUE}
\color{black}\sendprot{}{\var+2}{\TRUE}
\protend\\}
{&
\color{darkred}\sendprot{}{20}{\TRUE}
\color{darkgreen}\recvprot{(\var:\integer)}{\var}{\TRUE}
\color{black}\sendprot{}{\var+2}{\TRUE}
\protend}
\end{array}\]
This gives
$\interp{\chan'}{
\sendprot{}{20}{\TRUE}{
\recvprot{(\var:\integer)}{\var}{\TRUE}{
\sendprot{}{\var+2}{\TRUE}
\protend}}}
$.
Since the first send (with value 20) is independent of the variable
$\var$ bound by the receive, the subprotocol relation follows immediately from
the swapping property.
Note that it is \emph{not} possible to swap the second send
(with value $\var+2$) ahead of the receive, as it does in fact depend on variable
$\var$ bound by the receive.

In addition to allowing the verification of a larger class of programs,
\lname 2.0's subprotocols also provide a more extensional approach to reasoning about \pname.
This is beneficial whenever we want to reuse existing specifications that might
use a syntactically different protocol, but that nonetheless logically
entail each another.
For example, the ordering of \binders can be changed using the subprotocol
relation:
\[
  \sendprot{(\var:\integer) (\varB:\integer)}{(\var,\varB)}{\TRUE}\prot
  \subprotop
  \sendprot{(\varB:\integer) (\var:\integer)}{(\var,\varB)}{\TRUE}\prot
\]
Since the subprotocol relation is a first-class logical proposition of
Actris 2.0, it also allows the manipulation of separation logic resources,
such as moving in ownership.
For example,
we can show the following \emph{conditional} subprotocol relation:
\[
\begin{array}[b]{@{} l @{}}
\loc_1' \mapsto 20 \  \wand \\
  \sendprot {(\loc_1,\loc_2:\Loc)} {(\loc_1,\loc_2)}
    {\loc_1 \mapsto 20 * \loc_2 \mapsto 22} \prot
    \subprotop
   \sendprot {(\loc_2:\Loc)} {(\loc_1',\loc_2)}
    {\loc_2 \mapsto 22} \prot
\end{array}
\]
Here, we move the ownership of $\loc_1' \mapsto 20$ into the protocol,
to resolve the eventual obligation of sending it, while instantiating the
\binder $\loc_1$ with $\loc_1'$.

In addition to the demonstrated features, in the rest of this paper we show
that \lname 2.0's subprotocol relation is capable of moving resources from one
message to another.
This gives rise to a principle similar to \emph{framing}, known from conventional
separation logic, but applied to \pname.
Lastly, inspired by the work of Brandt and Henglein~\cite{brandt-1998FI}, the
subprotocol relation is defined coinductively, allowing us to use the principle of L\"ob induction
to prove subprotocol relations for recursive protocols.

\subsection{Formal correspondence to session types}

Even though \lname's notion of \pname is influenced by binary session types, this paper does not
provide a formal correspondence between the two systems.
However, since \lname is built on top of Iris, it forms a suitable foundation
for building logical relation models of type systems.
In related work by Hinrichsen \etal~\cite{actris-logrel}, \lname has been used to define a logical
relations model of binary session types, with support for various forms of
polymorphism and recursion, asynchronous subtyping, references, and locks/mutexes.
Similar to RustBelt \cite{jung-POPL2018,jung-CACM2021}, the work by Hinrichsen \etal~\cite{actris-logrel} gives rise
to an extensible approach for proving type safety, which can be
used to manually prove the typing judgements of racy, but safe, programs that cannot
be type checked using only the rules of the type system.

\subsection{Contributions and outline}

This paper introduces \textbf{\lname 2.0}: a higher-order impredicative concurrent
separation logic built on top of the Iris framework for reasoning about
functional correctness of programs with asynchronous message-passing that combine
higher-order functions, higher-order references, fork-based concurrency, and locks.
Concretely, this paper makes the following contributions:

\begin{itemize}[align=left] 
\item We introduce \emph{\pname} inspired by affine binary session types to
  model the transfer of resources (including higher-order functions) between
  channel endpoints.
  We show that they can be used to handle choice, recursion, and delegation
  (\Cref{sec:language,sec:iris_logic,sec:actris_logic,sec:tour}).
\item We introduce \emph{subprotocols} inspired by asynchronous session
  subtyping.
  This notion relaxes duality, allowing channels to send
  messages before receiving others, and gives rise to a more extensional
  approach to reasoning about \pname, providing more flexibility in the
  design and reuse of protocols.
  We moreover show how L\"ob induction is used to reason about recursive subprotocols
  (\Cref{sec:subprotocols}).
\item We demonstrate the benefits obtained from building \lname on top of Iris
  by showing how Iris's support for ghost state and locks can be used to prove
  functional correctness of programs using manifest
  sharing, \ie channel endpoints shared by multiple parties (\Cref{sec:integration}).
\item We provide a case study on \lname and its mechanisation in Coq by proving
  functional correctness of a variant of the
  map-reduce model by Dean and Ghemawat~\cite{dean-OSDI2004} (\Cref{sec:map_reduce}).
\item We give a model of \pname in the Iris framework to prove
  safety and postcondition
  validity of our Hoare triples (\Cref{sec:model}).
\item We provide a full mechanisation of \lname~\cite{actris_coq}
  using the interactive theorem prover Coq.
  On top of our Coq mechanisation, we provide custom tactics,
  which we use to mechanise all examples in the paper (\Cref{sec:coq}).
\end{itemize}

\subsection{Differences from the conference version}

This paper is an extension of the paper
``Actris: Session-type based reasoning in separation logic'' presented at
the POPL'20 conference~\cite{hinrichsen-POPL2020}.
In this paper we present \lname 2.0, which extends \lname 1.0 with the notion of
subprotocols.
This extension introduces new logical connectives and proof rules, but also
involves a significant overhaul of the original model and
its Coq mechanisation.
We extend the presentation of the programming language semantics, model and mechanisation
substantially, with additional details, considerations, and examples, to give a better
understanding of how \lname works and how it can be used.
Concretely, this paper includes the following extensions compared to the
conference version:

\begin{itemize}[align=left] 
\item An overview of subprotocols in the introduction (\Cref{sec:subprotocol_intro}).
\item A new background section on the programming language semantics (\Cref{sec:language}) and Iris (\Cref{sec:iris_logic}).
\item A section with an expanded overview of \lname (\Cref{sec:actris_logic}, moved from \Cref{sec:tour}).
\item A new section on \lname 2.0's notion of subprotocols
  (\Cref{sec:subprotocols}).
\item An updated and expanded description of the model of \lname in
  Iris (\Cref{sec:model}).
\item An extension of the section on the Coq mechanisation with sample proofs
  (\Cref{sec:coq}).
\end{itemize}


\newcommand{\heaplangsemfig}{
\begin{figure}
  \textbf{Call-by-value evaluation contexts:}
  \begin{align*}
  \lctx \in \Lctx \bnfdef{}&
    \bullet \mid
    \expr\ \lctx \mid
    \lctx\ \val \mid
    (\expr_1, \lctx) \mid (\lctx, \val_2) \mid \Fst\ (\lctx) \mid \Snd\ (\lctx) \mid
    \\ &
    \If \lctx then {\expr_1} \Else {\expr_2} \mid
    \Inj 1 (\lctx) \mid \Inj 2 (\lctx) \mid \\ &
    (\Match {\lctx} with {(\Inj 1 {\var})} => {\expr_2}
                           | {(\Inj 2 {\var})} => {\expr_3} end) \mid \\ &
    \newref(\lctx) \mid
    \deref \lctx \mid
    \expr \gets \lctx \mid
    \lctx \gets \val \mid & \hspace{-4em} \text{(Mutable state)}
    \\ &
    \CAS\ \expr_1\ \expr_2\ \lctx \mid
    \CAS\ \expr_1\ \lctx\ \val_2 \mid
    \CAS\ \lctx\ \val_1\ \val_2 \mid
    \ldots & \hspace{-4em} \text{(Concurrency)}
  \end{align*}
  
  \medskip
  \textbf{Head reductions of \heaplang:}
  \begin{align*}
    (\cfg{(\Rec f \var = \expr)(\val)}{\istate})
    \quad \hstep & \quad
    (\cfg{\subst{\subst{\expr}{\var}{\val}}{f}{(\Rec f \var = \expr)}}
     {\istate};\listnil)\\
    (\cfg{\Fst\ (\val_1, \val_2)}{\istate})
    \quad \hstep & \quad
    (\cfg{\val_1}{\istate};\listnil) \\
    (\cfg{\Snd\ (\val_1, \val_2)}{\istate})
    \quad \hstep & \quad
    (\cfg{\val_2}{\istate};\listnil) \\
    (\cfg{\If \True then {\expr_1} \Else {\expr_2}}{\istate})
    \quad \hstep & \quad
    (\cfg{\expr_1}{\istate};\listnil)\\
    (\cfg{\If \False then {\expr_1} \Else {\expr_2}}{\istate})
    \quad \hstep & \quad
    (\cfg{\expr_2}{\istate};\listnil)\\
    \left(\cfg{\MatchMLT {(\Inj i \val)} with {(\Inj 1 {\var})} => {\expr_1}
                         | {(\Inj 2 {\var})} => {\expr_2} end {}}{\istate}\right)
    \quad \hstep & \quad
    (\cfg{\subst{\expr_i}{\var}{\val}}{\istate};\listnil)
    & \quad \text{if}\ i \in \set{ 1,2 } \\
    (\cfg{\newref{\val}}{\istate})
    \quad \hstep & \quad
    (\cfg{\loc}{\mapinsert{\loc}{\val}{\istate}};\listnil)
    & \quad \text{if}\ \istate(\loc) =\ \perp \\
    (\cfg{\deref\loc}{\mapinsert{\loc}{\val}{\istate}})
    \quad \hstep & \quad
    (\cfg{\val}{\mapinsert{\loc}{\val}{\istate}};\listnil) \\
    (\cfg{\loc \gets \valB}{\mapinsert{\loc}{\val}{\istate}})
    \quad \hstep & \quad
    (\cfg{\TT}{\mapinsert{\loc}{\valB}{\istate}};\listnil) \\
    (\cfg{\CAS\ \loc\ \val'\ \valB}{\mapinsert{\loc}{\val}{\istate}})
    \quad \hstep & \quad
    (\cfg{\True}{\mapinsert{\loc}{\valB}{\istate}};\listnil)
    & \quad \text{if}\ \val = \val' \\
    (\cfg{\CAS\ \loc\ \val'\ \valB}{\mapinsert{\loc}{\val}{\istate}})
    \quad \hstep & \quad
    (\cfg{\False}{\mapinsert{\loc}{\val}{\istate}};\listnil)
    & \quad \text{if}\ \val \neq \val' \\
    (\cfg{\Fork{\expr}}{\istate})
    \quad \hstep & \quad
    (\cfg{\TT}{\istate};\lbrack\expr\rbrack)
  \end{align*}
  
  \medskip
  \textbf{Thread-local and threadpool reductions of \heaplang:}
  \begin{mathpar}
    \infer
    {\cfg{\expr_1}{\istate_1} \hstep \cfg{\expr_2}{\istate_2};\vec{\expr}}
    {\cfg{\fillctx\lctx[{\expr_1}]}{\istate_1} \step
      \cfg{\fillctx\lctx[{\expr_2}]}{\istate_2};\vec{\expr}}
    \and
    \infer
    {\cfg{\expr_1}{\istate_1} \step \cfg{\expr_2}{\istate_2};\vec{\expr}}
    {\cfg{\tpool \cdot [\expr_1] \cdot \tpool'}{\istate_1}
      \tpstep
      \cfg{\tpool \cdot [\expr_2] \cdot \tpool' \cdot \vec\expr}{\istate_2}}
  \end{mathpar}
\caption{The operational semantics of \heaplang.}
\label{fig:heap_lang}
\end{figure}
}

\section{Programming language semantics}
\label{sec:language}

The Iris program logic is parametric in the programming language that is used.
As a result there are multiple approaches to extend Iris with support for channels:

\begin{itemize}
\item Instantiate Iris with a language that has native support
  for channels.
  This approach was carried out in the original Iris paper~\cite{jung-POPL2015} and by Tassarotti \etal~\cite{tassarotti-ESOP2017}.
\item Instantiate Iris with a language that has low-level concurrency
  primitives, but no native support for channels, and implement channels as a
  library in that language.
  This approach was carried out by Bizjak \etal~\cite{bizjak-PACMPL2019} for a lock-free
  implementation of channels.
\end{itemize}

\noindent
In this paper we take the second approach.
We implement channels in \heaplang---the default
programming language that is shipped with Iris's Coq development~\cite{iris_coq}.
\heaplang is an untyped functional language with high-level features such as
higher-order functions,
higher-order mutable references, fork-based concurrency, and garbage collection.
Due to these high-level features, programs written in \heaplang are reminiscent of those written in high-level programming languages with message passing like Go or Erlang.

Since \heaplang is an untyped language, safety of a program is not obtained by establishing a typing judgement, but by proving a Hoare triple in the Iris/\lname logic.
Hinrichsen \etal~\cite{actris-logrel} show how logical relations in \lname can be used to define and
prove sound a session type system for \heaplang extended with message passing.

We proceed by describing \heaplang's syntax (\Cref{sec:heaplang}) and operational
semantics (\Cref{sec:opsem}).
We then present \heaplang's standard library for spin locks (\Cref{sec:lock_library}).
We use this lock library to implement channels (\Cref{sec:channel_implementation}), and to write programs that
combine message passing with lock-based concurrency (\Cref{sec:integration}).

\subsection{Syntax}
\label{sec:heaplang}

The syntax of \heaplang is as follows:
\begin{align*}
\val \in \Val \bnfdef{}&
  \TT \mid
  i \mid
  b \mid
  \loc \mid
  \Rec f x = \expr \mid &
  \hspace{-5em}
  (i \in \integer, b \in \bool, \loc \in \Loc) \\ &
  (\val_1, \val_2) \mid
  \Inj 1 \val \mid
  \Inj 2 \val \mid
  \ldots \\
\expr \in \Expr \bnfdef{}&
  \val \mid
  \var \mid
  \expr_1\ \expr_2 \mid
  (\expr_1, \expr_2) \mid
  \Fst\ \expr \mid \Snd\ \expr \mid \\ &
  \If {\expr_1} then {\expr_2} \Else {\expr_3} \mid
  \Inj 1 {\expr} \mid \Inj 2 {\expr} \mid \\ &
  (\Match {\expr_1} with {(\Inj 1 {\var})} => {\expr_2}
                         | {(\Inj 2 {\var})} => {\expr_3} end) \mid \\ &
  \newref\expr \mid
  \deref \expr \mid
  \expr_1 \gets \expr_2 \mid &
  \hspace{-5em}
  \text{(Mutable state)} \\ &
  \Fork \expr \mid \CAS\ \expr_1\ \expr_2\ \expr_3 \mid
  \ldots &
  \hspace{-5em}
  \text{(Concurrency)}
\end{align*}

\noindent
We elide the standard boolean and arithmetic operators such as equality, addition,
subtraction, and multiplication.
We define various notions as syntactic sugar (\ie as definitions in the
meta language by use of $\mdef$):
\begin{align*}
  \Lam \var. \expr \mdef{} & \Rec \_ \var = \expr
  & \expr_1;\ \expr_2 \mdef{} & \Let {\_} = {\expr_1} in {\expr_2}\\
  \Let \var = \expr_1 in \expr_2 \mdef{} & (\Lam \var. \expr_2)\ \expr_1
  & \SkipN \mdef{} & \Rec {\goname} {\var} =
                   \begin{array}[t]{@{} l @{}}
                     \If {0 < \var}\\ then
                     {\goname\ (\var - 1)} \Else \TT
                   \end{array}
\end{align*}
We use $\_$ as the anonymous binder that is not used in the body of the binding expression.
The $\SkipN$ operation, which performs a given number of no-op program steps, is used in the implementation of channels (\Cref{sec:channel_implementation}) for proof-related reasons (explained in \Cref{sec:chan_own_model}).
We often write definitions as
$\fvar\ \var_1 \cdots\ \var_n \langdef \expr$ rather than
$\fvar \mdef \Rec f\ {\var_1 \cdots\ \var_n}\! = \expr$.
For example, we write
$\SkipN\ \var \langdef{} \If {0 < \var} then {\SkipN\ (\var - 1)} \Else \TT$.

\heaplang includes the usual operations for ML-style references.
New references can be allocated using $\newref \expr$,
dereferenced using $\deref \expr$,
and updated using $\expr_1 \gets \expr_2$.
Concurrency is supported via $\Fork \expr$, which spawns a new thread $\expr$
that is executed in the background.
The language also supports atomic operations like compare-and-set ($\CAS$),
which are used to implement lock-free data structures and
synchronisation primitives, such as the locks (\Cref{sec:lock_library}).
\heaplang is garbage collected and thus does not have a deallocation operation.

\subsection{Operational semantics}
\label{sec:opsem}

\heaplangsemfig

The small-step operational semantics of \heaplang is presented in \Cref{fig:heap_lang}.
The type of program states $\State$ is defined as:
\[
\istate \in \State \eqdef \Loc \fpfn \Val
\]
That is, program states are finite partial maps from allocated locations to their stored values.

The \emph{head reduction} $(\cfg{\expr_1}{\istate_1} \hstep \cfg{\expr_2}{\istate_2};\vec\expr)$
describes how an expression $\expr_1 \in \Expr$ in an initial program state $\istate_1 \in \State$ reduces to a new expression $\expr_2 \in \Expr$ in a possibly updated program state $\istate_2 \in \State$.
Additionally, it keeps track of a list of newly spawned threads $\vec \expr \in \List\ \Expr$.
The reduction rule
$(\cfg{\Fork{\expr}}{\istate}) \hstep (\cfg{\TT}{\istate};\lbrack\expr\rbrack)$
describes how a new thread $\expr$ is spawned by adding it to the list of newly spawned threads
$\lbrack\expr\rbrack$.
Conversely, the list of newly spawned threads is empty for all of the other
reduction rules.

The \emph{thread-local reduction} $(\cfg{\expr_1}{\istate_1} \step
\cfg{\expr_2}{\istate_2};\vec{\expr})$ lifts the head reduction to whole expressions.
It decomposes the initial expression $\expr_1$ into $\fillctx{\lctx}[\expr_1']$, where $\lctx$ is a \emph{call-by-value evaluation context} \cite{felleisen_hieb} and a head expression $\expr_1'$.
The head expression $\expr_1'$ is then reduced, using $(\cfg{\expr_1'}{\istate_1} \hstep \cfg{\expr_2'}{\istate_2};\vec\expr)$, and the final expression $\expr_2$ is set to $\fillctx{\lctx}[\expr_2']$.
Evaluation contexts (shown in \Cref{fig:heap_lang})
provide a deterministic reduction order of sub-expressions.
\heaplang reduces right-to-left, meaning that in expressions such as
${\expr_1} \gets {\expr_2}$ the expression $\expr_2$ reduces before $\expr_1$.
This is determined by the corresponding evaluation contexts $\expr \gets \lctx$ and
$\lctx \gets \val$, which state that we only evaluate sub-expressions of the
target location, once the term to store is a value.
More precisely, we would initially get
$\fillctx{(\expr_1 \gets \bullet)}[\expr_2]$.
If $\expr_2$ reduces to a value $\val_2$ the context syntax dictates that
the hole then moves to $\expr_1$ yielding
$\fillctx{(\bullet \gets \val_2)}[\expr_1]$.
If $\expr_1$ reduces to a value $\val_1$ we finally end up with
the expression $\val_1 \gets \val_2$, as there is no context syntax where both constituents
are values, and this expression can be reduced using a standard head reduction.

Finally, the \emph{threadpool reduction} $(\cfg{\vec\expr_1}{\istate_1} \tpstep \cfg{\vec{\expr_2}}{\istate_2})$ is the top-level reduction relation that describes the interleaving of threads.
It describes how a concurrently running list of threads $\vec\expr_1$,
in an initial program state $\istate_1$,
reduce to a new list of threads $\vec{\expr_2}$ in an updated program state $\istate_2$.
At each step a thread $\expr_1$ is picked non-deterministically from $\vec\expr_1$
and reduced one step to $\expr_2$ via the thread-local reduction $(\cfg{\expr_1}{\istate_1} \step \cfg{\expr_2}{\istate_2};\vec{\expr})$.
The final list of threads $\vec{\expr_2}$ is obtained from $\vec{\expr_1}$ by replacing the expression $\expr_1$ with $\expr_2$ and appending the list $\vec\expr$ of newly spawned threads to the end.

We refer the interested reader
to Iris Development Team~\cite[\texttt{docs/heap\_lang.md}]{iris_coq}
for more details on the semantics of \heaplang,
and to Jung \etal~\cite[\S 6.1]{jung-JFP2018}
for details on the language-parametric aspects of Iris.

\subsection{Implementation of locks}
\label{sec:lock_library}

\begin{figure}
\begin{align*}
  \newlock \langdef{} & \newref \False \\
  \tryacquire{\lockvar} \langdef{} & \CAS\ \lockvar\ \False\ \True\\
  \acquire{\lockvar} \langdef{} &
  \If (\tryacquire\ \lockvar) then \TT \Else \acquire{\lockvar}\\
  \release{\lockvar} \langdef{} & \lockvar \gets \False
\end{align*}
\caption{Implementation of locks in \heaplang.}
\label{fig:lock_implementation}
\end{figure}

Using \heaplang it is possible to implement various kinds of locks/mutexes.
We consider the simplest kind of lock---a spin lock---whose implementation from the \heaplang standard library is shown in \Cref{fig:lock_implementation}.

A spin lock implemented using a reference to a boolean, which is $\False$ if the lock is unlocked, and $\True$ if the lock is locked.
The $\newlock$ operation creates a new lock $\lockvar$, which is initially unlocked (\ie $\False$).
The operation $\acquirelock \lockvar$ will atomically (using compare-and-set)
take the lock, or loop if the lock is already taken.
The $\releaselock \lockvar$ operation releases the lock so that it may be acquired
by other threads.

\subsection{Implementation of channels}
\label{sec:channel_implementation}

Following the literature on asynchronous session types, the message-passing semantics of our channels is
\emph{binary} (communication is between two parties),
\emph{asynchronous} (sending messages does not block),
\emph{bidirectional} (messages can be in transit in both directions simultaneously),
\emph{reliable} (messages are never dropped), and
\emph{order preserving} (messages always arrive in the order that they were sent).

\newcommand{\chanencfig}{
  \begin{figure}
    \begin{align*}
      \newchan \langdef &
        \begin{array}[t]{l}
          \Let (\listvar,\listvarB,\lockvar) = (\llistnil\ \TT, \llistnil\ \TT, \newlock) in \\
          ((\listvar,\listvarB,\lockvar),(\listvarB,\listvar,\lockvar))
        \end{array}\\
      \send{\chan}{\val} \langdef &
        \begin{array}[t]{l}
          \Let (\listvar,\listvarB,\lockvar) = \chan in\\
          \acquire{\lockvar};\\
          \quad \llistsnoc\ l\ \val;\\
          \quad \SkipN\ (\llistlength\ r);\\
          \release{\lockvar}
        \end{array}\\
      \tryrecv{\chan} \langdef &
        \begin{array}[t]{l}
          \Let (\listvar,\listvarB,\lockvar) = \chan in\\
          \acquire{\lockvar};\\
          \quad \Let \mathit{ret} = (\If (\llistisnil\ \listvarB) then {(\none)} \Else
          (\some{(\llistpop\ \listvarB)}))
          in\\
          \release{\lockvar};\ \mathit{ret}\\
        \end{array}\\
      \recv{\chan} \langdef & \
          \MatchML (\tryrecv{\chan}) with
          \none => \recv{\chan}
          | \some{\val} => \val
          end \
    \end{align*}
  \caption{Implementation of bidirectional channels in \heaplang.}
  \label{fig:channel_implementation}
\end{figure}
}
\chanencfig

The implementation of our channels in \heaplang is displayed in
\Cref{fig:channel_implementation}.
It uses locks (\Cref{sec:lock_library}) and a linked list library.
This list library provides functions for creating a empty list ($\llistnil$), testing
if a list is empty ($\llistisnil$),
computing the length of a list ($\llistlength$),
adding an element to the back ($\llistsnoc$),
and popping an element of front ($\llistpop$).
The last two functions mutate the list, instead of creating a copy.
The implementation of the list library is standard, and hence elided.

Intuitively, the channels can be thought of as a pair of buffers
$(\vec\val_1,\vec\val_2)$ of unbounded size.
The $\newchan$ operation creates a new channel
whose buffers are empty, and returns a tuple of endpoints $(\chan_1,\chan_2)$.
Bidirectionality is obtained by having one endpoint receive from the others
send buffer and \viceversa.
As such, the $\send {\chan_i} \val$ operation enqueues the value $\val$ in its own
buffer, \ie $\vec\val_i$,
and the $\recv {\chan_i}$ operation dequeues a value from the other buffer,
\ie from $\vec\val_{2}$ if $i=1$ and from $\vec\val_{1}$ if $i=2$.
The message passing is asynchronous, as $\send \chan \val$ will always
reduce, while $\recv \chan$ will loop as long as the receiving buffer is empty.

More specifically, the \newchanname function creates new channels
by allocating two empty mutable linked lists $\listvar$ and $\listvarB$ using
$\llistnil\ \TT$,
along with a lock $\lockvar$ using $\newlock$, and returns the tuples
$(\listvar,\listvarB,\lockvar)$ and $(\listvarB,\listvar,\lockvar)$,
where the order of the linked
lists $l$ and $r$ determines the side of the endpoints.
We refer to the list in the left position as the endpoint's own buffer,
and the list in the right position as the other endpoint's buffer.

The \sendname function sends a value $\val$ over a given channel endpoint
$(\listvar,\listvarB,\lockvar)$, by enqueueing it in the $\listvar$ buffer.
The function operates in an atomic fashion by first acquiring the lock via
$\acquire{\lockvar}$, thereby entering the critical section, after which the value
is enqueued (\ie appended to the end) of the endpoint's own buffer using
the function $\llistsnoc\ \listvar\ \val$.
The $\SkipN\ (\llistlength\ r)$ instruction is a no-op that is inserted to
aid the proof.
We come back to the reason why this instruction is needed
in \Cref{sec:chan_own_model}.

The \recvname function receives a value over a channel endpoint
$(\listvar,\listvarB,\lockvar)$, by dequeueing the first value in the $\listvarB$
buffer.
It does so by performing a loop that repeatedly calls the helper function
\tryrecvname.
This helper function attempts to receive a value atomically, and fails if there
is no value in the other endpoint's buffer.
The function \tryrecvname acquires the lock with
$\acquire{\lockvar}$, and then checks whether the other endpoint's buffer is empty
using $\llistisnil\ \listvarB$.
If it is empty, nothing is returned (\ie $\none$), while otherwise the value is
dequeued and returned (\ie $\some{(\llistpop\ \listvarB)}$).

Throughout the paper, we often use a combined operation for starting a thread and creating a channel between the parent and child thread:
\[
\start{f}
  \langdef
  \Let{(\chan,\chan')} = \newchan in \Fork{f\ \chan'};\ \chan
\]

\section{The Iris logic}
\label{sec:iris_logic}

\newcommand{\irisfig}{
\begin{figure}
\centering
\textbf{Grammar:}
\begin{align*}
\type, \typeB \bnfdef{} &
  \var \mid 0 \mid 1 \mid \bool \mid \nat \mid \integer \mid \Type \mid
  \All \var : \type.\typeB \mid \\
 & \Loc \mid \Val \mid \Expr \mid \iProp \mid \List\ \type \mid \ldots \\
\term , \termB , \iprop , \ipropB \bnfdef{}&
   \var \mid \Lam \var : \type . \term \mid
    \term (\termB) \mid \term (\type) \mid
    \tag{Polymorphic lambda-calculus}
    \\
  & \TRUE \mid \FALSE \mid \iprop \land \ipropB \mid
     \iprop \vee \ipropB \mid \iprop \Ra \ipropB \mid
     \tag{Propositional logic}
     \\
  & \All \var : \type. \iprop \mid \Exists \var : \type. \iprop \mid
     \term = \termB \mid
     \tag{Higher-order logic with equality}
     \\
  & \iprop * \ipropB \mid \iprop \wand \ipropB \mid \loc \mapsto \val \mid
     \hoare{\iprop}{\expr}{\Ret{\val}. \ipropB} \mid
     \tag{Separation logic}
     \\
  & \MU \var : \type . \term \mid \later \iprop \mid \ldots
    \tag{Guarded recursion and step indexing}
\end{align*}

\bigskip
\textbf{Basic affine separation logic:}
\begin{mathpar}
\inferH{Affine}{\iprop * \ipropB}{\iprop}
\and
\inferH{Ht-frame}
  {\hoare{\iprop}\expr{\Ret\valB. \ipropB}}
  {\hoare{\iprop * \ipropC}\expr{\Ret\valB. \ipropB * \ipropC}}
\and
\axiomH{Ht-val}
  {\hoare\TRUE\val{\Ret \valB. \valB = \val}}
\and
\inferH{Ht-fork}
  {\hoare{\iprop}{\expr}{\TRUE}}
  {\hoare{\iprop}{\Fork\expr}{\Ret \valB. \valB = \TT}}
\and
\inferrule*[lab=\textlabel{Ht-bind}{Ht-bind},right=\textnormal{$\lctx$ a call-by-value evaluation context}]
  {\hoare{\iprop}{\expr}{\Ret\val. \ipropB} \and
    \All \val. \hoare{\ipropB}{\fillctx \lctx[\val]}{\Ret\valB. \ipropC}}
  {\hoare{\iprop}{\fillctx \lctx[\expr]}{\Ret\valB. \ipropC}}
\end{mathpar}

\bigskip
\textbf{Heap manipulation:}
\begin{mathpar}
\axiomH{Ht-alloc} {\hoare{\TRUE}{\newref{\val}}{\Ret \loc. \loc \mapsto \val}}
\and
\axiomH{Ht-load} {\hoare{\loc \mapsto \val}{\deref\loc}
  {\Ret \valB. (\valB = \val) * \loc \mapsto \val}}
\and
\axiomH{Ht-store} {\hoare{\loc \mapsto \val}{\loc\gets\valB}{\loc \mapsto \valB}}
\end{mathpar}

\bigskip
\textbf{Guarded recursion and step indexing:}
\begin{mathpar}
\inferH{Ht-rec}
  {\hoare{\iprop}{\subst {\subst \expr x \val} f {\Rec f x = \expr}}{\Ret \valB. \ipropB}}
  {\hoare{\later\iprop}{(\Rec f x = \expr)\; \val}{\Ret \valB. \ipropB}}
\and
\inferhref{$\later$-intro}{Later-intro}
  {\iprop}
  {\later\iprop}
\and
\inferhref{$\later$-mono}{later-mono}
  {\iprop \wand \ipropB}
  {\later \iprop \wand \later \ipropB}
\and
\inferhref{L\"ob}{Loeb}
  {\later\iprop\Ra\iprop}
  {\iprop}
\and
\inferhref{$\mu$-unfold}{rec-unfold}
  {}
  {(\MU \var.\term) = \subst \term \var {\MU \var.\term}}
\end{mathpar}

\caption{The grammar and a selection of rules of Iris.}
\label{fig:iris_logic}
\end{figure}
}

We give a brief introduction to the features of Iris that play an important
role in \lname:
its support for basic separation logic (\Cref{sec:iris:separation}),
higher-order impredicative separation logic (\Cref{sec:iris:ho-impredicative}),
guarded recursion and step-indexing (\Cref{sec:iris:step-index}),
and Iris's adequacy theorem (\Cref{sec:iris:adequacy}).
This section does not present new material, so readers that are already
familiar with Iris can skip it.
An extensive overview of Iris can be found in \cite{jung-JFP2018}, and a
tutorial-style introduction can be found in \cite{lecturenotes}.

\irisfig

\subsection{Basic separation logic}
\label{sec:iris:separation}

Propositions in separation logic describe ownership of resources, and can thus intuitively be thought of as predicates over resources.
The propositions of Iris $\iprop,\ipropB \in \iProp$ range over an extensible set of resources,
which includes the program state.
Iris is a higher-order separation logic, so it has the usual
logical connectives such as conjunction ($\iprop \land \ipropB$), implication
($\iprop \Ra \ipropB$), universal
($\All {\var\compactcolon\type}. \iprop$) and existential
($\Exists {\var\compactcolon\type}. \iprop$) quantification, as well as
the connectives of separation logic:

\begin{itemize}
\item The \emph{points-to connective} $(\loc \mapsto \val)$ asserts exclusive
  resource ownership of a location $\loc \in \Loc$ in the program state,
  stating that it holds
  the value $\val \in \Val$.
\item The \emph{separating conjunction} ($\iprop * \ipropB$) states that $\iprop$ and
  $\ipropB$ holds for disjoint sets of resources.
\item The \emph{separating implication}
  ($\iprop \wand \ipropB$) states that by giving up ownership of the resources
  described by $\iprop$, we obtain ownership of the resources described by $\ipropB$.
  Separating implication is used similarly to
  implication since ($\iprop$ entails $\ipropB \wand \ipropC$)
  iff ($\iprop * \ipropB$ entails $\ipropC$).
\item The \emph{Hoare triple} $\hoare \iprop \expr {\Ret \valB. \ipropB}$
  states that if the initial program state satisfies the precondition $\iprop$,
  then (1) the expression $\expr$ is safe (\ie does not go wrong),
  and, (2) if $\expr$ reduces to a value $\val$, then the final program
  state satisfies the postcondition $\subst \ipropB \valB \val$.
  We often omit the binder $\valB$ in the postcondition if the result is the
  unit value $\TT$.
\end{itemize}

We say that an Iris proposition $\iprop$ \emph{is valid} iff it holds for all
resources, \ie $\iprop$ is valid iff $\TRUE$ entails~$\iprop$.
Note that $\iprop \wand \ipropB$ is valid iff $\iprop$ entails $\ipropB$, so
we often use the separating implication ($\wand$) in place of entailment. 
For readability, we use inference-style rules to denote
separation logic rules
$(\iprop_1 \ast \dotsb \ast \iprop_n) \wand \ipropB$ as:
\[
  \infer{\iprop_1 \and \dots \and \iprop_n}{\ipropB}
\]
Iris is an affine separation logic, which means that propositions are upwards
closed in the resources, \ie $\iprop * \ipropB$ entails $\iprop$ (rule
\ruleref{Affine}).
Affinity matches up with the use of a garbage-collected
programming language---one can simply dispose of an unused points-to connective
$\loc \mapsto \val$ using rule \ruleref{Affine} when a location $\loc$ is no
longer referenced.

While many propositions of separation logic assert exclusive ownership of
resources (\eg $\loc \mapsto \val$), others do not (\eg $\term = \termB$).
Propositions that do not assert exclusive ownership enjoy some useful laws.
Separation conjunction ($\iprop * \ipropB$) is logically equivalent to
regular conjunction ($\iprop \land \ipropB$) if at least one conjunct does not
assert exclusive ownership, and separating implication
($\iprop \wand \ipropB$) is logically equivalent to regular implication
($\iprop \Ra \ipropB$) if the premise $\iprop$ does not
assert exclusive ownership.%
\footnote{Formally, these equivalences hold for the class of
\emph{persistent} propositions, see \cite[\S 2.3]{jung-JFP2018}.}
For example, $(\term = \termB) * \ipropB$ and $(\term = \termB) \land \ipropB$
are logically equivalent.
Since separating conjunction/implication is omnipresent in Iris, we prefer the
use of separating conjunction/implication over regular conjunction/implication
if both can be used.
This is also the convention used in the Iris Coq development.

Iris's notion of resources is not limited to locations in the program state
(\ie $\loc \mapsto \val$), but can
be extended with user-defined \emph{ghost} resources.
We use ghost resources to define \lname's connective
$\interp{\chan}{\prot}$ for exclusive ownership of the channel endpoint $\chan$
with protocol $\prot$ (\Cref{sec:model}), and
to reason about programs with non-trivial sharing (\Cref{sec:integration}).

The rules for Hoare triples are mostly standard, but it is worth pointing out
the rule for \ruleref{Ht-bind}.
This rule enables reductions of an expression
$\expr$, in some evaluation context $\lctx$, based on the precedence enforced by
the evaluation contexts presented in \Cref{sec:opsem}.

\subsection{Higher-order impredicative separation logic}
\label{sec:iris:ho-impredicative}

The Iris logic is:

\begin{itemize}
\item \emph{Higher-order:}
  Using Iris's quantifiers $\All {\var\compactcolon\type}. \iprop$ and
  $\Exists {\var\compactcolon\type}. \iprop$ it is not only possible to quantify over
  first-order types (like $\integer$ and $\List\ \integer$), but over any type, including
  functions (like $\integer \to \integer$),
  higher-order functions (like $(\integer \to \integer) \to \integer$),
  polymorphic functions (like $\All \tvar. \List\ \tvar \to \nat$),
  Iris propositions ($\Prop$), and Iris predicates (like $\integer \to \Prop$).
\item \emph{Impredicative:}
  Iris's logical connectives can be nested arbitrarily.
  Notably, $\All \iprop : \Prop. \ipropB$ is an Iris proposition, and not an Iris
  proposition in a higher universe.
  Similarly, Hoare triples
  $\hoare \iprop \expr {\Ret \val. \ipropB}$ and other Iris connectives
  like $\islock \lockvar \ipropC$ for lock ownership (\Cref{sec:integration})
  are first-class Iris propositions themselves.
\end{itemize}

As we will see in this paper, \lname expands on Iris's support
for higher-order impredicative separation logic by allowing the variables
$\xdots$ in the \pname $\sendprot {\xdots} \val \iprop \prot$ and
$\recvprot {\xdots} \val \iprop \prot$ to range over any type (including
\lname's type of protocols $\iProto$), and the proposition $\iprop$ to contain
any Iris/Actris connective (including the \lname connective $\interp \chan \prot$
for channel ownership).
This is particularly useful to reason about message-passing programs
that transfer functions (\Cref{sec:functions}) and channels (\Cref{sec:delegation}).

To define (pure) functions and predicates used in program specifications, Iris
embeds the polymorphic lambda calculus.
In the Coq development of Iris, this lambda calculus is obtained via a shallow
embedding, and thus comprises the usual Coq data types and
functions.\footnote{Coq, Iris, and
\lname have a predicative $\Type$ hierarchy, while propositions are impredicative.
For brevity's sake, we omit details about predicativity of $\Type$, as they are
standard.}
We should stress that Iris's lambda calculus is different from our
programming language (\heaplang)---the former is typed and pure, whereas the
latter is untyped and impure.
Consequently, there are two kinds of lambda abstraction
($\Lam \var : \type . \term$ for Iris and $\Lam \var. \expr$ for \heaplang).
It should be clear from context which of the lambda abstractions is used.

\Cref{fig:iris_logic} includes a subset of the Iris grammar.
The typing judgement is mostly standard and can be derived from the use of
meta variables---we use the meta variables $\iprop$ and $\ipropB$ for
propositions (type $\Prop$),
the meta variable $\val$ for values (type $\Val$),
and the meta variables $\term$ and $\termB$ for general terms of any type.
Similar to Coq, $\Lam \var : \type. \term$ is used for both term and
type abstraction, and we write $\type \to \typeB$ for $\All \var:\type. \typeB$
if $\var$ is free in $\typeB$.

\subsection{Guarded recursion and step-indexing}
\label{sec:iris:step-index}

Iris is step-indexed~\cite{appel-mcallester,ahmed_thesis},
meaning that propositions are indexed by a natural number---referred to as
the \emph{step-index}---which is used to stratify a number of semantically
cyclic constructs and reasoning principles.
Iris employs the logical account of step-indexing~\cite{appel_very_model,dreyer_logical}
where the step-index is implicit, and internalised in the logic through the
\emph{later modality} ($\later$)~\cite{nakano-LICS2000}.
Actris and Iris use step-indexing as follows:

\begin{itemize}
\item The principle of L\"ob induction (rule \ruleref{Loeb}) is used to reason
  about (among others) recursive functions.
  When proving $\iprop$, L\"ob induction lets us assume that a proposition holds
  \emph{later}, denoted $\later \iprop$.
  The proposition $\later \iprop$ is strictly weaker than $\iprop$, since $\iprop$
  entails $\later \iprop$ (rule \ruleref{Later-intro}), while the reverse does
  not hold.
  The later modality ($\later$) can be eliminated by taking a program step, which is
  formalised by the Iris proof rule \ruleref{Ht-rec}.
  In \lname we use L\"ob induction to reason about infinite protocols
  (\Cref{sec:subprotocol_recursion}).
\item The guarded recursion operator ($\MU \var:\type. \term$) lets us construct
  recursive predicates without a restriction on the variance of $\var$ in $\term$.
  Instead, the variable $\var$ should be \emph{guarded}, which means that
  it should appear under a \emph{contractive} term construct.
  The prime example of a contractive construct is the later modality ($\later$).
  The rule \ruleref{rec-unfold} says that $\MU \var:\type. \term$ is in fact
  a fixpoint of $\term$.
  \lname's \pname $\sendprot {\xdots} \val \iprop \prot$ and
  $\recvprot {\xdots} \val \iprop \prot$
  are contractive in the tail argument $\prot$, and thereby make it possible
  to use Iris's guarded recursion operator to define recursive protocols (\Cref{sec:rec}).
\item Iris's support for higher-order ghost state~\cite{jung-ICFP06} is used
  to provide a model of \lname in Iris (\Cref{sec:model}).
  Additionally, higher-order ghost state is used by Iris to obtain
  impredicative invariants~\cite{svendsen-ESOP2014}, which in turn are used
  to prove the specification of locks~\cite{hobor2008oracle-semantics}
  used in \Cref{sec:integration}.
\end{itemize}

\subsection{Adequacy of Iris}
\label{sec:iris:adequacy}

The adequacy theorem of Iris connects the derivation of Hoare triples to
the operational semantics of the programming language.
A closed proof of a Hoare triple gives rise to safety and
postcondition validity.
By safety we mean that the program cannot go wrong, \eg by resolving
an illegal function application (\eg $\True + 42$), or accessing an
invalid location (\ie $\deref \loc$ with $\loc \notin \dom(\istate)$).
Safety is defined formally as:
\[
\begin{array}[t]{@{} l @{}}
\safe{\expr} \eqdef
  \begin{array}[t]{@{} l @{}}
    \All \istate, \tpool, \istate'.
    \begin{array}[t]{@{} l @{}}
      (\cfg{\lbrack\expr\rbrack}{\istate} \tpstep^{\ast}
      \cfg{\tpool}{\istate'})\\
      \text{implies}\
      \All \expr' \in \tpool.
      \begin{array}[t]{@{} l @{}}
        (\expr' \in \Val)\
        \text{or}\ \\
        (\Exists \expr'', \istate'', \vec{\expr}. \cfg{\expr'}{\istate'} \step
        \cfg{\expr''}{\istate''; \vec{\expr}})
      \end{array}
    \end{array}
  \end{array}
\end{array}
\]
This definition is not concerned with whether a program
terminates (total correctness).

Postcondition validity means that if the main thread terminates with a value
$\val$, then the postcondition holds for that value.
This is defined formally as:
\[
  \postvalid{\expr}{\fpred} \eqdef
  \begin{array}[t]{@{} l @{}}
    \All \istate,\val, \tpool, \istate'.
  \begin{array}[t]{@{} l @{}}
    (\cfg{\expr}{\istate} \tpstep^{\ast}
    \cfg{\lbrack\val\rbrack\cdot\tpool}{\istate'})\\
  \text{implies}\ (\fpred\ \val)
  \end{array}
  \end{array}
\]

\begin{thm}[Adequacy of Iris]
\label{thm:adequacy_pre}
Let $\fpred \in \Val \to \mProp$ be a meta-level (\ie Coq) predicate over values and
suppose $\hoare \TRUE \expr {\Ret \val. \fpred\ \val}$ is
derivable in Iris, then
$\safe{\expr}$ and
$\postvalid{\expr}{\fpred}$.
\end{thm}

\section{The \lname logic}
\label{sec:actris_logic}

\newcommand{\actrisfig}{
\begin{figure}
\centering
\textbf{Grammar:}
\begin{align*}
\type, \typeB \bnfdef{} &
  \ldots \mid \iProto \mid \ldots \\
\term , \termB , \iprop , \ipropB, \prot \bnfdef{}&
  \ldots \mid
    \sendprot{\xdots}{\val}{\iprop}{\prot} \mid
    \recvprot{\xdots}{\val}{\iprop}{\prot} \mid \protend \mid \\
    & \dual{\prot} \mid \protapp{\prot_1}{\prot_2} \mid
    \interp{\chan}{\prot} \mid \ldots
\end{align*}

\bigskip
\textbf{\Pname:}
\[\begin{aligned}
  \dual{\sendprot{\xdots}{\val}{\iprop}{\prot}} ={}& \recvprot{\xdots}{\val}{\iprop}{\dual{\prot}} \\
  \dual{\recvprot{\xdots}{\val}{\iprop}{\prot}} ={}& \sendprot{\xdots}{\val}{\iprop}{\dual{\prot}} \\
  \protapp{(\sendprot{\xdots}{\val}{\iprop}{\prot_1})}{\prot_2} ={}&
  \sendprot{\xdots}{\val}{\iprop}{(\protapp{\prot_1}{\prot_2})} \\
  \protapp{(\recvprot{\xdots}{\val}{\iprop}{\prot_1})}{\prot_2} ={}&
  \recvprot{\xdots}{\val}{\iprop}{(\protapp{\prot_1}{\prot_2})} \\
  \protapp{\prot_1}{(\protapp{\prot_2}{\prot_3})} ={}& \protapp{(\protapp{\prot_1}{\prot_2})}{\prot_3}\\
\end{aligned}
\quad
\begin{aligned}
  \dual{\protend} ={}& \protend \\
  \dual{\dual{\prot}} ={}& \prot \\
  \protapp\prot\protend ={}& \prot \\
  \protapp\protend\prot ={}& \prot \\
  \dual{\protapp{\prot_1}{\prot_2}} ={}& \protapp{\dual{\prot_1}}{\dual{\prot_2}}
\end{aligned}\]

\bigskip
\textbf{Message passing:}
\begin{mathpar}
\axiomH{Ht-new}
  {\hoare
    {\TRUE}
    {\newchan}
    {\Ret \valB. \Exists \chan_1,\chan_2.
      \valB = (\chan_1,\chan_2) \ast
      \interp{\chan_1}{\prot} \ast \interp{\chan_2}{\dual{\prot}}}}
\and
\axiomH{Ht-send}
  {\hoare
    {\interp \chan {\sendprot{\xdots}{\val}{\iprop}{\prot}} *
     \subst\iprop{\vec \var}{\vec\term}}
    {\send{\chan}{(\subst\val{\vec\var}{\vec\term})}}
    {\interp{\chan}{\subst\prot{\vec\var}{\vec\term}}}}
\and
\axiomH{Ht-recv}
  {\hoare
    {\interp \chan {\recvprot{\xdots}{\val}{\iprop}{\prot}}}
    {\recv{\chan}}
    {\Ret \valB. \Exists \vec\varB.
      \valB = \subst\val{\vec \var}{\vec\varB} *
      \interp{\chan}{\subst\prot{\vec \var}{\vec\varB}} *
      \subst\iprop{\vec \var}{\vec\varB}}}
\end{mathpar}

\caption{The primitive constructs and proof rules of \lname 1.0.}
\label{fig:actris_logic}
\end{figure}
}

\actrisfig

This section describes the core features of \lname 1.0: its \pname
mechanism (\Cref{sec:dsp}), proof rules (\Cref{sec:mp_rules}), and
its adequacy result (\Cref{sec:adequacy_pre_actris}).
\lname inherits all features of Iris, which is achieved by defining \lname as an
embedded logic in Iris.
This means that all of \lname's primitive constructs are defined in Iris,
and all of \lname's primitive proof rules are in fact lemmas in Iris.
We show how \lname is embedded in Iris in \Cref{sec:model}.

\subsection{\Pname}
\label{sec:dsp}

The key feature of \lname is its session-type like \pname mechanism.
\Pname $\prot$ are streams of
$\sendprot {\xdots} \val \iprop \prot$ and
$\recvprot {\xdots} \val \iprop \prot$ constructors that are either infinite or finite.
The finite streams are ultimately terminated by an $\protend$ constructor.
The value $\val$ denotes the message that is being sent ($\SEND$) or received
($\RECV$), the Iris proposition $\iprop$ denotes the ownership that is transferred
along the message, and $\prot$ denotes the protocol that describes the subsequent
messages.
The logical variables $\xdots$ can be used to bind variables in $\val$, $\iprop$,
and $\prot$.
For example, the following \pname expresses that a pair of a boolean and an
integer reference whose value is at least 10 is sent:%
\footnote{Note that $\loc \mapsto i \ast 10 < i$ is logically equivalent to
$\loc \mapsto i \land 10 < i$ as $10 < i$ does not describe ownership.
As discussed in \Cref{sec:iris:separation}, we prefer the version with
separation conjunction.}
\[
\sendprot{(b:\bool)\,(\loc:\Loc)\,(i:\nat)}{(b,\loc)}
  {\loc \mapsto i * 10 < i} \prot
\]
We often omit the proposition $\curlybracket{\iprop}$, which simply means it is
$\TRUE$.

Apart from the constructors for \pname, \lname provides two primitive
operations, $\dual{\prot}$ and $\protapp{\prot_1}{\prot_2}$.
The $\dual{\prot}$ operator denotes the \emph{dual} of a protocol.
Similar to conventional session types, it transforms the protocol by changing all
sends ($\SEND$) into receives ($\RECV$), and \viceversa.
Taking the dual twice thus results in the original protocol.
The operator $\protapp{\prot_1}{\prot_2}$ \emph{appends} the protocols $\prot_1$
and $\prot_2$, which is achieved by substituting any $\protend$ in $\prot_1$
with $\prot_2$.

Channel endpoints are ascribed with \pname using the channel endpoint ownership
connective $\interp \chan \prot$, which captures unique
ownership of the channel endpoint $\chan$ and states that the endpoint follows
the protocol $\prot$.

\subsection{\lname's proof rules for message passing}
\label{sec:mp_rules}

\lname provides proof rules for the three message passing operations
$\newchanname$, $\sendname$, and $\recvname$ (see \Cref{sec:channel_implementation}
for the definition of these operations).
The rule \ruleref{Ht-new} allows ascribing any protocol to newly created
channels using $\newchan$, obtaining ownership of $\interp {\chan_1} \prot$ and
$\interp {\chan_2} {\dual \prot}$ for the respective endpoints.
The duality of the protocol guarantees that any receive ($\RECV$) is matched with a
send ($\SEND$) by the dual endpoint, which is crucial for establishing safety.

The rule \ruleref{Ht-send} for $\send \chan \valB$ requires the head of the
\pnameSingular of $\chan$ to be a send ($\SEND$) constructor, and
the value $\valB$ that is sent to match up with the ascribed value.
To send a message $\valB$, we need to give up ownership of
$\interp{\chan}{\sendprot{\xdots}{\val}{\iprop}{\prot}}$,
pick an appropriate instantiation $\vec\term$ for the variables
$\xdots$ so that $\valB = \subst\val{\vec\var}{\vec\term}$,
give up ownership of the associated resources $\subst\iprop{\vec \var}{\vec\term}$,
and finally regain ownership of the protocol tail
$\interp{\chan}{\subst\prot{\vec\var}{\vec\term}}$.

The rule \ruleref{Ht-recv} for $\recv \chan$ is essentially dual to the rule
\ruleref{Ht-send}.
We need to give up ownership of
$\interp{\chan}{\recvprot{\xdots}{\val}{\iprop}{\prot}}$, and
in return acquire the resources $\subst\iprop{\vec \var}{\vec\varB}$,
the return value $\valB$ where $\valB = \subst\val{\vec\var}{\vec\varB}$,
and finally the ownership of the protocol tail
$\interp{\chan}{\subst\prot{\vec \var}{\vec\varB}}$,
where $\vec\varB$ is some instantiation of the protocol variables.

Finally, we derive the following specification for the
$\startname$ construct from \lname's rule \ruleref{Ht-new} and Iris's rule \ruleref{Ht-fork}:
\begin{mathpar}
  \inferH{Ht-start}
      {\All \chan_2. \hoare{\interp{\chan_2}{\dual{\prot}}}{\fvar\; \chan_2}{\TRUE}}
      {\hoare{\TRUE}{\start{\fvar}}{\Ret \chan_1.
          \interp{\chan_1}{\prot}}}
\end{mathpar}

\subsection{Adequacy of \lname}
\label{sec:adequacy_pre_actris}

By virtue of being an extension of Iris, \lname inherits Iris's adequacy
theorem (\Cref{sec:iris:adequacy}), which says that a closed proof of a Hoare
triple gives rise to safety (programs cannot go wrong) and postcondition validity.
In \lname this means that the implementation of the message passing operations
(\Cref{sec:channel_implementation}) cannot go wrong, and that transferred
messages cannot cause the program to go wrong down the line.

Many conventional session-type systems additionally ensure deadlock freedom---which means
that program execution cannot result in a state where all threads are waiting
on a message to be sent.
Deadlock freedom is ensured through a linear type system and combining
thread and channel creation into a $\startname$ primitive.
\lname is affine (instead of linear), has a $\langkw{fork}$ and
$\newchanname$ primitive (instead of a $\startname$ primitive), and supports
locks for channel sharing.
\lname thus provides more flexibility in terms of what programs can be written
and verified
(there exist programs that are deadlock free, but cannot be type-checked using
conventional session types, while they can be verified using \lname).
On the flip side, using \lname one can prove Hoare triples for programs that
deadlock, for example:
\[
\hoare \TRUE {\Let (\chan,\chan') := \newchan in \recv{\chan}} \TRUE
\]
Indeed, in our operational semantics programs such as the above are safe.
The semantics of both lock acquisition ($\acquirelockname$) and message reception
($\recvname$) is that the thread
loops until it succeeds.
Loops are considered safe in Iris (and thus also \lname), as the threads in question
will continue to take steps, although they will never terminate.

\section{A tour of \lname}
\label{sec:tour}

This section demonstrates the core features of \lname.
We introduce and iteratively extend a simple channel-based merge sort algorithm to
demonstrate the main features of \lname (\Cref{sec:basics}--\Cref{sec:dependent}).
Note that as the point of the sorting algorithms is to showcase the features
of \lname, they are intentionally kept simple and no effort has been made
to make them efficient (\eg to avoid spawning threads for small jobs).

\subsection{Basic protocols}
\label{sec:basics}

\newcommand{\sortfigure}{
\begin{figure}
\begin{equation*}
\begin{array}{ll}
\begin{array}[t]{@{}l@{\qquad}}
\listsortservicename\ \cmpvar\ \chan\ \langdef\\
\quad \Let \listvar = \recv \chan in\\
\quad \If \listlength{\listvar} \le 1 then \send \chan \TT \Else\\
\quad \Let \listvar' = \llistsplit\ \listvar in\\
\quad \Let \chan_1 = \start{(\listsortservicename\ \cmpvar)} in\\
\quad \Let \chan_2 = \start{(\listsortservicename\ \cmpvar)} in\\
\quad \send {\chan_1} \listvar;\ \send{\chan_2}{\listvar'};\\
\quad \recv {\chan_1};\ \recv {\chan_2};\\
\quad \llistmerge\ \cmpvar\ \listvar\ \listvar';\ \send \chan \TT
\end{array}
&
\begin{array}[t]{@{}l}
\listsortclientname\ \cmpvar\ \listvar \langdef \\
\quad \Let \chan = \\
\qquad \start{(\listsortservicename\ \cmpvar)} in \\
\quad \send \chan \listvar;\\
\quad \recv \chan
\end{array}
\end{array}
\end{equation*}
\caption{A channel-based merge sort algorithm
(the code for $\llistmerge$ and $\llistsplit$ is standard and thus elided).}
\label{fig:sort}
\end{figure}}

\sortfigure

We first prove functional
correctness of a simple channel-based merge sort algorithm, whose code is shown
in \Cref{fig:sort}.
The function $\listsortclientname\ \cmpvar\ \listvar$ takes a comparison function
$\cmpvar$ and a linked list $\listvar$ that will be sorted.
The function mutates the linked list $\listvar$, so it returns a unit value $\TT$ when done.
The bulk of the work is done by the $\listsortservicename\ \cmpvar\ \chan$ function, which
takes a channel endpoint $\chan$ over which it receives a linked list, and
over which it sends back $\unittt$ to inform the sender that the list has been sorted.
The function $\listsortservicename$ is implemented as follows.
If the received list is an empty or singleton list, which both are trivially sorted, the
function immediately sends back $\TT$.
Otherwise, the list is split into two partitions using $\llistsplit\ \listvar$,
which mutates the list $\listvar$ to contain the first partition, while
returning $\listvar'$ containing the second partition.
These partitions are recursively sorted using two newly started instances of
$\listsortservicename$.
The results of the processes are then requested and merged using
$\llistmerge\ \cmpvar\ \listvar\ \listvar'$,
which mutates the list $\listvar$ to contain the merged list.
Finally, the unit value $\TT$ is sent back along the original
channel endpoint $\chan$.

In order to verify the correctness of the sorting algorithm we first need a
specification for the comparison function $\cmpvar$, which must satisfy the
following specification:
\begin{equation*}
\begin{array}{l}
\cmpspecname\ (\interpvar : \tvar \to \Val \to \iProp)
  \ (\relvar : \tvar \to \tvar \to \bool)\ (\cmpvar : \Val) \eqdef \\
\quad\quad\quad(\All \var_1\,\var_2. \relvar\ \var_1\ \var_2 \lor \relvar\ \var_2\ \var_1) \ast{} \\
\quad\quad\quad(\All \var_1\,\var_2\,\val_1\,\val_2.
  \hoare{\interpvar\ \var_1\ \val_1 * \interpvar\ \var_2\ \val_2}
    {\cmpvar\ \val_1\ \val_2}
    {\Ret r. r = \relvar\ \var_1\ \var_2 * I\ \var_1\ \val_1 * I\ \var_2\ \val_2})
\end{array}
\end{equation*}

\noindent
This definition is polymorphic in type $\tvar$.
Here, $\relvar$ is a total relation in type $\tvar$,
and $\interpvar$ is an interpretation predicate that relates language
values to elements of type $\tvar$.
While the relation $\relvar$ dictates the ordering, the interpretation predicate
$\interpvar$ allows for flexibility about what is ordered.
Setting $\interpvar$ to \eg $\Lam \var\;\val. \val \mapsto \var$
orders references by what they point to in memory, rather than the memory address
itself.
To specify how lists are laid out in memory we use the following notation:
\begin{equation*}
\llistrefI \interpvar \loc {\vec \var} \eqdef
  \begin{cases}
  \loc \mapsto \Inl\;\TT & \textnormal{if $\vec\var = \nil$} \\
  \Exists \val_1\;\loc_2.
    \loc \mapsto \Inr\;(\val_1,\loc_2) * \interpvar\ \var_1\ \val_1 *
    \llistrefI \interpvar {\loc_2} {\vec\var_2}
    & \textnormal{if $\vec\var = [\var_1] \cdot \vec\var_2$}
  \end{cases}
\end{equation*}

\noindent
The channel endpoint $\chan$ adheres to the following \pnameSingular:
\begin{equation*}
\begin{array}{l}
\listsortprotname\ (\interpvar : \tvar \to \Val \to \Prop)
\ (\relvar : \tvar \to \tvar \to \bool) \eqdef \\
\qquad
  \sendprot{(\vec{\var}:\List\ \tvar)\;(\loc : \Loc)}{\loc}
  {\llistrefI \interpvar \loc {\vec\var}}{
\recvprot{\vec{\varB}}{\TT}
  {\llistrefI \interpvar \loc {\vec \varB} *
   \sortedof {\vec\varB} {\vec\var} \relvar}{
      \protend}}
\end{array}
\end{equation*}

\noindent
The protocol describes the interaction of first sending a linked list, and then
receiving a unit value $\TT$ once the list is sorted.
The predicate $\sortedof {\vec\varB} {\vec\var} \relvar$
is true iff $\vec\varB$ is a sorted version of $\vec\var$ with respect to
the relation $\relvar$.
We prove the following specifications of the service and the client:
\[
\begin{array}{c !{\quad} c}
\begin{array}{l}
  \hoareV
    {\cmpspecname\ \interpvar\ \relvar\ \cmpvar *
     \interp{\chan}{\protapp{\dual{\listsortprotname\ \interpvar\ \relvar}} \prot}}
    {\listsortservicename\ \cmpvar\ \chan}
    {\interp{\chan}{\prot}}
\end{array}
  &
\begin{array}{l}
  \hoareV
    {\cmpspecname\ \interpvar\ \relvar\ \cmpvar *
     \llistrefI \interpvar \loc {\vec\var}}
    {\listsortclientname\ \cmpvar\ \loc}
    {\Exists \vec\varB.
     \sortedof {\vec\varB} {\vec\var} \relvar *
     \llistrefI \interpvar \loc {\vec\varB}}
\end{array}
\end{array}
\]
\noindent
There are two important things to note about these specifications.
First, the protocol $\listsortprotname\ \interpvar\ \relvar$ is written from the
point of view of the client.
As such, the precondition for $\listsortservicename$ requires that $\chan$
follows the dual $\dual{\listsortprotname\ \interpvar\ \relvar}$.
Second, the pre- and postcondition of $\listsortservicename$ are generalised
to have an arbitrary protocol $\prot$ appended at the end.
It is important to write specifications this way, so they can be embedded in
other protocols.
We will see examples of such an embedding in \Cref{sec:rec} and~\Cref{sec:delegation}.

The proof of these specifications is almost entirely performed by symbolic
execution using the rules \ruleref{Ht-new}, \ruleref{Ht-send},
\ruleref{Ht-recv}, and the standard separation logic rules.

Now that we have proven Hoare triples for $\listsortservicename$ and
$\listsortclientname$, we can use them to prove Hoare triples of other programs
that use these functions.
Recall that if we use them to prove a Hoare triple of a closed program, we
obtain safety and postcondition validity by virtue of \lname's
adequacy theorem (\Cref{sec:iris:adequacy}).

\subsection{Transferring functions}
\label{sec:functions}

\newcommand{\sendfuncfig}{
\begin{figure}
\begin{equation*}
\begin{array}{@{}l@{\qquad}l@{}}
\begin{array}[t]{@{}l@{}}
  \listsortcmpservicename\ \chan \langdef \\
  \quad \Let \cmpvar = \recv \chan in\\
  \quad \listsortservicename\ \cmpvar\ \chan
\end{array}
&
\begin{array}[t]{@{}l@{}}
  \listsortcmpclientname\ \cmpvar\ \listvar \langdef \\
  \quad
  \Let \chan = \start\listsortcmpservicename in\\
  \quad
  \send \chan \cmpvar;\
  \send \chan \listvar;\ \recv \chan
\end{array}
\end{array}
\end{equation*}
\caption{A version of the sort service that receives the comparison function
over the channel.}
\label{fig:sendFunction}
\end{figure}}

\sendfuncfig

The channel-based $\listsortservicename$ from the previous section
(\Cref{fig:sort}) is parametric on a comparison function.
To demonstrate \lname's support for reasoning about functions transferred over
channels, we verify the correctness of the function
$\listsortcmpservicename\ \chan$ in \Cref{fig:sendFunction}.
This function takes a channel endpoint $\chan$, over which it receives the comparison
function $\cmpvar$ (instead of via a function argument), followed by the
list to sort.
Similar to the service in \Cref{sec:basics}, it mutates the list,
and sends back $\TT$ when done.
To verify this program, we extend the protocol $\listsortprotname$ from
\Cref{sec:basics} as follows:
\begin{align*}
\listsortcmpprotname \eqdef{}& \SEND
    (\tvar : \Type)\ (\interpvar : \tvar \to \Val \to \Prop)\
    (\relvar : \tvar \to T \to \bool)\ (\cmpvar: \Val)\\
  & \quad \langle \cmpvar \rangle
  \curlybracket{\cmpspecname\ \interpvar\ \relvar\ \cmpvar}.\,
  \listsortprotname\ \interpvar\ \relvar
\end{align*}
The new protocol specifies that we first send a comparison function $\cmpvar$.
It includes binders for the polymorphic type $\tvar$, the
interpretation predicate $\interpvar$, and the relation $\relvar$.
The specifications are much the same as before, with the proofs being similar
besides the addition of a symbolic execution step to resolve the sending and
receiving of the comparison function:
\[
\begin{array}{c !{\quad} c}
\begin{array}{l}
  \hoareV
    {\interp \chan {\protapp {\dual{\listsortcmpprotname}} \prot}}
    {\listsortcmpservicename\ \chan}
    {\interp \chan \prot}
\end{array}
  &
\begin{array}{l}
\hoareV
  {\cmpspecname\ \interpvar\ \relvar\ \cmpvar *
   \llistrefI \interpvar \loc {\vec\var}}
  {\listsortcmpclientname\ \cmpvar\ \loc}
  {\Exists \vec\varB.
   \llistrefI \interpvar \loc {\vec\varB} *
   \sortedof {\vec\varB} {\vec\var} \relvar}
\end{array}
\end{array}
\]

\subsection{Choice}
\label{sec:choice}

Branching communication is commonly modelled using the \emph{choice}
session types $\branchop$ for branching and $\selectop$ for selection.
We show that corresponding \pname can readily be encoded in \lname.
At the level of the programming language, the instructions for choice are
encoded by sending and receiving a boolean value that is matched using an if-then-else construct:
\begin{align*}
\select \expr \expr' \eqdef{}&
  \send \expr \expr' \\
\branchA \expr {\expr_1} {\expr_2} \eqdef{}&
  \If \recv \expr then \expr_1 \Else \expr_2
\end{align*}
The instructions are syntactic sugar, \ie defined in the meta language (using $\eqdef$), which effectively
means that the arguments are evaluated lazily.
We define syntactic sugar $\leftname \eqdef \True$ and $\rightname \eqdef \False$ to be used
together with \selectname for readability's sake.

Due to the higher-order nature of \lname, the usual protocol specifications for
choice from session types can be encoded as regular logical branching within
the protocols:
\begin{align*}
\selectprotprop{\prot_1}{\ipropB_1}{\ipropB_2}{\prot_2} \eqdef{}&
  \sendprot {(b:\bool)} b {\If b then \ipropB_1 \Else \ipropB_2}
                {\If b then \prot_1 \Else \prot_2} \\
\branchprotprop{\prot_1}{\ipropB_1}{\ipropB_2}{\prot_2} \eqdef{}&
  \recvprot {(b:\bool)} b {\If b then \ipropB_1 \Else \ipropB_2}
                {\If b then \prot_1 \Else \prot_2}
\end{align*}

\noindent
We often omit the conditions $\ipropB_1$ and $\ipropB_2$, which simply means
that they are $\TRUE$.
The following rules can be directly derived from the rules
\ruleref{Ht-send} and \ruleref{Ht-recv}:
\begin{mathparpagebreakable}
\axiomH{Ht-select}{\hoare
  {\begin{array}{@{}l@{}}
     \interp{\chan}{\selectprotprop{\prot_1\!}{\ipropB_1}{\ipropB_2}{\!\prot_2}}\ * \\\ite{b}{\ipropB_1}{\ipropB_2}
     \end{array}}
  {\select{\chan}{b}}
  {\interp{\chan}{\ite{b}{\prot_1}{\prot_2}}}}
\and
\inferH{Ht-branch}
  {\hoare
    {\iprop * \ipropB_1 * \interp{\chan}{\prot_1}}
    {\expr_1}
    {\Ret \val. \ipropC} \and
   \hoare
    {\iprop * \ipropB_2 * \interp{\chan}{\prot_2}}
    {\expr_2}
    {\Ret \val. \ipropC}}
  {\hoare
    {\iprop * \interp{\chan}{\branchprotprop{\prot_1}{\ipropB_1}{\ipropB_2}{\prot_2}}}
    {\branchA{\chan}{\expr_1}{\expr_2}}
    {\Ret \val. \ipropC}}
\end{mathparpagebreakable}

\noindent
Apart from branching on boolean values, \pname can be used to encode
choice on any enumeration type (\eg lists, natural numbers, days of the
week, \etc).
These encodings follow the same scheme.

\subsection{Recursive protocols}
\label{sec:rec}

\newcommand{\loopsortfig}{
\begin{figure}[t!]
\begin{equation*}
\begin{array}{l l}
\begin{array}[t]{@{}l@{}}
\listsortloopservicename\ \cmpvar\ \chan \langdef \\
\quad\begin{array}[t]{@{}l@{}}
  \mbranchA \chan
  {\begin{array}[t]{@{} l}
     \listsortservicename\ \cmpvar\ \chan;\ \\
     \listsortloopservicename\ \cmpvar\ \chan
   \end{array}}
  \TT \\
\end{array}
\end{array}
&
\begin{array}[t]{@{}l@{}}
  \listsortloopclientname\ \cmpvar\ \listvar \langdef \\
  \quad \Let \chan = \start {(\listsortloopservicename\ \cmpvar)} in \\
  \quad \llistiter\ (\Lam \listvar'. \select \chan \leftname;\ \send \chan {\listvar'};\ \recv \chan)\ \listvar; \\
  \quad \select \chan \rightname
\end{array}
\end{array}
\end{equation*}
\caption{A recursive version of the sort service that can perform multiple jobs in sequence
(the code for the function $\llistiter$, which applies a function to each
element of the list, is standard and has been elided).}
\label{fig:loopservice}
\end{figure}
}

\loopsortfig

We now use choice and recursion to verify the correctness
of a sorting service that supports performing multiple sorting jobs in sequence.
The code of the sorting service $\listsortloopservicename$
and a possible
client $\listsortloopclientname$ are displayed
in \Cref{fig:loopservice}.
The service $\listsortloopservicename\ \cmpvar\ \chan$ takes a comparison
function $\cmpvar$ and a channel endpoint $\chan$, and returns $\TT$.
It contains a loop in which choice is used to either terminate the service,
or to sort an individual list using the
channel-based merge sort algorithm $\listsortservicename$
from \Cref{sec:basics}.
The client $\listsortloopclientname\ \cmpvar\ \listvar$ takes a comparison
function $\cmpvar$ and a nested linked list of linked lists $\listvar$,
and returns $\TT$.
It starts a single instance of the service at channel endpoint $\chan$,
and then sequentially sends requests to sort each inner linked list $\listvar'$
in $\listvar$.
Finally, the client selects the terminating branch to end the communication with
the service.
A protocol for interacting with the sorting service can be defined as follows:
\begin{equation*}
\begin{array}{l}
\listsortloopprotname\ (\interpvar : \tvar \to \Val \to \Prop)
\ (\relvar : \tvar \to \tvar \to \bool) \eqdef{} \\
\qquad\qquad\qquad\qquad \MU (\recvar : \iProto).
  \selectprot
    {(\protapp {\listsortprotname\ \interpvar\ \relvar} \recvar)}
    \protend
\end{array}
\end{equation*}

\noindent
The protocol uses the choice operator $\selectop$ to specify that the client
may either request the service to perform a sorting job, or terminate
communication with the service.
After the job has been finished the protocol proceeds
recursively.

We use Iris's operator $\MU \var:\type.\term$ for guarded recursion
(\Cref{sec:iris:step-index}) to define recursive protocols.
It is important to recall that---as is usual in logics with guarded
recursion---the variable $\var$ should appear under a
\emph{contractive} term construct in the body $\term$ of $\MU \var:\type.\term$.
In our protocol, the recursive variable $\recvar$ appears under the argument of
$\selectop$, which is defined in terms of $\sendprot {\xdots} \val \iprop \prot$,
which, similarly to $\recvprot {\xdots} \val \iprop \prot$, is contractive in the
tail protocol $\prot$.
We can then prove the following specifications of the service and the client:
\[
\begin{array}{c !{\quad} c}
\begin{array}{l}
\hoareV
  {\begin{array}{@{} l}
   \cmpspecname\ \interpvar\ \relvar\ \cmpvar\ *\\
   \interp \chan {\protapp {\dual{\listsortloopprotname\ \interpvar\ \relvar}} \prot}
   \end{array}}
  {\listsortloopservicename\ \cmpvar\ \chan}
  {\interp\chan \prot}
\end{array}
&
\begin{array}{l}
\hoareV
  {\cmpspecname\ \interpvar\ \relvar\ \cmpvar *
   \llistrefI \interpvarB \loc {\vec{\vec\var}}}
  {\listsortloopclientname\ \cmpvar\ \loc}
  {
    \Exists \vec{\vec\varB}.
    \listlength {\vec{\vec\varB}} = \listlength {\vec{\vec\var}} *
    \llistrefI \interpvarB \loc {\vec{\vec\varB}} *
    (\All i < \listlength {\vec{\vec\var}}.
      \sortedof{\vec{\vec\varB}_i} {\vec{\vec\var}_i} \relvar)}
\end{array}
\end{array}
\]

\noindent
We let $\interpvarB \eqdef \Lam \loc'\,\vec\varB. \llistrefI \interpvar {\loc'} {\vec\varB}$
to express that $\loc$ points to a list of lists $\vec{\vec\var}$.
The proof of the service follows naturally by symbolic execution
using the induction hypothesis (obtained from \ruleref{Loeb}), the
rules \ruleref{Ht-branch} and \ruleref{Ht-select}, and the specification
of $\listsortservicename$.
Note that we rely on the specification of $\listsortservicename$ having an
arbitrary protocol as its suffix.

It is worth pointing out that protocols in \lname provide a lot of flexibility.
Using just minor changes, we can extend the protocol to support transferring a
comparison function over the channel, like the extension made in
$\listsortcmpclientname$,
or in a way such that a different comparison function can be used for each sorting job.

\subsection{Higher-order protocols}
\label{sec:delegation}

\newcommand{\delegationfig}{
\begin{figure}[t!]
\begin{equation*}
\begin{array}{@{}l@{\quad}l@{}}
\begin{array}[t]{@{}l@{}}
\listsortdelservicename\ \cmpvar\ \chan\ \langdef \\
  \quad \mbranchE
     \chan
     {\begin{array}[t]{@{}l@{}}
         \Let \chan' = \\
         \quad \start{(\listsortservicename\ \cmpvar)} in \\
      \send \chan {\chan'}; \\
      \listsortdelservicename\ \cmpvar\ \chan
      \end{array}}
     \TT
\end{array}
&
\begin{array}[t]{@{}l@{}}
  \listsortdelclientname\ \cmpvar\ \listvar \langdef \\
  \quad \Let \chan = \start{(\listsortdelservicename\ \cmpvar)} in \\
  \quad \Let k = \llistnil\ \TT in \\
  \quad \begin{array}{@{}l @{} l @{}}
      \llistiter\ (\Lam \listvar'.{} & \select \chan \leftname; \\
      & \Let \chan' = \recv \chan in \\
      & \send {\chan'} {\listvar'};\ \llistcons\ c'\ k)\ l
      \end{array} \\
  \quad \select \chan \rightname; \\
  \quad \llistiter\ \recvname\ k\\
\end{array}
\end{array}
\end{equation*}
\caption{A recursive version of the sort service that uses delegation to perform
multiple jobs in parallel (the code for the function $\llistcons$,
which pushes an element to the head of a list, has been elided).}
\label{fig:delegation}
\end{figure}
}

Higher-order communication is a common feature within communication protocols,
and particularly the session-types community---it is the concept of transferring
a channel endpoint over a channel, often called delegation.
Due to the impredicativity of \pname in \lname, higher-order reasoning about
programs with delegation is readily available.
The protocols $\sendprot {\xdots} \val \iprop \prot$ and
$\recvprot {\xdots} \val \iprop \prot$ can simply refer to the channel endpoint
ownership $\interp \chan {\prot'}$ in the proposition $\iprop$.

\delegationfig

An example of a program that uses delegation is the $\listsortdelservicename$
variant of the recursive sorting service in \Cref{fig:delegation}, which allows
multiple sorting jobs to be performed in parallel.
The function $\listsortdelservicename\ \cmpvar\ \chan$ takes a comparison
function $\cmpvar$, a channel endpoint $\chan$, and returns $\TT$.
Using the channel endpoint $\chan$, a client can request the service to start a new inner
sorting service $\chan'$, which the service delegates over channel endpoint $\chan$.

Similar to the client in \Cref{sec:rec},
the client $\listsortdelclientname\ \cmpvar\ \listvar$ takes a comparison
function $\cmpvar$ and a nested linked list of linked lists $\listvar$,
and returns $\TT$.
The client starts a connection $\chan$ to the service, and for each inner
list $\listvar'$, it acquires a delegated channel endpoint $\chan'$, over which it
sends the inner list $\listvar'$ that should be sorted.
The client keeps track of all channels to delegated services in a linked
list $k$ so that it can wait for all of them to finish (using
$\llistiter\ \recvname$).

A protocol for the delegation service can be defined as follows,
denoting that the client can select whether to acquire a
connection to a new delegated service or to terminate:
\begin{equation*}
\begin{array}{l}
\listsortdelprotname\ (\interpvar : \tvar \to \Val \to \Prop)
  \ (\relvar : \tvar \to \tvar \to \bool) \eqdef{} \\
\qquad \MU (\recvar : \iProto).
  \selectprot
    {(\recvprot {(\chan:\Val)} \chan {\interp \chan
       {\listsortprotname\ \interpvar\ \relvar}} \recvar)}
    \protend
\end{array}
\end{equation*}

\noindent
We can then prove the following specifications of the service and the client:
\[
\begin{array}{c !{\quad} c}
\begin{array}{l}
\hoareV
  {\begin{array}{@{} l}
   \cmpspecname\ \interpvar\ \relvar\ \cmpvar\ *\\
   \interp \chan {\protapp {\dual{\listsortdelprotname\ \interpvar\ \relvar}} \prot}
   \end{array}}
  {\listsortdelservicename\ \cmpvar\ \chan}
  {\interp\chan \prot}
\end{array}
&
\begin{array}{l}
\hoareV
  {\cmpspecname\ \interpvar\ \relvar\ \cmpvar *
   \llistrefI \interpvarB \loc {\vec{\vec\var}}}
  {\listsortdelclientname\ \cmpvar\ \loc}
  {
    \Exists \vec{\vec\varB}.
    \listlength {\vec{\vec\varB}} = \listlength {\vec{\vec\var}} *
    \llistrefI \interpvarB \loc {\vec{\vec\varB}} *
    (\All i < \listlength {\vec{\vec\var}}.
      \sortedof{\vec{\vec\varB}_i} {\vec{\vec\var}_i} \relvar)}
\end{array}
\end{array}
\]

\noindent
As before, we let $\interpvarB \eqdef \Lam \loc'\,\vec\varB. \llistrefI \interpvar {\loc'} {\vec\varB}$
to express that $\loc$ points to a list of lists $\vec{\vec\var}$.
Once again the proofs are straightforward, as they are simply a combination of
recursive reasoning combined with the application of \lname's rules for channels.

\subsection{Dependent protocols}
\label{sec:dependent}

\newcommand{\dependentfig}{
\begin{figure}
\begin{equation*}
\begin{array}{@{}l@{\quad}l@{}}
\begin{array}[t]{@{}l@{}}
\listsortelemservicename\ \cmpvar\ \chan\ \langdef\\
\quad \mbranchD \chan
  {\begin{array}[t]{@{}l@{}}
    \Let \var_1 = \recv\chan in\\
    \mbranchD{\chan}
    {\begin{array}[t]{@{} l @{}}
      \Let \var_2 = \recv \chan in \\
      \Let \chan_1 = \start{(\listsortelemservicename\ \cmpvar)} in \\
      \Let \chan_2 = \start{(\listsortelemservicename\ \cmpvar)} in \\
      \select {\chan_1} \leftname;\ \send {\chan_1} \var_1; \\
      \select {\chan_2} \leftname;\ \send {\chan_2} \var_2; \\
      \listsortelemservicesplitname\ \chan\ \chan_1\ \chan_2;\
      \listsortelemservicemergename\ \cmpvar\ \chan\ \chan_1\ \chan_2
     \end{array}}
    {\begin{array}[t]{@{} l @{}}
       \select \chan \leftname;\
       \send \chan \var_1; \\
       \select \chan \rightname
     \end{array}}
    \end{array}}
  {\select \chan \rightname} \\
~ \\[-0.7em]
\listsortelemservicesplitname\ \chan\ \chan_1\ \chan_2\ \langdef\\
\quad \mbranchD \chan
  {\begin{array}[t]{@{}l@{}}
     \Let \var = \recv \chan in \\
     \select {\chan_1} \leftname;\ \send {\chan_1} \var;\\
     \listsortelemservicesplitname\ \chan\ \chan_2\ \chan_1
   \end{array}}
  {\begin{array}[t]{@{}l@{}}
     \select {\chan_1} \rightname;\\ \select {\chan_2} \rightname
   \end{array}}
\end{array}
&
\begin{array}[t]{@{} l @{}}
\listsortelemservicemergename\ \cmpvar\ \chan\ \chan_1\ \chan_2 \langdef\\
  \quad \mbranchD {\chan_1}
     {\begin{array}[t]{@{}l@{}}
        \Let \var = \recv {\chan_1} in \\
        \listsortelemservicemergerecname\ \cmpvar\ \chan\ \var\ \chan_1\ \chan_2
      \end{array}}
     {\Assert\ \False} \\
~ \\[-0.7em]
\listsortelemservicemergerecname\ \cmpvar\ \chan\ \var\ \chan_1\ \chan_2 \langdef\\
\quad \mbranchD {\chan_2}
  {\begin{array}[t]{@{}l@{}}
     \Let \varB = \recv{\chan_2} in \\
     \If {\cmpvar\ \var\ \varB} then \\
       \quad \select \chan \leftname;\ \send \chan \var; \\
       \quad \listsortelemservicemergerecname\ \cmpvar\ \chan\ \varB\ \chan_2\ \chan_1 \\
     \Else \\
       \quad \select \chan \leftname;\ \send \chan \varB; \\
       \quad \listsortelemservicemergerecname\ \cmpvar\ \chan\ \var\ \chan_1\ \chan_2
   \end{array}}
  {\begin{array}[t]{@{}l@{}}
     \select{\chan}{\leftname}; \send{\chan}{\var_1}; \\
     \listsortelemservicetransfername\ \chan_1\ \chan
   \end{array}} \\
~ \\[-0.7em]
\listsortelemclientname\ \cmpvar\ \listvar\ \langdef\\
  \quad \Let \chan =\\
  \qquad \start{(\listsortelemservicename\ \cmpvar)} in\\
  \quad \sendallname\ \chan\ \listvar;\
  \recvallname\ \chan\ \listvar
\end{array}
\end{array}
\end{equation*}
\caption{A fine-grained version of the sort service that transfers elements one
by one (the code for the functions $\listsortelemservicetransfername$,
$\sendallname$, and $\recvallname$ has been elided).}
\label{fig:dependent}
\end{figure}
}

The protocols we have seen so far have only made limited use of \lname's support
for recursion.
We now demonstrate \lname's support for dependent protocols, which make it
possible to keep track of the history of what messages have been sent and
received.
We demonstrate this feature by considering a fine-grained version of the
channel-based merge-sort service as shown in \Cref{fig:dependent}.
Like previous versions, the function $\listsortelemservicename\ \cmpvar\ \chan$
takes a comparison function $\cmpvar$ and a channel endpoint $\chan$, and returns $\TT$.
However, unlike previous versions, the input list should be transferred
element by element over the channel endpoint $\chan$ to the service,
and when done, the service sends back the
sorted list element by element.
We use choice to indicate whether the whole list has been
sent ($\rightname$) or another element remains to be sent ($\leftname$).

\dependentfig

The structure of $\listsortelemservicename$
is somewhat similar to the coarse-grained merge-sort algorithm that we have
seen before.
The base cases of the empty or the singleton list are handled initially.
This is achieved by waiting for at least two values before starting
the recursive sub-services $\chan_1$ and $\chan_2$.
In the base cases the values are sent back immediately, as they are trivially
sorted.
The inductive case is handled by starting two sub-services at the
channel endpoints $\chan_1$ and $\chan_2$.
First, each of the channel endpoints are sent one of the two initially
received elements.
The remaining elements are then received by the parent service on $\chan$,
and forwarded to the sub-services alternatingly on $\chan_1$ and $\chan_2$,
using the function $\listsortelemservicesplitname\ \chan\ \chan_1\ \chan_2$.
Once the $\rightname$ flag is received, the $\listsortelemservicesplitname$
function terminates, and the algorithm moves to the second phase.

In the second phase, the function $\listsortelemservicemergename\ \cmpvar\ \chan\ \chan_1\ \chan_2$
is used to merge the stream of elements returned by the sub-services
on $\chan_1$ and $\chan_2$ and forwards them to the parent service on $\chan$.
It initially acquires the first
element $\var$ from the first sub-service on $\chan_1$,
which it passes to the auxiliary function
$\listsortelemservicemergerecname$ as the current largest value.
The auxiliary function
$\listsortelemservicemergerecname\ \cmpvar\ \chan\ \var\ \chan_1\ \chan_2$
recursively requests a value $\varB$ from the sub-service from which the current
largest value was not acquired from (initially $\chan_2$).
It then compares $\var$ and $\varB$ using the comparison function $\cmpvar$,
and forwards the smallest element on $\chan$.
This is repeated until the $\rightname$ flag is received from either sub-service,
after which the remaining values of the other sub-service are forwarded to the
parent service on $\chan$ using
$\listsortelemservicetransfername\ \chan_1\ \chan$.

The interface of the client $\listsortelemclientname\ \cmpvar\ \listvar$
is similar to the one from \Cref{sec:basics,sec:functions}.
It takes a comparison function $\cmpvar$ and a linked lists $\listvar$,
sorts the linked list $\listvar$, and returns $\TT$ when done.
The client sorts the list $\listvar$ by sending its elements to the sort
service using the $\sendallname\ \chan\ \listvar$ function (which mutates the
list $\listvar$ by removing all of its values and sending them over the
channel~$\chan$),
and puts the received values back into the linked list using the
$\recvallname\ \chan\ \listvar$ function
(which also mutates the list $\listvar$).
A suitable protocol for proving functional correctness of the fine-grained
sorting service is as follows:
\begin{align*}
\listsortelemprotname\; (\interpvar : \tvar \to \Val \to \Prop)
  \ (\relvar : \tvar \to \tvar \to \bool) \eqdef{}&
  \listsortelemprotheadname\ \interpvar\ \relvar\ \nil \\[0.1em]
\listsortelemprotheadname\; (\interpvar : \tvar \to \Val \to \Prop)
  \ (\relvar : \tvar \to \tvar \to \bool) \eqdef{}&
  \MU (\recvar : \List\ \tvar \to \iProto).\\[-0.2em]
  &\hspace{-22em}
  \Lam \vec\var. \selectprot
    {(\sendprot {(\var : \tvar)\ (\val : \Val)} \val
       {\interpvar\ \var\ \val}
       {\recvar\ (\snoc{\var}{\vec\var})}) \ }
    {\ \listsortelemprottailname\ \interpvar\ \relvar\ \vec{\var}\ \nil} \\[0.1em]
\listsortelemprottailname\ (\interpvar : \tvar \to \Val \to \Prop)
  \ (\relvar : \tvar \to \tvar \to \bool) \eqdef{}&
  \MU (\recvar : \List\ \tvar \to \List\ \tvar \to \iProto). \\[-0.2em]
  &\hspace{-22em}
  \Lam \vec\var\ \vec\varB. \branchprotpropr
  {(\recvprot{(\varB : \tvar)\ (\val : \Val)} \val
     {(\All i < \listlength {\vec\varB}. \relvar\ \vec\varB_i\ \varB) * \interpvar\ \varB\ \val}
     {\recvar\ \vec\var\ (\snoc{\varB}{\vec\varB})}) \ }
  {\vec\var \perm \vec\varB}
  {\ \protend}
\end{align*}

\noindent
The protocol is split into two phases $\listsortelemprotheadname$ and
$\listsortelemprottailname$, mimicking the behaviour of the program.
The $\listsortelemprotheadname$ phase is indexed by the values $\vec\var$ that
have been sent so far.
The protocol describes that one can either send another value and proceed
recursively, or stop, which moves the protocol to the next phase.

The $\listsortelemprottailname$ phase is dependent on the list of values $\vec\var$
received in the first phase, and the list of values $\vec\varB$ returned so
far.
The condition $(\All i < \listlength {\vec\varB}. \relvar\ \vec\varB_i\ \varB)$
states that the received element is larger than any of the elements that have
previously been returned,
which maintains the invariant that the stream of received elements is sorted.
When the $\rightname$ flag is received $\vec\var \perm \vec\varB$ shows that the
received values $\vec\varB$ are a permutation of the ones $\vec\var$ that were sent,
making sure that all of the sent elements have been accounted for.

We can then prove top-level specifications for the service and client that are
similar to the coarse-grained version of the channel-based merge sort:
\[
\begin{array}{c !{\quad} c}
\begin{array}{l}
  \hoareV
    {\cmpspecname\ \interpvar\ \relvar\ \cmpvar *
     \interp \chan {\protapp {\dual{\listsortelemprotname\ \interpvar\ \relvar}} \prot}}
    {\listsortelemprotname\ \chan}
    {\interp \chan \prot}
\end{array}
  &
\begin{array}{l}
\hoareV
  {\cmpspecname\ \interpvar\ \relvar\ \cmpvar *
   \llistrefI \interpvar \loc {\vec\var}}
  {\listsortelemclientname\ \cmpvar\ \loc}
  {
  \Exists \vec\varB.
   \llistrefI \interpvar \loc {\vec\varB} *
   \sortedof {\vec\varB} {\vec\var} \relvar}
\end{array}
\end{array}
\]

\noindent
Proving these specifications requires one to pick appropriate specifications for
the auxiliary functions to capture the required invariants with regard to
sorting.
After having picked these specifications, the parts of the proofs that involve
communication are mostly straightforward, but require a number of trivial
auxiliary results
about sorting and permutations.

\newcommand{\subprotfig}{
\begin{figure}
  \centering
  \textbf{Grammar:}
  \begin{align*}
    \term , \termB , \iprop , \ipropB, \prot \bnfdef{}&
    \ldots \mid \subprot{\prot_1}{\prot_2} \mid \ldots
  \end{align*}

  \bigskip
  \textbf{\Binder manipulation and resource transfer:}
  \begin{mathpar}
  \inferrule*[lab=\textlabel{SP-send-elim}{$\subprotop$-send-out},
      right=\textnormal{$\prot_1 \neq \protend$}]
    {\All \xdots. \iprop \wand \big(
      \subprot {\prot_1} {\sendprotT {} \val {\prot_2}}\big)}
    {\subprot {\prot_1} {\sendprot \xdots \val \iprop {\prot_2}}}
    \and
  \inferhref{$\subprotop$-send-in}{SP-send-intro}
    {\subst \iprop {\vec\var} {\vec\term}}
    {\subprot
      {\sendprot \xdots \val \iprop {\prot}}
      {\sendprotT{}
        {\subst \val {\vec\var} {\vec\term}}
        {\subst \prot {\vec\var} {\vec\term}}}}
    \and
  \inferrule*[lab=\textlabel{SP-recv-elim}{$\subprotop$-recv-out},
      right=\textnormal{$\prot_2 \neq \protend$}]
    {\All \xdots. \iprop \wand \big(
     \subprot {\recvprotT {} \val {\prot_1}} {\prot_2} \big)}
     {\subprot {\recvprot \xdots \val \iprop {\prot_1}} {\prot_2}}
    \and
  \inferhref{$\subprotop$-recv-in}{SP-recv-intro}
    {\subst \iprop {\vec\var} {\vec\term}}
    {\subprot
      {\recvprotT {}
        {\subst \val {\vec\var} {\vec\term}}
        {\subst \prot {\vec\var} {\vec\term}}}
      {\recvprot \xdots \val \iprop \prot}}
  \end{mathpar}

  \bigskip
  \textbf{Monotonicity and swapping:}
  \begin{mathpar}
  \inferhref{$\subprotop$-send-mono}{SP-send-mono}
    {\later (\subprot {\prot_1} {\prot_2})}
    {\subprot
      {\sendprotT {} {\val} {\prot_1}}
      {\sendprotT {} {\val} {\prot_2}}}
  \quad
  \inferhref{$\subprotop$-recv-mono}{SP-recv-mono}
    {\later (\subprot {\prot_1} {\prot_2})}
    {\subprot
      {\recvprotT {} {\val} {\prot_1}}
      {\recvprotT {} {\val} {\prot_2}}}
  \quad
  \inferhref{$\subprotop$-swap}{SP-swap}
    {}
    {\subprot
      {\recvprotT {} {\val} {\sendprotT {} {\valB} \prot}}
      {\sendprotT {} {\valB} {\recvprotT {} {\val} \prot}}}
  \end{mathpar}

  \medskip
  \textbf{Reflexivity and transitivity:}
  \begin{mathpar}
    \inferhref{$\subprotop$-refl}{SP-refl}
    {}
    {\subprot {\prot} {\prot}}
    \and
    \inferhref{$\subprotop$-trans}{SP-trans}
    {\subprot {\prot_1} {\prot_2} \and \subprot {\prot_2} {\prot_3}}
    {\subprot {\prot_1} {\prot_3}}
  \end{mathpar}

  \bigskip
  \textbf{Dual and append:}
  \begin{mathpar}
  \inferhref{$\subprotop$-dual}{SP-dual}
    {\subprot{\prot_2}{\prot_1}}
    {\subprot{\dual{\prot_1}}{\dual{\prot_2}}}
  \and
  \inferhref{$\subprotop$-append}{SP-append}
    {\subprot{\prot_1}{\prot_2} \and \subprot{\prot_3}{\prot_4}}
    {\subprot{\protapp{\prot_1}{\prot_3}}{\protapp{\prot_2}{\prot_4}}}
  \end{mathpar}

  \bigskip
  \textbf{Channel ownership:}
  \begin{mathpar}
  \inferhref{$\subprotop$-chan-mono}{SP-chan-mono}
    {\interp \chan {\prot_1} \and {\subprot {\prot_1} {\prot_2}}}
    {\interp \chan {\prot_2}}
  \end{mathpar}
  \caption{The grammar and primitive rules of \lname 2.0 for subprotocols.}
  \label{fig:subprotocol_rules}
\end{figure}}

\section{Subprotocols}
\label{sec:subprotocols}

\newcommand{\swapexamplesessiontype}{
  \begin{array}{@{}l@{}}
    \sendtypepoly{\tvarB_1, \tvarC_1}{(\arrtype{\tvarB_1}{\tvarC_1})}
    \sendtype{\tvarB_1}
    \sendtypepoly{\tvarB_2, \tvarC_2}{(\arrtype{\tvarB_2}{\tvarC_2})}
    \sendtype{\tvarB_2}
    \recvtype{\tvarC_1} \recvtype{\tvarC_2}\\
    \polymappertype
  \end{array}
}

This section describes \textbf{\lname 2.0}, which extends
\lname 1.0---as presented in the conference version of this
paper~\cite{hinrichsen-POPL2020}---with \emph{subprotocols}, inspired by
asynchronous subtyping of session types \cite{mostrous-ESOP2009,mostrous-IaC2015}.
The intention of both of these relations is to capture
protocol-preserving changes,
that allow for some internal flexibility of how an endpoint fulfills a protocol,
while being indistinguishable by the other endpoint.
In particular, subprotocols have two key features.
First, they exploit the asynchronous semantics of channels by relaxing the
notion of duality, thereby making it possible
to prove functional correctness of a larger class of programs.
Second, they give rise to a more extensional approach to reasoning about \pname,
as we can work up to the subprotocol relation rather than equality,
thereby providing more flexibility in the design and reuse of protocols.

We first introduce \lname 2.0's subprotocol relation and its proof rules
(\Cref{sec:subprotocol_relation}).
These should (similar to the \lname 1.0 logic, presented in~\Cref{sec:actris_logic})
be considered to be primitives of \lname;
in \Cref{sec:subprotocol_model} we define and prove them in Iris.
We then show how subprotocols can be employed to prove a mapper service,
which handles requests one at a time, while its client may send
multiple requests up front (\Cref{sec:subprotocol_swapping}).
Next, we demonstrate how the subprotocol relation allows for the composition of
slightly differing protocols, by composing a list reversal service whose protocol
is based on a list predicate that does not carry ownership, with a client whose
protocol is based on a list predicate that does carry ownership
(\Cref{sec:subprotocol_reuse}).
Finally, we show that the subprotocol relation is coinductive, and, when
combined with \ruleref{Loeb} induction, can be used to reason about recursive
protocols (\Cref{sec:subprotocol_recursion}).

\subsection{The subprotocol relation}
\label{sec:subprotocol_relation}

The \pname of channel endpoints are picked on channel creation (using the rule
\ruleref{Ht-new} shown in \Cref{fig:actris_logic}),
which then determines how the channel endpoints should interact.
To ensure safe communication, \lname adapts the notion of duality from session
types, which requires every send ($\SEND$) of one endpoint to be paired with a receive
($\RECV$) for the other endpoint, and \viceversa.
However, working with a channel's protocol and its dual is more restrictive than
strictly necessary.
Some variations from the original protocol preserve the
externally observed interaction, as the other endpoint is
agnostic to the variations in question, which will be made clear momentarily.
We capture some of these so-called protocol-preserving variations via a new
notion---the
\textit{subprotocol relation}---denoted as follows:
\[
  \subprot{\prot_1}{\prot_2}
\]
The subprotocol relation describes that protocol $\prot_1$ is \textit{stronger} than $\prot_2$,
or conversely, that protocol $\prot_2$ is \textit{weaker} than $\prot_1$.
More specifically, this means that $\prot_2$ can be used \textit{in place of}
$\prot_1$ whenever such a protocol is expected during verification.
This property is captured by the following monotonicity rule for channel
ownership:
\begin{mathpar}
\infer
  {\interp \chan {\prot_1} \and {\subprot {\prot_1} {\prot_2}}}
  {\interp \chan {\prot_2}}
\end{mathpar}
The subprotocol relation is inspired by asynchronous subtyping for session types
\cite{mostrous-ESOP2009,mostrous-IaC2015}, which allows (1) sending subtypes (contravariance),
(2) receiving supertypes (covariance), and (3) swapping sends ahead of receives.
These variations preserve the protocol, as (1) the originally expected type that
is to be sent can be derived
from the subtype, (2) the originally expected type to be received can be derived
from the supertype, and (3) sends do not block because channels are buffered in both directions, so messages can be enqueued ahead of time.
These variations, including the swapping property,
are generalised to \pname using the following proof rules:
\begin{mathpar}
\inferhref{$\subprotop$-send-mono'}{SP-send-mono'}
  {\All \xdots. \iprop_2 \wand \iprop_1 \and
   \All \xdots. \subprot {\prot_1} {\prot_2}}
  {\subprot
  {\sendprot \xdots {\val} {\iprop_1} {\prot_1}}
  {\sendprot \xdots {\val} {\iprop_2} {\prot_2}}}
  \and
\inferhref{$\subprotop$-recv-mono'}{SP-recv-mono'}
  {\All \xdots. \iprop_1 \wand \iprop_2 \and
   \All \xdots. \subprot {\prot_1} {\prot_2}}
  {\subprot
  {\recvprot \xdots {\val} {\iprop_1} {\prot_1}}
  {\recvprot \xdots {\val} {\iprop_2} {\prot_2}}}
  \and
\inferhref{$\subprotop$-swap'}{SP-swap'}
  {}
  {\subprot
  {\recvprot \xdots {\val} {\iprop} {\sendprot \xdotsB {\valB} {\ipropB} \prot}}
  {\sendprot \xdotsB {\valB} {\ipropB} {\recvprot \xdots {\val} {\iprop} \prot}}}
\end{mathpar}
The rules \ruleref{SP-send-mono'} and \ruleref{SP-recv-mono'}
use \emph{separation implication} $\iprop \wand \ipropB$---which states that
ownership of $\ipropB$ can be obtained by giving up ownership
of $\iprop$---to mimic the contra- and covariance of session subtyping.
The rule \ruleref{SP-swap'} states that sends can be swapped ahead of receives.
To be well-formed, this rule has the implicit side condition that $\xdots$ does not
bind into $\valB$ and $\ipropB$, and that $\xdotsB$ does not bind into $\val$
and $\iprop$.

To give an intuition behind the protocol-consistent changes that the above rules
capture, consider the following subprotocol derivation:
\[
\begin{array}{@{} l @{\ } l @{\qquad } l @{}}
    &\recvprot{(i:\integer)}{i}{i < 42}
      \sendprot{(j:\integer)}{j}{j > 42}
      \prot & \ruleref{SP-send-mono'}\\
    \subprotop
    &\recvprot{(i:\integer)}{i}{i < 42}
      \sendprot{(j:\integer)}{j}{j > 50}
      \prot & \ruleref{SP-recv-mono'} \\
    \subprotop
    &\recvprot{(i:\integer)}{i}{i < 40}
      \sendprot{(j:\integer)}{j}{j > 50}
      \prot & \ruleref{SP-swap'} \\
    \subprotop
    & \sendprot{(j:\integer)}{j}{j > 50}
      \recvprot{(i:\integer)}{i}{i < 40}
      \prot &
\end{array}
\]
Here, we first strengthen the proposition of the send (by increasing the bound
from $j > 42$ to $j > 50$), then weaken the proposition of the receive
(by reducing the bound from $i < 42$ to $i < 40$),
and finally swap the send ahead of the receive.

While the aforementioned rules cover the intuition behind \lname's subprotocol
relation, \lname's actual subprotocol rules provide a number of additional
features:

\subprotfig

\begin{enumerate}
\item They can be used to manipulate the \binders $\xdots$ that appear in protocols.
\item They can be used to transfer ownership of resources in and out of messages.
\item They can be used to reason about recursive protocols defined using
  \ruleref{Loeb} induction.
\end{enumerate}
The full set of primitive rules for subprotocols is shown in \Cref{fig:subprotocol_rules}.
The first four rules account for \binder manipulation and resource transfer:
Rules \ruleref{SP-send-elim} and \ruleref{SP-recv-elim}
generalise over the \binders $\xdots$ and transfer ownership of
$\iprop$ out of the weaker sending protocol $\sendprot \xdots \val \iprop {\prot}$,
and stronger receiving protocol $\recvprot \xdots \val \iprop {\prot}$,
respectively.
Rule \ruleref{SP-send-intro} weakens a sending protocol
$\sendprot \xdots \val \iprop {\prot}$ by instantiating the
\binders $\xdots$ and transferring ownership of $\subst{\iprop}{\vec{\var}}{\vec{\term}}$ into the protocol.
Dually, the rule \ruleref{SP-recv-intro} strengthens a receiving protocol
$\recvprot \xdots \val \iprop {\prot}$ by instantiating the
\binders $\xdots$ and transferring ownership of $\subst{\iprop}{\vec{\var}}{\vec{\term}}$ into the protocol.

To demonstrate the intuition behind these rules consider the following proof of
the subprotocol relation presented in \Cref{sec:subprotocol_intro},
where we transfer ownership of $\loc_1' \mapsto 20$ into a protocol,
while instantiating the \binder $\loc_1$ with $\loc_1'$:
\newcommand{\linetext}[1]{\smash{\raisebox{-6pt}{\rlap{\ #1}}}}
\[
\def\arraystretch{1.3}
\hspace{-5.5em}
\begin{array}[b]{@{} l @{\ } c @{\ } l @{} l @{}}
  \hline
  \loc_1' \mapsto 20 \ast \loc_2 \mapsto 22 \wand
  \loc_1' \mapsto 20 \ast \loc_2 \mapsto 22
  & &  &  \linetext{\ruleref{SP-send-intro}} \\ \hline
  \begin{array}[b]{@{}l@{}}
    \loc_1' \mapsto 20 \ast \loc_2 \mapsto 22 \wand \\[-0.3em]
    \quad\quad\sendprot {(\loc_1,\loc_2\!:\!\Loc)} {({\loc_1},\loc_2)}
      {{\loc_1 \!\mapsto\! 20 * \loc_2 \!\mapsto\! 22}} \prot
  \end{array}
  & \subprotop &
   \sendprotT {} {(\loc_1',\loc_2)}
    \prot
   & \linetext{\ruleref{SP-send-elim}} \\ \hline
  \begin{array}[b]{@{}l@{}}
    \loc_1' \mapsto 20 \wand \\[-0.3em]
    \quad\quad\sendprot {(\loc_1,\loc_2\!:\!\Loc)} {(\loc_1,\loc_2)}
      {\loc_1 \!\mapsto\! 20 * \loc_2 \!\mapsto\! 22} \prot
  \end{array}
    & \subprotop \\[-0.3em]
   \quad\quad\sendprot {(\loc_2\!:\!\Loc)} {(\loc_1',\loc_2)}
    {{\loc_2 \!\mapsto\! 22}} \prot
\end{array}
\]
We first use rule \ruleref{SP-send-elim} to
generalise over the \binder $\loc_2$ and transfer
ownership of $\loc_2 \mapsto 22$ out of the weaker protocol (\ie the send on the
RHS), and then use \ruleref{SP-send-intro} to instantiate the \binders $\loc_1'$
and $\loc_2$ and transfer ownership of $\loc_1' \mapsto 20$ and $\loc_2 \mapsto 22$
into the stronger protocol (\ie the send on the LHS).

The rules for monotonicity (\ruleref{SP-send-mono} and \ruleref{SP-recv-mono})
and swapping (\ruleref{SP-swap}) in \Cref{fig:subprotocol_rules} differ in two aspects
from the rules for monotonicity (\ruleref{SP-send-mono'} and \ruleref{SP-recv-mono'})
and swapping (\ruleref{SP-swap'}) that we have seen in the beginning of this
section.
First, the actual rules only apply to protocols whose head does not have
\binders $\xdots$ and resources $\iprop$, \ie protocols of the shape
$\sendprotT {} {\val} {\prot}$ or $\recvprotT {} {\val} {\prot}$, instead of
those of the shape $\sendprot \xdots {\val} {\iprop} {\prot}$ or
$\recvprot \xdots {\val} {\iprop} {\prot}$.
While this restriction might seem to make the rules more restrictive, the more general rules
for monotonicity (\ruleref{SP-send-mono'} and \ruleref{SP-recv-mono'})
and swapping (\ruleref{SP-swap'}) are derivable from these simpler rules.
This is done using the rules for \binder manipulation and resource
transfer.
Second, the actual rules for monotonicity have a later modality ($\later$) in
their premise.
The later modality makes these rules stronger (by \ruleref{Later-intro}
we have that $\iprop$ entails $\later \iprop$), and thereby internalizes its
coinductive nature into the \lname logic so \ruleref{Loeb} induction can be used to
prove subprotocol relations for recursive protocols (\Cref{sec:subprotocol_recursion}).

The remaining rules in \Cref{fig:subprotocol_rules}
express that the subprotocol relation is
reflexive (\ruleref{SP-refl}) and transitive (\ruleref{SP-trans}),
as well as that the dual operation is anti-monotone (\ruleref{SP-dual}) and
the append operation is monotone (\ruleref{SP-append}).

Let us consider the following subprotocol relation to provide some further insight
into the expressivity of our rules, (where \binders are omitted for simplicity):
\[
  \subprot
  { \sendprot{}{\val}{ \iprop }
   \recvprot{}{\valB}{ \ipropB }
   \prot}
  {\sendprot{}{\val}{ {\iprop \ast \ipropC} }
   \recvprot{}{\valB}{ \ipropB \ast \ipropC }
   \prot}
\]
Here we extend the protocol
$\sendprot{}{\val}{ \iprop } \recvprot{}{\valB}{ \ipropB } \prot$
with a \emph{frame} $\ipropC$.
The proposition $\ipropC$ describes resources that can be sent along with the
originally
expected resources $\iprop$, and which are reacquired along with the resources
$\ipropB$ that are sent back.
We demonstrate the usefulness of this notion of framing at the protocol level
in~\Cref{sec:subprotocol_reuse}.

The above subprotocol relation mimics the frame rule of separation logic
(\ruleref{Ht-frame}), which makes it possible to apply specifications while
maintaining a \textit{frame} of resources $\ipropC$:
\begin{mathpar}
\infer
  {\hoare{\iprop}\expr{\Ret\valB. \ipropB}}
  {\hoare{\iprop * \ipropC}\expr{\Ret\valB. \ipropB * \ipropC}}
\end{mathpar}
The frame-like subprotocol relation is proven as follows:
\[
\def\arraystretch{1.3}
\hspace{-10em}
\begin{array}{@{} r @{\ } r @{\ } c @{\ } l @{} l @{}}
& & & & \linetext{\ruleref{SP-recv-intro}} \\ \hline
\ipropB \ast \ipropC \wand & \subprot
  {\recvprotT{}{\valB} \prot &}
  {& \recvprot{}{\valB}{\ipropB \ast \ipropC} \prot}
  & \linetext{\ruleref{SP-recv-elim}} \\ \hline
\ipropC \wand & \subprot
  {\recvprot{}{\valB}{\ipropB} \prot &}
  { & \recvprot{}{\valB}{\ipropB \ast \ipropC} \prot}
  & \linetext{\ruleref{SP-send-mono}, \ruleref{Later-intro}} \\ \hline
\ipropC \wand & \subprot
  { \sendprotT{}{\val} \recvprot{}{\valB}{\ipropB} \prot &}
  {& \sendprotT{}{\val} \recvprot{}{\valB}{\ipropB \ast \ipropC} \prot}
  & \linetext{\ruleref{SP-send-intro}, \ruleref{SP-trans}} \\ \hline
\iprop \ast \ipropC \wand & \subprot
  {\sendprot{}{\val}{\iprop} \recvprot{}{\valB}{\ipropB} \prot &}
  {& \sendprotT{}{\val} \recvprot{}{\valB}{\ipropB \ast \ipropC} \prot}
  & \linetext{\ruleref{SP-send-elim}} \\ \hline
& \subprot
  {\sendprot{}{\val}{\iprop} \recvprot{}{\valB}{\ipropB} \prot &}
  {& \sendprot{}{\val}{\iprop \ast \ipropC} \recvprot{}{\valB}{\ipropB \ast \ipropC} \prot}
\end{array}
\]
We use rule \ruleref{SP-send-elim} to transfer $\iprop$ and the frame $\ipropC$ out of the
weaker protocol (\ie the send on the RHS), and then use rule \ruleref{SP-send-intro}
to transfer $\iprop$ into the stronger protocol (\ie the send on the LHS), leaving us
with a context in which we still own the frame $\ipropC$.
We then use rule \ruleref{SP-send-mono} to proceed with the receiving
part of the protocol in a dual fashion---we use rule
\ruleref{SP-recv-elim} to transfer out $\ipropB$ of the stronger protocol (\ie the
receive on the LHS), and use rule \ruleref{SP-recv-intro} to transfer $\ipropB$ and the frame $\ipropC$
into the weaker protocol (\ie the receive on the RHS).

\subsection{Swapping}
\label{sec:subprotocol_swapping}

\newcommand{\recvallnamefixed}{\mathtt{recvN}}

\newcommand{\mapperfig}{
\begin{figure}[t!]
\begin{equation*}
\begin{array}[t]{@{} l @{}}
  \mapperservicename\ \vmapvar\ \chan \langdef\\
  \quad
  \begin{array}[t]{@{} l @{}}
    \mbranchA{\chan}
    {\begin{array}[t]{@{} l @{}}
        \Let \var = \recv \chan in\\
        \Let \varB = \vmapvar\ \var in\\
        \send \chan {\varB};\\
        \mapperservicename\ \vmapvar\ \chan
     \end{array}}
    {\TT}
  \end{array}
\end{array}
\qquad\qquad
\begin{array}[t]{@{} l @{}}
  \mapperclientname\ \vmapvar\ \listvar\ \langdef\\
  \quad
  \begin{array}[t]{@{} l @{}}
    \Let {\chan} = {\start {(\mapperservicename\ \vmapvar)}} in\\
    \Let {n} = \listlength{l} in\\
    \sendallname\ \chan\ \listvar;\\
    \recvallnamefixed\ \chan\ \listvar\ n;\\
    \select{\chan}{\rightname};
  \end{array}
\end{array}
\end{equation*}
\caption{A mapper service whose verification relies on swapping
(the code for the functions $\sendallname$ and $\recvallnamefixed$ has been elided).}
\label{fig:mapper}
\end{figure}
}
\mapperfig
Subprotocols make it possible to verify message-passing
programs whose order of sends and receives does not match up w.r.t.\
duality.
As an example of such a program, let us consider the mapper service and client in
\Cref{fig:mapper}.
The service $\mapperservicename\ \vmapvar\ \chan$ is a loop, which
iteratively receives an element over channel endpoint $\chan$,
maps the function $\vmapvar$ over that element, and sends the resulting value back.
Conversely, the client $\mapperclientname\ \vmapvar\ \listvar$
sends all of the elements of the list
$\listvar$ up front, and only requests the mapped results back once all elements have
been sent.
Since the service interleaves the sends and receives, while the client does not,
the \pname for the service and client cannot be dual of each other.
However, the communication between the service and client is in fact safe as
messages are buffered.
We now show that using subprotocols we can prove that this is indeed the case.
We define the protocol based on the communication where sends and receives are
interleaved:
\[
\begin{array}{@{} l @{}}
  \mapperprotname\
  (\interpvar_\tvar : \tvar \to \Val \to \iProp)\
  (\interpvar_\tvarB : \tvarB \to \Val \to \iProp)\
  (\mapvar : \tvar \to \tvarB) \eqdef\\
  \quad
  \begin{array}{@{} l @{}}
    \MU (\recvar : \iProto).
    \selectprot
    {(
    \sendprot{(\var : \tvar)\ (\val : \Val)}
    \val
    {\interpvar_\tvar\ \var\ \val}
    \recvprot{(\valB : \Val)}{\valB}
    {\interpvar_{\tvarB}\ {(\mapvar\ \var)}\ \valB}
    \recvar)}
    {\protend}
  \end{array}
\end{array}
\]
The protocol is parameterised by representation predicates $\interpvar_\tvar$
and $\interpvar_\tvarB$ that relate \heaplang values to elements of type $\tvar$
and $\tvarB$ in the Iris/\lname logic, and a function $\mapvar : \tvar \to \tvarB$
in Iris/\lname that specifies the behaviour of the \heaplang function $\vmapvar$.
The connection between $\mapvar$ and $\vmapvar$ is formalised as:
\[
\begin{array}{l}
\mapspecname\
  (\interpvar_\tvar : \tvar \to \Val \to \iProp)\
  (\interpvar_\tvarB : \tvarB \to \Val \to \iProp)\
  (\mapvar : \tvar \to \tvarB)\
  (\vmapvar : \Val) \eqdef \\
\quad\All \var\,\val.
  \hoare{\interpvar_\tvar\ \var\ \val}
  {\vmapvar\ \val}
  {\Ret \valB.
  \interpvar_\tvarB\ {(\mapvar\ \var)}\ \valB
  }
\end{array}
\]
Since $\mapperprotname$ describes an interleaved sequence of transactions,
$\mapperservicename$ can be readily verified against the protocol
$\dual \mapperprotname$ using just the symbolic execution rules of
\lname 1.0 as presented in \Cref{sec:mp_rules}.
However, to verify $\mapperclientname$ against the protocol
$\mapperprotname$, we need to weaken the protocol using the rules for
subprotocols of \lname 2.0.
Given a list of $n$ elements, the subprotocol relation (together with an
intermediate step) that describes this weakening is:
\[
\begin{array}[t]{@{} l @{\ } l @{} @{\qquad} l}
&\mapperprotname\ \interpvar_\tvar\ \interpvar_\tvarB\ \mapvar \\
\subprotop
&
  \begin{array}[t]{@{} l @{}}
    \sendprotT{}{\leftname}
    \sendprot{(\var_1 : \tvar)\ (\val_1 : \Val)}
      {\val_1}
      {\interpvar_\tvar\ \var_1\ \val_1} \\
    \recvprot{(\varB_1 : \tvarB)}{\varB_1}
      {\interpvar_{\tvarB}\ {(\mapvar\ \var_1)}\ \varB_1}
    \cdots\\
    \sendprotT{}{\leftname}
    \sendprot{(\var_n : \tvar)\ (\val_n : \Val)}
      {\val_n}
      {\interpvar_\tvar\ \var_n\ \val_n}\\
    \recvprot{(\varB_n : \tvarB)}{\varB_n}
      {\interpvar_{\tvarB}\ {(\mapvar\ \var_n)}\ \varB_n}
    \\
    \mapperprotname\ \interpvar_\tvar\ \interpvar_\tvarB\ \mapvar
  \end{array} &
  \begin{array}[t]{@{} l}
  \textnormal{$n$ times \ruleref{rec-unfold} and}\\
  \textnormal{weaken $\selectop$ into $\genprotTHead{\SEND}{}{\leftname}$}
  \end{array} \\
\subprotop
&
  \begin{array}[t]{@{} l @{}}
    \sendprotT{}{\leftname}
    \sendprot{(\var_1 : \tvar)\ (\val_1 : \Val)}
      {\val_1}
      {\interpvar_\tvar\ \var_1\ \val_1}
      \cdots\\
    \sendprotT{}{\leftname}
    \sendprot{(\var_n : \tvar)\ (\val_n : \Val)}
      {\val_n}
      {\interpvar_\tvar\ \var_n\ \val_n}\\
    \recvprot{(\varB_1 : \tvarB)}{\varB_1}
      {\interpvar_{\tvarB}\ {(\mapvar\ \var_1)}\ \varB_1}
      \cdots\\
    \recvprot{(\varB_n : \tvarB)}{\varB_n}
     {\interpvar_{\tvarB}\ {(\mapvar\ \var_n)}\ \varB_n}
    \\
    \mapperprotname\ \interpvar_\tvar\ \interpvar_\tvarB\ \mapvar
  \end{array} & \textnormal{$n$ times \ruleref{SP-swap'}}
\end{array}
\]
Both steps are proven by induction on $n$.
In the first step, we unfold the recursive protocol $n$ times using
\ruleref{rec-unfold}, and use the derived rule
$\subprot {(\prot_1 \selectop \prot_2)} {\sendprotT{}{\leftname}{\prot_1}}$
to weaken the choices to the left choice $\leftname$.
Recall from \Cref{sec:choice} that $\selectop$
is defined in terms of the send protocol ($\SEND$).
This allows us to prove the derived rule
$\subprot {(\prot_1 \selectop \prot_2)} {\sendprotT{}{\leftname}{\prot_1}}$
using \ruleref{SP-send-elim} and \ruleref{SP-send-intro}.
The second step involves swapping all sends ahead of the receives using the rule
\ruleref{SP-swap'}.

The weakened protocol that we have obtained follows the behaviour of the client,
making its verification straightforward using \lname's rules for symbolic execution.
Concretely, we prove the following specifications for the service and the client:
\newcommand{\mapname}{\defemph{map}}
\[
\begin{array}{c !{\ } c}
\begin{array}{l}
  \hoareV
    {
  \begin{array}{@{} l @{}}
    \mapspecname\
     \interpvar_\tvar\
     \interpvar_\tvarB\
     \mapvar\ \vmapvar\ *
  \interp \chan
  {\protapp{\dual{\mapperprotname\ \interpvar_\tvar\ \interpvar_\tvarB\ \mapvar}}
    {\prot}}
    \end{array}}
  { \mapperservicename\ \vmapvar\ \chan}
    {\interp \chan {\prot}}
\end{array}
  &
\begin{array}{@{} l @{}}
  \hoareV
    {
     \mapspecname\
     \interpvar_\tvar\
     \interpvar_\tvarB\
     \mapvar\ \vmapvar\ *
  \llistrefI {\interpvar_\tvar} \loc {\vec\var}}
  { \mapperclientname\ \vmapvar\ \loc}
    { \llistrefI {\interpvar_\tvarB} \loc {\mapname\ \mapvar\ \vec\var}}
\end{array}
\end{array}
\]

\subsection{Protocol compositionality}
\label{sec:subprotocol_reuse}

\newcommand{\tvarF}{R}
\newcommand{\listrevname}{\mathtt{list\_rev}}
\newcommand{\listrevservicename}{\listrevname\_{\mathtt{service}}}
\newcommand{\listrevclientname}{\listrevname\_{\mathtt{client}}}
\newcommand{\listrevprotname}{\logdefemph{list\_rev\_prot}}

\newcommand{\listrevfig}{
\begin{figure}[t!]
\begin{equation*}
\begin{array}[t]{@{} l @{}}
  \listrevservicename\ \chan \langdef\\
  \quad
  \Let \listvar = \recv{\chan} in\\
  \quad \llistreverse\ \listvar;\ \send \chan {\TT}
\end{array}
\qquad
\begin{array}[t]{@{} l @{}}
  \listrevclientname\ \listvar\ \langdef\\
  \quad
  \begin{array}[t]{@{} l @{}}
    \Let {\chan} = {\start {\listrevservicename}} in\\
    \send \chan \listvar;\ \recv{\chan}
  \end{array}
\end{array}
\end{equation*}
\caption{A list reversing service
(the code for the function $\llistreverse$ has been elided).}
\label{fig:listrevprog}
\end{figure}
}

\listrevfig

An essential feature of separation logic is the ability to compose specifications
of different libraries, so that each library can be defined and verified once
against its own specification, while being used in the context of slightly
differing specifications and proofs of other libraries.
To achieve a similar property for our dependent separation protocols we would
similarly like to be able to compose compatible protocols.

A key ingredient that enables such compositionality in traditional separation logic is
the frame rule (\ruleref{Ht-frame}).
In \Cref{sec:subprotocol_relation} we demonstrated how subprotocols allow for
similar framing in our protocols.
In this section we give a more detailed example of such framing in our protocols
by considering the service $\listrevservicename\ \chan$ in \Cref{fig:listrevprog},
which receives a linked list over channel endpoint $\chan$, reverses it, and
sends it back over $\chan$.

To specify this service, we could use a protocol similar to the sorting service
in \Cref{sec:basics}, defined in terms of the
representation predicate $\llistrefI{\interpvar_\tvar}{\loc}{\vec{\var}}$ for
linked lists:
\[
\listrevprotname_{\interpvar_\tvar} \eqdef{}
  \sendprot{(\loc:\Loc)(\vec{\var}:\List\ \tvar)}{\loc}
    {\llistrefI{\interpvar_\tvar}{\loc}{\vec{\var}}}
  \recvprot{}{\TT}
    {\llistrefI{\interpvar_\tvar}{\loc}{\listrev{\vec{\var}}}}
  \protend
\]
Although it is possible to verify the service against the protocol
$\dual {\listrevprotname_{\interpvar_\tvar}}$, this approach is not quite satisfactory.
Unlike the sorting service, the reversal service does not access the list
elements, but only changes the structure of the list.
Hence, there is no need to keep track of the ownership of the elements through
the predicate $\interpvar_\tvar$.
A self-contained and simpler protocol for this service would instead be the
following:
\[
\listrevprotname \eqdef{}
  \sendprot{(\loc:\Loc)(\vec{\val}:\List\ \Val)}{\loc}{\llistref{\loc}{\vec{\val}}}
  \recvprot{}{\TT}{\llistref{\loc}{\listrev{\vec{\val}}}}
  \protend
\]
Here, $\llistref{\loc}{\vec{\val}}$ is a version of the list representation
predicate that does not keep track of the resources of the elements, but only
describes the structure of the list.
It is defined as:
\begin{equation*}
\llistref \loc {\vec \val} \eqdef
  \begin{cases}
  \loc \mapsto \Inl\;\TT & \textnormal{if $\vec\val = \nil$} \\
  \Exists \loc_2.
    \loc \mapsto \Inr\;(\val_1,\loc_2) \ast
    \llistref {\loc_2} {\vec\val_2}
    & \textnormal{if $\vec\val = [\val_1] \cdot \vec\val_2$}
  \end{cases}
\end{equation*}
However, once we have verified the service against the simple protocol,
the proof of a client might prefer to interact with the list reversal
service through the general protocol $\listrevprotname_{\interpvar_\tvar}$.
Doing so can be achieved by proving the subprotocol relation
$\subprot \listrevprotname {\listrevprotname_{\interpvar_\tvar}}$.
To prove this subprotocol relation, we first establish the following relation
between the two versions of the list representation predicate:
\[
  \textstyle
  \llistrefI{\interpvar_\tvar}{\loc}{\vec{\var}} \ \wandIff \
  (\Exists \vec{\val}.
  \llistref{\loc}{\vec{\val}} \ast
  \Sep_{(\var,\val) \in (\vec{\var},\vec{\val})}.
  \interpvar_\tvar\ \var\ \val)
  \tagH{list-rel}
\]
Here, $\Sep_{(\var,\val) \in (\vec{\var},\vec{\val})}$ is the pairwise iterated
separation conjunction over two lists of equal length,
and $\wandIff$ is a bi-directional separation implication.
The above result thus states that $\llistrefI{\interpvar_\tvar}{\loc}{\vec{\var}}$
can be split into two parts, ownership of the links of the list $\llistref{\loc}{\vec\val}$,
and a range of interpretation predicates $\interpvar_\tvar$
for each element of the list, and \viceversa.
With this result at hand, the proof of the desired subprotocol relation is
carried out as follows:
\[
\begin{array}{@{} l @{\ } l @{}}
  &\listrevprotname \\
={} &
  \sendprot{(\loc:\Loc)(\vec{\val}:\List\ \Val)}{\loc}{\llistref{\loc}{\vec{\val}}}
    \recvprot{}{\TT}{\llistref{\loc}{\listrev{\vec{\val}}}}
    \protend \\
\subprotop &
  \sendprot{(\loc:\Loc)(\vec{\val}:\List\ \Val)(\vec{\var}:\List\ \tvar)}{\loc}
    {\llistref{\loc}{\vec{\val}} \ast
     \Sep_{(\var,\val) \in (\vec{\var},\vec{\val})}. \interpvar_\tvar\ \var\ \val} \\
  &
  \recvprot{}{\TT}
    {\llistref{\loc}{(\listrev\ \vec{\val})} \ast
     \Sep_{(\var,\val) \in (\vec{\var},\vec{\val})}. \interpvar_\tvar\ \var\ \val}
  \protend \\
\subprotop &
  \sendprot{(\loc:\Loc)(\vec{\var}:\List\ \tvar)}{\loc}
    {\llistrefI{\interpvar_\tvar}{\loc}{\vec{\var}}}
  \recvprot{}{\TT}
    {\llistrefI{\interpvar_\tvar}{\loc}{\listrev{\vec{\var}}}}
  \protend \\
={}& \listrevprotname_{\interpvar_\tvar}
\end{array}
\]
We first frame the range of interpretation predicates owned by the list
$\Sep_{(\var,\val) \in (\vec{\var},\vec{\val})}. \interpvar_\tvar\ \var\ \val$,
using an approach similar to the frame example in \Cref{sec:subprotocol_relation},
and then use \ruleref{list-rel} to combine it with $\llistref{\loc}{\vec{\val}}$ and
$\llistref{\loc}{\listrev{\vec{\val}}}$ for the sending and receiving step,
to turn them into $\llistrefI{\interpvar_\tvar}{\loc}{\vec{\var}}$ and
$\llistrefI{\interpvar_\tvar}{\loc}{\listrev{\vec{\var}}}$, respectively.
Note that the \binder $\vec{\val}$ is changed into $\vec{\var}$,
using the subprotocol rules for \binder manipulation.
With this subprotocol relation at hand, it is possible to prove the following
specifications for the service and client:
\[
\begin{array}{c !{\quad} c}
\begin{array}{l}
  \hoareV
  { \interp{\chan}{\protapp {\dual{\listrevprotname}} {\prot}} }
  { \listrevservicename\ \chan }
  { \interp{\chan}{\prot} }
\end{array}
\qquad
\begin{array}{l}
  \hoareV
  { \llistrefI {\interpvar_{\tvar}} \loc {\vec\var} }
  { \listrevclientname\ \loc}
  { \llistrefI {\interpvar_{\tvar}} \loc {\listrev \vec\var} }
\end{array}
\end{array}
\]

\subsection{Subprotocols and recursion}
\label{sec:subprotocol_recursion}

\newcommand{\listrevrecprotname}{\listrevprotname_{\prot}}

We conclude this section by showing how subprotocol relations involving recursive
protocols can be proven using \ruleref{Loeb} induction.
Recall from \Cref{sec:tour} that the principle of \ruleref{Loeb} induction
is as follows:
\begin{align*}
\infer
  {\later\iprop\Ra\iprop}
  {\iprop}
\end{align*}
By letting $\iprop$ be $\subprot {\prot_1} {\prot_2}$, we can
prove $\subprot {\prot_1} {\prot_2}$ using the induction hypothesis
$\later (\subprot {\prot_1} {\prot_2})$.
The later modality ($\later$) ensures that the induction hypothesis is not
used immediately, but a monotonicity rule for
send (\ruleref{SP-send-mono}) or receive (\ruleref{SP-recv-mono}) is applied
first.
This is done typically after unfolding the recursion operator using \ruleref{rec-unfold}.
The monotonicity rules \ruleref{SP-send-mono} or \ruleref{SP-recv-mono} contain
a later modality ($\later$) in their premise, which makes it possible to strip off the
later of the induction hypotheses (by rule \ruleref{later-mono} for monotonicity of $\later$).

Our approach for proving subprotocol relations using \ruleref{Loeb} induction
is similar to the approach of Brandt and Henglein~\cite{brandt-1998FI} for proving
subtyping relations for recursive types using coinduction.
Brand and Henglein~\cite{brandt-1998FI} however have a syntactic restriction on
proofs to ensure that the induction hypothesis is not used immediately (\ie is
used in a \emph{contractive} fashion), while
we use the later modality ($\later$) of Iris to achieve that.

To demonstrate how our approach works, we prove $\subprot {\prot_1} {\prot_2}$,
where:
\begin{align*}
\prot_1 \eqdef{}&
  \MU (\recvar:\iProto). \selectprot {(\protapp \listrevprotname \recvar)} \protend \\
\prot_2 \eqdef{}&
  \MU (\recvar:\iProto). \selectprot {(\protapp {\listrevprotname_{\interpvar_\tvar}} \recvar)} \protend
\end{align*}
Here, $\listrevprotname$ and $\listrevprotname_{\interpvar_\tvar}$ are the
protocols from \Cref{sec:subprotocol_reuse}, for which we have already
proven $\subprot \listrevprotname {\listrevprotname_{\interpvar_\tvar}}$.
The proof of $\subprot {\prot_1} {\prot_2}$ is as follows:
\[
\def\arraystretch{1.3}
\hspace{-5em}
\begin{array}{@{} r @{\ } c @{\ } r @{\ } c @{\ } l @{} l @{}}
  \hline
  \subprot {\prot_1} {\prot_2} & \wand &
  \subprot {\prot_1 &} {&\prot_2}
  & \linetext{\ruleref{SP-append}} \\ \hline
  \subprot {\prot_1} {\prot_2} & \wand &
  \subprot
    {\protapp \listrevprotname {\prot_1} &}
    {& \protapp {\listrevprotname_{\interpvar_\tvar}} {\prot_2}}
    & \linetext{\ruleref{later-mono}} \\ \hline

  \later (\subprot {\prot_1} {\prot_2}) & \wand  &
  \later (\subprot
    {\protapp \listrevprotname {\prot_1} &}
    {& \protapp {\listrevprotname_{\interpvar_\tvar}} {\prot_2}})
  & \linetext{\ruleref{select-mono}} \\ \hline
  \later (\subprot { \prot_1} {\prot_2 }) & \wand &
  \subprot
    {\selectprot {(\protapp \listrevprotname {\prot_1})} \protend &}
    {&\selectprot {(\protapp {\listrevprotname_{\interpvar_\tvar}} {\prot_2})} \protend}
  & \linetext{\ruleref{rec-unfold}} \\ \hline
  \later (\subprot {\prot_1} {\prot_2}) & \wand &
    \subprot {\prot_1 &} {& \prot_2}
  & \linetext{\ruleref{Loeb}} \\ \hline
  && \subprot {\prot_1 &} {& \prot_2}
\end{array}
\]
The proof starts with the \ruleref{Loeb} rule, followed by unfolding the
recursive types with \ruleref{rec-unfold}.
We then proceed with the following derived rule for monotonicity of selection ($\selectop$):
\begin{mathpar}
\inferhref{$\selectop$-mono}{select-mono}
  {\later (\subprot{\prot_1}{\prot_2} \wedge \subprot{\prot_3}{\prot_4})}
  {\subprot{(\selectprot{\prot_1}{\prot_3})}{(\selectprot{\prot_2}{\prot_4})}}
\end{mathpar}
Due to the regular conjunction in the premise, the same resources can be used to
prove both branches of $\selectop$.
This is sound because only one branch of $\selectop$ will be chosen.
The rule \ruleref{select-mono} follows from \ruleref{SP-send-mono} as selection ($\selectop$)
is defined in terms of send ($\SEND$).

We continue the main proof with monotonicity of the later modality
(\ruleref{later-mono}),
which lets us strip
off the later of the induction hypothesis $\later (\subprot {\prot_1} {\prot_2})$.
We then use \ruleref{SP-append}, along with
$\subprot \listrevprotname {\listrevprotname_{\interpvar_\tvar}}$,
which we have proven in \Cref{sec:subprotocol_reuse}.
The remaining proof obligation $\subprot{\prot_1}{\prot_2}$ follows
from the induction hypothesis.

While the protocols in the prior examples are similar in structure, our approach
scales to protocols for which that is not the case.
For example, consider $\subprot {\prot_1} {\prot_2}$, where:
\begin{align*}
\prot_1 \eqdef{}&
  \MU (\recvar:\iProto).
    \sendprotT{(\var:\integer)}{\var}
    \recvprotT{}{\var+2}
    \recvar \\
\prot_2 \eqdef{}&
  \MU (\recvar:\iProto).
    \sendprotT{(\var:\integer)}{\var}
    \sendprotT{(\varB:\integer)}{\varB}
    \recvprotT{}{\var+2}
    \recvprotT{}{\varB+2}
    \recvar
\end{align*}
Intuitively, these protocols are related, as we can unfold the body of $\prot_1$
twice, the body of $\prot_2$ once, and swap the second receive over the first send.
The proof is as follows:
\[
\def\arraystretch{1.3}
\hspace{-10em}
\begin{array}{@{} r @{\ } c @{\ } r @{} l @{\ } l @{} l @{}}
  \hline
  \subprot {\prot_1} {\prot_2} & \wand &
  \subprot {\prot_1 &\ } {&\prot_2}
  & \linetext{\ruleref{SP-recv-mono}, \ruleref{Later-intro}} \\ \hline
  \subprot {\prot_1} {\prot_2} & \wand &
  \subprot
    {\recvprotT{}{\var\!+\!2}
     \recvprotT{}{\varB\!+\!2}
     \prot_1 &\ }
    {\\[-0.3em] && \recvprotT{}{\var\!+\!2}
     \recvprotT{}{\varB\!+\!2}
     \prot_2}
  &&& \linetext{\ruleref{SP-send-mono'}, \ruleref{Later-intro}} \\ \hline
  \subprot {\prot_1} {\prot_2} & \wand &
  \subprot
    {\sendprotT{\varB}{\varB}
     \recvprotT{}{\var\!+\!2}
     \recvprotT{}{\varB\!+\!2}
     \prot_1 &\ }
    {\\[-0.3em] && \sendprotT{\varB}{\varB}
     \recvprotT{}{\var\!+\!2}
     \recvprotT{}{\varB\!+\!2}
     \prot_2}
  &&& \linetext{\ruleref{SP-swap'}, \ruleref{SP-trans}} \\ \hline
    \subprot {\prot_1} {\prot_2} & \wand &
    \subprot
    {\recvprotT{}{\var\!+\!2}
     \sendprotT{\varB}{\varB}
     \recvprotT{}{\varB\!+\!2}
     \prot_1 &\ }
    {\\[-0.3em] && \sendprotT{\varB}{\varB}
     \recvprotT{}{\var\!+\!2}
     \recvprotT{}{\varB\!+\!2}
     \prot_2}
    &&& \linetext{\ruleref{later-mono}} \\ \hline

    \later(\subprot {\prot_1} {\prot_2}) & \wand &
    \later (
  \subprot
    {\recvprotT{}{\var\!+\!2}
     \sendprotT{\varB}{\varB}
     \recvprotT{}{\varB\!+\!2}
     \prot_1 &\ }
    {\\[-0.3em] && \sendprotT{\varB}{\varB}
     \recvprotT{}{\var\!+\!2}
     \recvprotT{}{\varB\!+\!2}
     \prot_2}
  &)&& \linetext{\ruleref{SP-send-mono'}} \\ \hline
  \later (\subprot {\prot_1} {\prot_2}) & \wand &
  \subprot
    {\sendprotT{\var}{\var}
     \recvprotT{}{\var\!+\!2}
     \sendprotT{\varB}{\varB}
     \recvprotT{}{\varB\!+\!2}
     \prot_1 &\ }
    {\\[-0.3em] && \sendprotT{\var}{\var}
     \sendprotT{\varB}{\varB}
     \recvprotT{}{\var\!+\!2}
     \recvprotT{}{\varB\!+\!2}
     \prot_2}
  &&& \linetext{\ruleref{rec-unfold}} \\ \hline
  \later (\subprot {\prot_1} {\prot_2}) & \wand &
    \subprot {\prot_1 &\ } {& \prot_2}
  & \linetext{\ruleref{Loeb}} \\ \hline
  && \subprot {\prot_1 &\ } {& \prot_2}
\end{array}
\]
After we use \ruleref{SP-send-mono'} for the first time, we strip
off the later of the induction hypothesis
$\later (\subprot {\prot_1} {\prot_2})$, using \ruleref{later-mono}.
Subsequently, when we use \ruleref{SP-send-mono'} and \ruleref{SP-recv-mono},
there are no more laters to strip.
We therefore instead introduce the laters using \ruleref{Later-intro} before
applying the appropriate subprotocol monotonicity rule.

\section{Manifest sharing via locks}
\label{sec:integration}

Since \pname and the connective $\interp \chan \prot$ for ownership of protocols
are first-class objects of the \lname logic, they can be used like any other
logical connective.
This means that protocols can be combined with any other
mechanism that \lname inherits from Iris.
In particular, they can be combined with Iris's generic invariant and ghost
state mechanism, and can be used in combination with Iris's
abstractions for reasoning about other concurrency connectives like locks,
barriers, lock-free data structures, \etc

In this section we demonstrate how \pname can be combined with lock-based
concurrency.
This combination allows us to prove functional correctness of programs that
make use of the notion of \emph{manifest sharing} \cite{balzer-PACMPL2017, balzer-ESOP2019},
where channel endpoints are shared between multiple parties.
Instead of having to extend \lname, we make use of the locks and ghost state that
\lname inherits from Iris.
We present the basic idea with a simple introductory example of sharing a channel
endpoint between two parties (\Cref{sec:locks}).
We then consider a more challenging example of a channel-based load-balancing
mapper (\Cref{sec:mapper}).

\subsection{Locks and ghost state}
\label{sec:locks}

\newcommand{\locksfig}{
\begin{figure}
\centering
\textbf{Grammar:}
\begin{align*}
  \term , \termB , \iprop , \ipropB, \prot \bnfdef{}&
  \ldots \mid \islock \lockvar \iprop \mid \ldots
\end{align*}
\medskip
\textbf{Locks:}
\begin{align*}
\hoare
  {\ipropC}
  {&\newlock}
  {\Ret\lockvar. \islock \lockvar \ipropC}
  \tagH{Ht-new-lock} \\
\hoare
  {\islock \lockvar \ipropC}
  {&\acquire \lockvar}
  \ipropC
  \tagH{Ht-acquire} \\
\hoare
  {\islock \lockvar \ipropC * \ipropC}
  {&\release \lockvar}
  \TRUE
  \tagH{Ht-release} \\
\islock \lockvar \ipropC \wand\ &
  \islock \lockvar \ipropC * \islock \lockvar \ipropC
  \tagH{Lock-dup}
\end{align*}
\caption{The grammar and rules of locks in Iris.}
\label{fig:locks}
\end{figure}}

\newcommand{\locksimplefig}{
\begin{figure}
\begin{align*}
\simpleexamplename \langdef{}&
  \LetNoIn \chan = \start {(\Lam \chan.
    \begin{array}[t]{@{}l@{}}
      \Let \lockvar = \newlock in \\
      \Fork { \acquirelock \lockvar;\ \send \chan {21};\ \releaselock \lockvar };\\
      \acquirelock \lockvar;\ \send \chan {21};\ \releaselock \lockvar)\ \In
    \end{array}
  } \\
  & \recv \chan + \recv \chan
\end{align*}
\caption{A sample program that combines locks and channels to
achieve manifest sharing.}
\label{fig:lock_simple}
\end{figure}}

\newcommand{\authfig}{
\begin{figure}
\begin{align*}
\TRUE \vs{}& \Exists\gname. \serverpredT \gname 0
  \tagH{Auth-init} \\
\serverpredT \gname n \vs{}&
  \clientpredT \gname * \serverpredT \gname {(1 + n)}
  \tagH{Auth-alloc} \\
\serverpredT \gname {(1 + n)} * \clientpredT \gname \vs{}&
  \serverpredT \gname n
  \tagH{Auth-dealloc} \\
\serverpredT \gname n * \clientpredT \gname \wand{}&
  n > 0
  \tagH{Auth-contrib-pos}
\end{align*}
\caption{The authoritative contribution ghost theory.}
\label{fig:auth}
\end{figure}}

As presented in \Cref{sec:language}, \heaplang includes a lock library,
with the operations $\newlock$, $\acquirelock{\lockvar}$, and
$\releaselock{\lockvar}$.
The operations satisfy the separation logic specifications shown
in \Cref{fig:locks}.

\locksfig

The specifications for locks make use of the representation
predicate $\islock \lockvar \ipropC$, which expresses that a lock $\lockvar$ guards
the resources described by the proposition $\ipropC$.
When creating a new lock one has to give up ownership of $\ipropC$, and in turn,
obtains the representation predicate $\islock \lockvar \ipropC$ (\ruleref{Ht-new-lock}).
The representation predicate can then be freely duplicated so it
can be shared between multiple threads (\ruleref{Lock-dup}).
When entering a critical section using $\acquire \lockvar$,
a thread gets exclusive
ownership of $\ipropC$ (\ruleref{Ht-acquire}), which has to be given up
when releasing the lock using $\release \lockvar$ (\ruleref{Ht-release}).
The resources $\ipropC$ that are protected by the lock are therefore invariant
in-between any of the critical sections.
The lock can only ever be acquired by one thread at a time, as
$\acquire \lockvar$ will loop until the lock is released.
The \ruleref{Ht-acquire} rule reflects this, as the exclusive resources $\ipropC$
are only obtained once the function terminates, \ie when the lock is available.

\locksimplefig

To show how locks can be used, consider the program $\simpleexamplename$ in
\Cref{fig:lock_simple}.
This program uses a lock to share a channel endpoint
between two threads, which each send the integer $21$ to the main thread.
The following dependent protocol specifies the expected interaction from the
point of view of the
main thread:
\[
\lockprotname \eqdef \MU (\recvar : \nat \rightarrow \iProto). \Lam n.
 \If (n=0) then \protend \Else \recvprotT{}{21}{\recvar\ (n-1)}
\]
Here, $n$ denotes the number of messages that should be exchanged.
In the example program, $n$ is initially 2.
Since $\interp \chan {\dual {\lockprotname\ n}}$ is an exclusive resource, we
need a lock to share it between the threads that send $21$.
For this we will use the following lock invariant:
\[
\islock \lockvar {(\Exists n.
   \serverpredT \gname n * \interp \chan {\dual {\lockprotname\ n}})}
\]

\noindent
The natural number $n$ is existentially quantified since it changes
whenever a message is exchanged.
To keep track of the number of exchanges that each thread is allowed to make
we then need to tie the number $n$ to some local resource.
We achieve this by using Iris's \emph{ghost theory} mechanism for creating
user-defined ghost state~\cite{jung-POPL2015,jung-JFP2018}.
In particular, we define two logical connectives $\serverpredT \gname n$ and
$\clientpredT \gname$ using Iris.\footnote{Defining a ghost theory in Iris
involves picking an appropriate \emph{resource algebra} with which one can
define a set of abstract predicates (here $\serverpredT \gname n$ and
$\clientpredT \gname$).
The details of resource algebras are beyond the scope of this paper and can be
found in Jung \etal~\cite{jung-JFP2018}.}

\authfig

The $\serverpredT \gname n$ fragment can be thought of as an authority that
keeps track of the number of ongoing contributions $n$, while each
$\clientpredT \gname$ is a token that witnesses that a contribution is still in
progress.
This intuition is made precise by the rules in \Cref{fig:auth}.
The rule \ruleref{Auth-init} expresses that an authority $\serverpredT \gname 0$
can always be created, capturing that 0 contributions are initially in progress.
A fresh ghost identifier $\gname$ is given, which is conceptually similar to how
we obtain fresh locations for newly allocated references on the physical heap.
Using the rules \ruleref{Auth-alloc} and \ruleref{Auth-dealloc}, we can
allocate and deallocate $\clientpredT \gname$ tokens as long as the number $n$
of ongoing contributions in $\serverpredT \gname n$ is updated accordingly.
The rule \ruleref{Auth-contrib-pos} expresses that ownership of a token
$\clientpredT \gname$ implies that the count $n$ of $\serverpredT \gname n$
must be positive.

Most of the rules in \Cref{fig:auth} involve Iris's \textit{view shift}
connective $\vs$ for performing ghost updates.
This is made precise by the structural rules
\ruleref{Vs-csq} and \ruleref{Vs-frame}, which establish
the connection between $\vs$ and Iris's Hoare triples:
\begin{mathpar}
\inferH{Vs-csq}
  {\iprop \vs \iprop' \and
   \hoare {\iprop'} \expr {\Ret\val. \ipropB'} \and
   \All\val. \ipropB' \vs \ipropB}
  {\hoare \iprop \expr {\Ret\val. \ipropB}}
\and
\inferH{Vs-frame}
  {\iprop \vs \ipropB}
  {\iprop * \ipropC \vs \ipropB * \ipropC}
\end{mathpar}

\noindent
With the ghost theory in place, we can now prove suitable specifications
for the program.
The specification of the top-level program is shown on the right, while the
left Hoare triple shows the auxiliary specification of both threads that send
the integer $21$:
\[
\begin{array}{l @{\qquad} l}
\hoareV
  {\clientpredT \gname *
   \islock \lockvar {(\Exists n.
   \serverpredT \gname n * \interp \chan {\dual {\lockprotname\ n}})}}
  {\acquirelock \lockvar;\ \send \chan {21};\ \releaselock\lockvar}
  \TRUE
&
  \hoareV \TRUE \simpleexamplename {\Ret\val. \val = 42}
\end{array}
\]
We use rule \ruleref{Ht-new} to assign protocol $\lockprotname\ 2$ to the
channel.
To establish the initial lock invariant, we use the rules \ruleref{Auth-init}
and \ruleref{Auth-alloc} to create the authority $\serverpredT \gname 2$ and
two $\clientpredT \gname$ tokens.
The $\clientpredT \gname$ tokens play a crucial role in the proofs of the
sending threads to establish that the existentially quantified variable $n$ is
positive (using \ruleref{Auth-contrib-pos}).
Knowing $n > 0$, these threads can establish that the protocol
$\dual {\lockprotname\ n}$ has not terminated yet (\ie is not $\protend$).
This is needed to use the rule \ruleref{Ht-send} to prove the correctness of
sending $21$, and thereby advancing the protocol from
$\dual {\lockprotname\ n}$ to $\dual {\lockprotname\ (n-1)}$.
Subsequently, the sending threads can deallocate the token $\clientpredT \gname$
(using \ruleref{Auth-dealloc}) to decrement the $n$ of $\serverpredT \gname n$
accordingly to restore the lock invariant.

\subsection{A channel-based load-balancing mapper}
\label{sec:mapper}

\newcommand{\parmapperworkerfig}{
\begin{figure}
\begin{equation*}
\begin{array}{@{} l @{}}
\parmapperworkername\ \vmapvar\ \lockvar\ \chan \langdef \\
\quad \begin{array}[t]{@{} l @{}}
  \acquirelock{\lockvar};\\
  \select \chan \leftname;\\
  \mbranch \chan
    \rightname {\release \lockvar}
    \leftname {
      \begin{array}[t]{@{} l !{\quad\quad}  l @{}}
      \Let \var = \recv\chan in
      \releaselock \lockvar; & \Comment{acquire work} \\
      \Let \varB = \vmapvar\ \var in & \Comment{map it} \\
      \acquirelock \lockvar;\\
      \quad \select \chan \rightname;\
      \send \chan \varB; & \Comment{send it back} \\
      \releaselock \lockvar; \\
      \parmapperworkername\ \vmapvar\ \lockvar\ \chan
      \end{array}}
  \end{array}
\end{array}
\end{equation*}
\caption{A worker of the channel-based mapper service.}
\label{fig:mapperworker}
\end{figure}}

\newcommand{\authMfig}{
\begin{figure}
\begin{align*}
\TRUE \vs{}&
  \Exists\gname. \serverpred \gname 0 \emptyset
  \tagH{AuthM-init} \\
\serverpred \gname n X \vs{}&
  \serverpred{\gname}{(1 + n)}{X} * \clientpred \gname \emptyset
  \tagH{AuthM-alloc} \\
\serverpred \gname n X * \clientpred \gname \emptyset \vs{}&
  \serverpred \gname {(n-1)} X
  \tagH{AuthM-dealloc} \\
\serverpred \gname n X * \clientpred{\gname}{Y} \vs{}&
  \serverpred \gname n {(X \uplus Z)} * \clientpred \gname {(Y \uplus Z)}
  \tagH{AuthM-add} \\
Z \subseteq Y * \serverpred \gname n X * \clientpred \gname Y \vs{}&
  \serverpred \gname n {(X \setminus Z)} * \clientpred \gname {(Y \setminus Z)}
  \tagH{AuthM-remove} \\
\serverpred \gname n X * \clientpred \gname Y \wand{}&
  n > 0 * Y \subseteq X
  \tagH{AuthM-contrib-agree} \\
\serverpred \gname 1 X * \clientpred \gname Y \wand{}&
  Y = X
  \tagH{AuthM-contrib-agree1}
\end{align*}
\caption{The authoritative contribution ghost theory extended with multisets.}
\label{fig:authM}
\end{figure}}

This section demonstrates a more interesting use of manifest sharing.
We show how \lname can be used to verify functional correctness
of a channel-based load-balancing mapper that maps the \heaplang function
$\vmapvar$ over a list.
Our channel-based mapper consists of one client that distributes the work, and a
number of workers that perform the function $\vmapvar$ on individual elements of
the list.
To enable communication between the client and the workers, we make use of
a single channel.
One endpoint is used by the client to distribute the work between the workers,
while the other endpoint is shared between all workers to request and
return work from the client.
The implementation of the workers $\parmapperworkername\ \vmapvar\ \lockvar\ \chan$,
which can be found in
\Cref{fig:mapperworker}, consists of a loop over three phases:
\begin{enumerate}
\item The worker notifies the client that it wants to perform work (using
  $\select \chan \leftname$), after which it is then notified (using
  \branchname) whether there is more work or all elements have been mapped.
  If there is more work, the worker receives an element $\var$ that needs to
  be mapped.
  Otherwise, the worker will terminate.
\item The worker maps the function $\vmapvar$ on $\var$.
\item The worker notifies the client that it wants to send back a result
  (using $\select \chan \rightname$), and subsequently sends back the
  result $\varB$ of mapping $\vmapvar$ on $\var$.
\end{enumerate}
The first and last phases are in a critical section guarded by a lock $\lockvar$
since they involve interaction over a shared channel endpoint.
As the sharing behaviour is encapsulated by the worker, we omit the code of the client
for brevity's sake.\footnote{The entire code is present in
the accompanied Coq development~\cite{actris_coq}.}
\parmapperworkerfig
\authMfig

A protocol that describes the interaction from the client's point of view is
as follows:
\[
\begin{array}{l}
\parmapperprotname\
  (\interpvar_\tvar : \tvar \to \Val \to \Prop)\
  (\interpvar_\tvarB : \tvarB \to \Val \to \Prop)\
  (\mapvar : \tvar \to \List\ \tvarB) \eqdef\\
\quad \MU (\recvar : \nat \to \MultiSet\ \tvar \to \iProto). \Lam n\ X. \\
\quad\quad \begin{array}{@{} l @{}}
  \If n=0 then \protend \Else\\
  \begin{array}[t]{@{} l @{}}
    \mbranchprotprop
      {\selectprot
        {(\sendprot{(\var : \tvar)\ (\val : \Val)}
          \val
          {\interpvar_\tvar\ \var\ \val}
          {\recvar\ n\ (X \uplus \{\var\})})}
        {\recvar\ (n-1)\ X}}
      {(n = 1) \wand (X = \emptyset)}
      \TRUE
      {\recvprot {(\var : \tvar)\ (\loc : \Loc)}
        \loc
        {\var \in X * \llistrefI{\interpvar_\tvarB} \loc {(\mapvar\ \var)}}
        {\recvar\ n\ (X \setminus \{ x \})}}
  \end{array}
\end{array}
\end{array}
\]
Similarly to $\mapperprotname$ from \Cref{sec:subprotocol_swapping}, the protocol is
parameterised by representation predicates $\interpvar_\tvar$
and $\interpvar_\tvarB$, and a function $\mapvar : \tvar \to \List\ \tvarB$ in
the Iris/\lname logic that will be related to $\vmapvar$
through a $\mapspecname$ specification.
Similar to the protocol $\lockprotname$ from \Cref{sec:locks}, the protocol
$\parmapperprotname$ is indexed by the number of remaining workers $n$.
On top of that, it carries a multiset $X$ describing the values currently
being processed by all the workers.
The multiset $X$ is used to make sure that the returned results are in
fact the result of mapping the function $\mapvar$.
The condition $(n = 1) \wand (X = \emptyset)$ on the branching operator ($\branchop$)
expresses that the last worker may only request more work if there are no
ongoing jobs.

To accommodate sharing of the channel endpoint between all workers using a lock
invariant, we extend the authoritative contribution ghost theory from
\Cref{sec:locks}.
We do this by adding multisets $X$ and $Y$ to the connectives
$\serverpred \gname n X$ and $\clientpred \gname Y$.
These multisets keep track of the values held by the workers.
The rules for the ghost theory extended with multisets are shown in
\Cref{fig:authM}.
The rules \ruleref{AuthM-init}, \ruleref{AuthM-alloc} and \ruleref{AuthM-dealloc}
are straightforward generalisations of the ones we have seen before.
The new rules \ruleref{AuthM-add} and \ruleref{AuthM-remove} determine that the
multiset $Y$ of $\clientpred \gname Y$ can be updated as long as it is done in
accordance with the multiset $X$ of $\serverpred \gname n X$.
Finally, the \ruleref{AuthM-contrib-agree} rule expresses that the
multiset $Y$ of $\clientpred \gname Y$ must be a subset of the
multiset $X$ of $\serverpred \gname n X$,
while the stricter rule \ruleref{AuthM-contrib-agree1} asserts equality between
$X$ and $Y$ when only one contribution remains.

We then prove the following specifications of $\parmapperworkername$
and a possible top-level client $\parmapperclientname$ that uses $n$
workers to map $\vmapvar$ over the linked list $\loc$:
\[
\begin{array}{l l @{}}
\hoareV
  {\begin{array}{@{\,} l @{}}
     \mapspecname\
     \interpvar_\tvar\
     \interpvar_\tvarB\
     \mapvar\ \vmapvar *
     \clientpred \gname \emptyset\ *\\
     \islock \lockvar {\left(
       \begin{array}{@{} l @{} l @{}}
         \Exists n\ X.
         \serverpred \gname n X\ * \\
         \ \interp \chan {\dual{\parmapperprotname\
         \interpvar_\tvar\ \interpvar_\tvarB\ \mapvar\ n\ X}}
       \end{array}\right)}
     \end{array}}
  {\parmapperworkername\ \vmapvar\ \lockvar\ \chan}
  \TRUE
&
\hoareV
  {\begin{array}{@{\,} l @{\;}}
     \mapspecname\
     \interpvar_\tvar\
     \interpvar_\tvarB\
     \mapvar\ \vmapvar\ * \\
     0 < n *
     \llistrefI {\interpvar_\tvar} \loc {\vec\var}
   \end{array}}
  {\parmapperclientname\ n\ \vmapvar\ \loc}
  {\Exists \vec\varB.
    \vec\varB \perm \flatmapname\ \mapvar\ \vec\var *
    \llistrefI{\interpvar_\tvarB} \loc {\vec\varB}}
\end{array}
\]

\noindent
The lock invariant and specification of $\parmapperworkername$ are similar to
those used in the simple example in \Cref{sec:locks}.
The specification of $\parmapperclientname\ n\ \vmapvar\ \loc$ simply states
that the resulting linked contains a permutation of performing the
map at the level of the logic.
To specify that, we make use of $\flatmapname :
  (\tvar \to \List\ \tvarB) \to (\List\ \tvar \to \List\ \tvarB)$, whose
definition is standard.

The proof of the client involves allocating the channel with the protocol
$\parmapperprotname$, with the initial number of workers $n$.
Subsequently, we use the rules \ruleref{AuthM-init} and \ruleref{AuthM-alloc}
to create the authority $\serverpred \gname n \emptyset$ and $n$ tokens
$\clientpred \gname \emptyset$, which allow us to establish the lock invariant
and to distribute the tokens among the mappers.
The proof of the mapper proceeds as usual.
After acquiring the lock, the mapper obtains ownership of the lock invariant.
Since the worker owns the token $\clientpred \gname \emptyset$, it knows that
the number of remaining workers $n$ is positive, which allows it to conclude
that the protocol has not terminated (\ie is not $\protend$).
After using the rules for channels, the rules \ruleref{AuthM-add} and
\ruleref{AuthM-remove} are used to update the authority, which is needed to
reestablish the lock invariant so the lock can be released.

\section{Case study: map-reduce}
\label{sec:map_reduce}

As a means of demonstrating the use of \lname for verifying more
realistic programs, we present a proof of functional correctness
of a simple channel-based load-balancing implementation of the map-reduce model
by Dean and Ghemawat~\cite{dean-OSDI2004}.

Since \lname is not concerned with distributed systems over networks, we
consider a version of map-reduce that delegates the work over forked-off
threads on a single machine.
This means that we do not consider mechanics like handling the failure,
restarting, and rescheduling of nodes that a version that operates on a network
has to consider.

In order to implement and verify our map-reduce version we make use of the
implementation and verification of the fine-grained channel-based merge sort
algorithm (\Cref{sec:dependent}) and the channel-based load-balancing
mapper (\Cref{sec:mapper}).
As such, our map-reduce implementation is mostly a suitable client that glues
together communication with these services.
The purpose of this section is to give a high-level description of the
implementation.
The actual code and proofs can be found in the accompanied
Coq development~\cite{actris_coq}.

\subsection{A functional specification of map-reduce}
\label{sec:map_reduce_functional}

\newcommand{\mapreducename}{\defemph{map\_reduce}}
\newcommand{\vmapreducename}{\valuefy{\mapreducename}}
\newcommand{\curryvar}{\defemph{curry}}
\newcommand{\redvar}{\logemph{g}}
\newcommand{\vredvar}{\valuefy{\redvar}}
\newcommand{\tvarK}{K}
\newcommand{\maptype}{\tvar \to \List\ (\tvarK * \tvarB)}
\newcommand{\redtype}{(\tvarK * \List\ \tvarB) \to \List\ \tvarC}
\newcommand{\groupname}{\defemph{group}}

The purpose of the map-reduce model is to transform an input set of type
$\List\ \tvar$ into an output set of type $\List\ \tvarC$
using two functions $\mapvar$ (often called ``map'') and $\redvar$ (often
called ``reduce''):
\[
\mapvar : \maptype
\quad\quad
\redvar : \redtype
\]

\noindent
An implementation of map-reduce performs the transformation in three steps:
\begin{enumerate}
\item First, the function $\mapvar$ is applied to each element of the input set.
  This results in lists of key/value pairs which are then flattened
  using a $\flatmapname$ operation (an operation that takes a list of lists and appends all nested lists):
  \[
  \flatmapname\ \mapvar \quad:\quad
  \List\ \tvar \to \List\ (\tvarK * \tvarB)
  \]
\item Second, the resulting lists of key/value pairs are grouped together by
  their key (this step is often called ``shuffling''):
  \[
  \groupname \quad:\quad
  \List\ (\tvarK * \tvarB) \to \List\ (\tvarK * \List\ \tvarB)
  \]
\item Finally, the grouped key/value pairs are passed on to the $\redvar$
  function, after which the results are flattened to aggregate the results.
  This is done using a $\flatmapname$ operation:
  \[
  \flatmapname\ \redvar \quad:\quad
  \List\ (\tvarK * \List\ \tvarB) \to \List\ \tvarC
  \]
\end{enumerate}

\noindent
The complete functionality of map-reduce is equivalent to applying the following
$\mapreducename$ function on the entire data set:
\[
  \mapreducename\quad
  :
  \quad
  \List\ \tvar \to \List\ \tvarC
  \quad \eqdef \quad
      (\flatmapname\ \redvar) \circ \groupname \circ (\flatmapname\ \mapvar)
\]

\noindent
A standard instance of map-reduce is counting word occurrences, where we let
$\tvar \eqdef \tvarK \eqdef \String$ and $\tvarB \eqdef \nat$ and
$\tvarC \eqdef \String * \nat$ with:
\begin{align*}
\mapvar : \String \to \List\ (\String * \nat) \eqdef{}&
  \Lam \var. [(\var,1)] \\
\redvar : (\String * \List\ \nat) \to \List\ (\String * \nat) \eqdef{}&
  \Lam (k,\vec n). [(k,\Sigma_{i<\listlength{\vec n}}.\ \vec n_i)]
\end{align*}

\subsection{Implementation of map-reduce}

The general distributed model of map-reduce is achieved by delegating the
phases of mapping, shuffling, and reducing, over a number of worker nodes
(\eg nodes of a cluster or individual CPUs).
To perform the computation in a delegated way, there is some work involved
in coordinating the jobs over these worker nodes,
which is usually done as follows:
\begin{enumerate}
\item Split the input data into chunks and delegate these chunks to worker
  nodes, that each apply the ``map'' function $\mapvar$ to their given data in
  parallel. We call these nodes the ``mappers''.
\item Collect the complete set of mapped results and ``shuffle'' them, \ie
  group them by key.
  The grouping is commonly implemented using a parallel sorting algorithm.
\item Split the shuffled data into chunks and delegate these chunks to worker
  nodes that each apply the ``reduce'' function $\redvar$ to their given
  data in parallel. We call these nodes the ``reducers''.
\item Collect and aggregate the complete set of result of the reducers.
\end{enumerate}

\noindent
Our variant of the map-reduce model is defined as a function
$\vmapreducename\ n\ m\ \vmapvar\ \vredvar\ \loc$ in \heaplang, which coordinates the work
for performing map-reduce on a linked list $\loc$ between $n$ mappers
applying the \heaplang ``map'' function $\vmapvar$,
and $m$ reducers applying the \heaplang ``reduce'' function $\vredvar$.
To make the implementation more interesting, we prevent storing intermediate
values locally by forwarding/returning them immediately as they are
available/requested.
The global structure is as follows:
\begin{enumerate}
\item Start $n$ instances of the load-balancing
  $\parmapperworkername$ from \Cref{sec:integration},
  parameterised with the $\vmapvar$ function, acting as the mappers.
  Additionally start an instance of $\listsortelemservicename$ from
  \Cref{sec:dependent}, parameterised by a concrete comparison
  function on the keys, corresponding to $\Lam (k_1,\_)\ (k_2,\_). k_1 < k_2$.
  Note that the type of keys are restricted to be integers
  for brevity's sake.
\item Perform a loop that handles communication with the mappers.
  If a mapper requests work, pop a value from the input list.
  If a mapper returns work, forward it to the sorting service.
  This process is repeated until all inputs have been mapped and forwarded.
\item Start $m$ instances of the $\parmapperworkername$, parameterised by $\vredvar$, acting as the reducers.
\item Perform a loop that handles communication with the mappers.
  If a mapper requests work, group elements returned by the sort service.
  If a mapper returns work, aggregate the returned value in a the linked
  list.
  Grouped elements are created by requesting and aggregating elements from the
  sorter until the key changes.
\end{enumerate}
The aggregated linked list then contains the fully mapped input set upon
completion.

\subsection{Functional correctness of map-reduce}

\newcommand{\tvarKey}{\integer}
\newcommand{\tvarZB}{{\tvarKey * \tvarB}}
\newcommand{\tvarZBs}{{\tvarKey * \List\ \tvarB}}

The specification of the map-reduce program that we prove is as follows:
\[
  \begin{array}{@{} l @{}}
    \hoareV
    {
    \begin{array}{@{} l @{}}
      0 < n\ *\
      0 < m\ *\
      \mapspecname\ \interpvar_\tvar\ \interpvar_\tvarZB\
      \mapvar\ \vmapvar\ * \
      \mapspecname\ \interpvar_\tvarZBs\
      \interpvar_\tvarC\ \redvar\ \vredvar\ * \
      \llistrefI{\interpvar_\tvar} \loc {\vec\var}
    \end{array}
    }
    {\vmapreducename\ n\ m\ \vmapvar\ \vredvar\ \loc}
    {
    \Exists{\vec{\varC}}.
    \vec{\varC} \equiv_p\ \mapreducename\ \mapvar\ \redvar\ \vec{\var} *
    \llistrefI{\interpvar_\tvarC}{\loc}{\vec\varC}
    }
  \end{array}
\]

\noindent
The $\mapspecname$ predicates (as introduced in \Cref{sec:subprotocol_swapping})
establish a connection between the functions $\mapvar$ and $\redvar$ in
Iris/\lname and the functions $\vmapvar$ and $\vredvar$ in \heaplang.
These make use of the various interpretation predicates $\interpvar_\tvar$,
$\interpvar_\tvarZB$, $\interpvar_{\tvarZBs}$, and $\interpvar_\tvarC$ for
the types in question.
Lastly, the $\llistrefI{\interpvar_\tvar} \loc {\vec\var}$ predicate determines
that the input is a linked list of the initial type $\tvar$.
The postcondition asserts that the result $\vec \varC$ is a permutation of the
original linked list $\vec \var$ applied to the functional specification
$\mapreducename$ of map-reduce from \Cref{sec:map_reduce_functional}.

\section{The model of \lname}
\label{sec:model}

\newcommand{\chanmapsto}[4]{\interp {(#1,#2)} {(#3, #4)}}
\newcommand{\protevalname}{\mathsf{prot\_consistent}}
\newcommand{\proteval}[4]{\protevalname\ #1\ #2\ #3\ #4}

We construct a model of \lname as a shallow embedding in
the Iris framework \cite{krebbers-ESOP2017,jung-POPL2015,jung-ICFP06,jung-JFP2018}.
This means that the type $\iProto$ of \pname,
the subprotocol relation $\subprot{\prot_1}{\prot_2}$,
and the connective $\interp \chan \prot$ for the channel ownership,
are definitions in Iris,
and the \lname proof rules are lemmas about these definitions in Iris.

In this section we describe the relevant aspects of the model of \lname.
We model the type $\iProto$ of \pname as the solution of a recursive domain equation, and
describe how the operations for dual and composition are defined
(\Cref{sec:protocol_model}).
We then define the subprotocol relation $\subprot {\prot_1} {\prot_2}$
and prove its proof rules as lemmas (\Cref{sec:subprotocol_model}).
To connect protocols to the endpoint channel buffers in the semantics
we define the \emph{protocol consistency relation},
which ensures that a pair of protocols is consistent with the messages in
their associated buffers (\Cref{sec:dual_consistency}).
On top of the protocol consistency relation, we define the \emph{\lname ghost theory} for
\pname (\Cref{sec:ghost_theory}),
which forms the key ingredient for defining the connective $\interp\chan\prot$ for
channel ownership (\Cref{sec:chan_own_model}) that links protocols to the
semantics of channels (\Cref{sec:channel_implementation}).
We then show how adequacy follows from the embedding in Iris (\Cref{sec:adequacy}).
Finally, we show how to solve the recursive domain equation for the type
$\iProto$ of \pname
(\Cref{sec:equation_solution}).

\subsection{The model of \pname}
\label{sec:protocol_model}

To construct a model of \pname, we first need to determine what they mean
semantically.
The challenging part involves the constructors
$\sendprot{\xdots}{\val}{\iprop}{\prot}$ and
$\recvprot{\xdots}{\val}{\iprop}{\prot}$, whose (higher-order and
impredicative) \binders $\xdots$ bind into the communicated value $\val$, the
transferred resources $\iprop$, and the tail protocol $\prot$.
We model these constructors as predicates over the communicated value
and the tail protocol.
To describe the transferred resources $\iprop$, we model these protocols as
Iris predicates (functions to $\iProp$).
This gives rise to the following recursive domain equation:
\begin{align*}
  \newaction \bnfdef{}& \actionsend \mid \actionrecv\\
\iProto \cong{}&
  1 + (\newaction \times (\Val \to \latert \iProto \to \iProp))
\intertext{The left part of the sum type (the unit type $\unit$) indicates that
the protocol has terminated, while the right part describes a message that is
exchanged, expressed as an Iris predicate.
Since the recursive occurrence of $\iProto$ in the predicate appears in negative
position,
we guard it using Iris's \emph{type-level later} ($\latert$) operator (whose only
constructor is $\latertinj : \tvar \to \latert\ \tvar$).
The exact way the solution is constructed is detailed in
\Cref{sec:equation_solution}.
For now, we assume a solution exists, and define the \pname constructors as:}
\protend \eqdef{}& \Inj 1 \TT \\
\sendprot{\xdots}{\val}{\iprop}{\prot} \eqdef{}&
  \Inj 2 (\actionsend, \Lam \valB\,\prot'.
    \Exists \xdots. (\val = \valB) * \iprop * (\prot' = \latertinj\ \prot)) \\
\recvprot{\xdots}{\val}{\iprop}{\prot} \eqdef{}&
  \Inj 2 (\actionrecv, \Lam \valB\,\prot'.
    \Exists \xdots. (\val = \valB) * \iprop * (\prot' = \latertinj\ \prot))
\end{align*}
The definitions of $\sendprot{\xdots}{\val}{\iprop}{\prot}$ and
$\recvprot{\xdots}{\val}{\iprop}{\prot}$ make use of the (higher-order and
impredicative) existential quantifiers of Iris
to constrain the actual message $\valB$ and tail $\prot'$ so that they agree
with the message $\val$ and tail $\prot$ prescribed by the protocol.

\paragraph{\textbf{Recursive protocols.}}

Iris's guarded recursion operator $\MU \var. \term$ requires the
recursion variable $\var$ to appear under a \emph{contractive} term construct in $\term$.
Hence, to use Iris's recursion operator to construct recursive protocols, it is
essential that the protocols $\sendprot{\xdots}{\val}{\iprop}{\prot}$ and
$\recvprot{\xdots}{\val}{\iprop}{\prot}$ are contractive in the tail $\prot$.
To show why this is the case, let us first define what it means for a function
$f : \tvar \to \tvarB$ to be contractive:
\[
\All \var,\varB. \later(\var = \varB) \Ra f\; \var = f\; \varB
\]
Examples of contractive functions are the later modality
$\later : \iProp \to \iProp$ and the constructor $\latertinj : \tvar \to \latert\ \tvar$.
The protocols $\sendprot{\xdots}{\val}{\iprop}{\prot}$ and
$\recvprot{\xdots}{\val}{\iprop}{\prot}$ are defined so that
$\prot$ appears below a $\latertinj$, and hence we can prove
that they are contractive in $\prot$.

\paragraph{\textbf{Operations.}}
With these definitions at hand, the dual $\dual{(\_)}$ and append
$(\protapp {\_} {\_})$ operations are defined using Iris's guarded
recursion operator ($\MU \var.\term$):
\begingroup
\allowdisplaybreaks
\begin{align*}
\dual{(\_)} \eqdef{}&
  \MU \recvar. \Lam \prot.
  \begin{cases}
    \Inj 1 \TT
    & \textnormal{if } \prot = \Inj 1 \TT \\
    \Inj 2 (\dual a,
    \begin{array}[t]{@{} l @{}}
      \Lam \valB\ \prot'. \Exists \prot''. \\ \quad
    \begin{array}[t]{@{} l @{}}
     \pred\ \valB\ (\latertinj\ \prot'')\ \ast \\
     \prot' = \latertinj\ (\recvar\ \prot''))
    \end{array}\end{array}
    & \textnormal{if } \prot = \Inj 2 (a, \pred)
  \end{cases} \\
(\protapp{\_}{\prot_2}) \eqdef{}&
  \MU \recvar. \Lam \prot_1.
  \begin{cases}
    \prot_2
    & \textnormal{if } \prot_1 = \Inj 1 \TT \\
    \Inj 2 (a,
    \begin{array}[t]{@{} l @{}}
      \Lam \valB\ \prot'. \Exists \prot''. \\ \quad
    \begin{array}[t]{@{} l @{}}
      \pred\ \valB\ (\latertinj\ \prot'')\ \ast \\
      \prot' = \latertinj\ (\recvar\ \prot''))
    \end{array} \end{array}
    & \textnormal{if }\prot_1 = \Inj 2 (a, \pred)
  \end{cases}
\end{align*}
\endgroup
In the above definitions, we let $\dual{\actionsend} \eqdef{} \actionrecv$
and $\dual{\actionrecv} = \actionsend$.

The base cases of both definitions are as expected.
In the recursive cases, we construct a new predicate, given the original
predicate $\pred$.
In these new predicates, we quantify over an original tail protocol $\prot''$ such
that $\pred\ \valB\ (\latertinj\ \prot'')$ holds, and unify the new tail
protocol $\prot'$ with the result of the recursive call $\recvar\ \prot''$.

The equational rules for dual $\dual{(\_)}$ and append $(\protapp {\_} {\_})$ from
\Cref{fig:actris_logic} are proven as lemmas in Iris using \ruleref{Loeb} induction.
This is possible as the recursive call $\recvar\ \prot''$ appears below a
$\latertinj$ constructor---since the $\latertinj$ constructor is contractive, we can
strip-off the later from the induction hypothesis when proving the equality for
the tail.

\paragraph{\textbf{Difference from the conference version.}}

In the conference version of this paper~\cite{hinrichsen-POPL2020},
we described two versions of the recursive domain equation for \pname:
an ``ideal'' version (as used in this paper), where $\iProto$ appears in negative
position, and an ``alternative'' version, where $\iProto$ appears in positive
position.
At that time, we were unable to construct a solution of the ``ideal''
version, so we used the ``alternative'' version.
In \Cref{sec:equation_solution} we show how we are now able to solve the ``ideal''
version.

In the conference version of this paper, the proposition $\iprop$
appeared under a later modality in the definitions of the protocols
$\sendprot{\xdots}{\val}{\iprop}{\prot}$ and $\recvprot{\xdots}{\val}{\iprop}{\prot}$,
making these protocols contractive in $\iprop$.
This choice was motivated by the ability to construct recursive protocols like
$\MU \recvar. \sendprot{(\chan:\Val)}{\chan}{\interp \chan \prot}{\prot'}$,
where the payload refers to the recursion variable $\recvar$.
In the current version (without the later modality) we can still construct such
protocols, because $\interp \chan \prot$ is contractive in $\prot$.
We removed the later modality because it is incompatible with the rules \ruleref{SP-send-elim}
and \ruleref{SP-recv-elim} for subprotocols.

\subsection{The model of the subprotocol relation}
\label{sec:subprotocol_model}

We now model the subprotocol relation $\subprot {\prot_1} {\prot_2}$ from
\Cref{sec:subprotocols}.
For legibility, we present it in the style of an inference system through its
constructors, whereas it is formally defined using Iris's
guarded recursion operator ($\MU \var.\term$):
\begin{mathpar}
\infer{}{\subprot {\Inj 1 \TT} {\Inj 1 \TT}}
  \\
\begin{array}{@{} c @{}}
  \All \val,\prot_2.
  \begin{array}[t]{@{} l @{}}
    \pred_2\; \val\; (\latertinj\; \prot_2) \wand\\
    \begin{array}{@{} l @{\;} l @{}}
      \Exists \prot_1. &
      \pred_1\; \val\; (\latertinj\; \prot_1)\ \ast\\
      \ & \later (\subprot {\prot_1} {\prot_2})
    \end{array}
  \end{array}
  \\ \hline
  \subprot{\Inj 2 {(\actionsend,\pred_1)}}{\Inj 2 {(\actionsend,\pred_2)}}
\end{array}
  \and
\begin{array}{@{} c @{}}
  \All \val,\prot_1.
  \begin{array}[t]{@{} l @{}}
    \pred_1\; \val\ (\latertinj\; \prot_1) \wand \\
    \begin{array}{@{} l @{\;} l @{}}
    \Exists \prot_2. & \pred_2\; \val\; (\latertinj\; \prot_2)\ \ast \\
    \ & \later (\subprot {\prot_1} {\prot_2})
    \end{array}
  \end{array}
  \\ \hline
  \subprot{\Inj 2 {(\actionrecv,\pred_1)}}{\Inj 2 {(\actionrecv,\pred_2)}}
\end{array}
  \\
\begin{array}{@{} c @{}}
  \begin{array}[t]{@{} r @{\ } l @{}}
    \All \val_1,\val_2,\prot_1,\prot_2. &
      (\pred_1\ \val_1\ (\latertinj\ \prot_1) \ast
      \pred_2\ \val_2\ (\latertinj\ \prot_2)) \wand \\
    & \Exists \prot.
      \later (\subprot {\prot_1} {
         \sendprotT {} {\val_2} {\prot}
         }) \ast
      \later (\subprot {
         \recvprotT {} {\val_1} {\prot}
         } {\prot_2})
  \end{array}
  \\ \hline
  \subprot{\Inj 2 {(\actionrecv,\pred_1)}}{\Inj 2 {(\actionsend,\pred_2)}}
\end{array}
\end{mathpar}
To be a well-formed guarded recursion definition, every
recursive occurrence of $\subprotop$ is guarded by the later modality ($\later$).
Aside from later being required for well-formedness,
these laters make it possible to reason about
the subprotocol relation using \ruleref{Loeb} induction; both to prove the
subprotocol rules from \Cref{fig:subprotocol_rules} as lemmas, and for \lname users to
reason about recursive protocols as shown in \Cref{sec:subprotocol_recursion}.
The relation is defined in a syntax directed fashion (\ie
there are no overlapping rules), and therefore all constructors need to be
defined so that they are closed under monotonicity and transitivity.

The first constructor states that terminating protocols ($\protend \eqdef
\Inj 1 \TT$) are related.
The other constructors concern the protocols
$\sendprot{\xdots}{\val}{\iprop}{\prot}$ and
$\recvprot{\xdots}{\val}{\iprop}{\prot}$, which are modelled
as $\Inj 2 {(\actionsend,\pred)}$ and $\Inj 2 {(\actionrecv,\pred)}$, where
$\pred : \Val \to \latert \iProto \to \iProp$
is a predicate over the communicated value and tail protocol.
While the actual constructors are somewhat intimidating because they are defined
in terms of these predicates in the model, they
essentially correspond to the following high-level versions:
\begin{mathpar}
\infer
  {\All \xdotsB. \iprop_2 \wand \Exists \xdots.
     (\val_1 = \val_2) * \iprop_1 *
     \later (\subprot {\prot_1} {\prot_2})}
  {\subprot
    {\sendprot \xdots {\val_1} {\iprop_1} {\prot_1}}
    {\sendprot \xdotsB {\val_2} {\iprop_2} {\prot_2}}}
\and
\infer
  {\All \xdots. \iprop_1 \wand \Exists \xdotsB.
     (\val_1 = \val_2) * \iprop_2 *
     \later (\subprot {\prot_1} {\prot_2})}
  {\subprot
    {\recvprot \xdots {\val_1} {\iprop_1} {\prot_1}}
    {\recvprot \xdotsB {\val_2} {\iprop_2} {\prot_2}}}
\and
\infer
  {\All \xdots,\xdotsB. (\iprop_1 * \iprop_2) \wand \Exists \prot.
    \later (\subprot {\prot_1} {\sendprotT {} {\val_2} {\prot}}) *
    \later (\subprot {\recvprotT {} {\val_1} {\prot}} {\prot_2})
  }
  {\subprot
    {\recvprot \xdots {\val_1} {\iprop_1} {\prot_1}}
    {\sendprot \xdotsB {\val_2} {\iprop_2} {\prot_2}}}
\end{mathpar}
To obtain syntax directed rules, the first rule combines \ruleref{SP-send-elim},
\ruleref{SP-send-intro}, and \ruleref{SP-send-mono}, and dually, the second
rule combines \ruleref{SP-recv-elim}, \ruleref{SP-recv-intro}, and
\ruleref{SP-recv-mono}.
The third rule combines \ruleref{SP-recv-elim}, \ruleref{SP-send-elim} and
\ruleref{SP-swap} and bakes in transitivity, instead of asserting
that $\prot_1$ and $\prot_2$ are equal to $\sendprotT {} {\val_2} {\prot}$ and
$\recvprotT {} {\val_1} {\prot}$, respectively.

The rules from the beginning of this section are defined by generalising
the high-level rules to arbitrary predicates.
For example, rule $\subprot{\Inj 2 {(\actionsend,\pred_1)}}{\Inj 2 {(\actionsend,\pred_2)}}$
requires that for any value $\val$ and tail protocol $\prot_2$ that
are allowed by the predicate $\pred_2$, there is a stronger tail protocol $\prot_1$
(\ie where $\subprot {\prot_1} {\prot_2}$),
so that the same value $\val$ and stronger tail protocol
$\prot_1$ are allowed by the predicate $\pred_1$.

The rules in \Cref{fig:subprotocol_rules} on page \pageref{fig:subprotocol_rules}
are proven as lemmas.
Those for \binder and resource manipulation (\ruleref{SP-send-elim},
\ruleref{SP-send-intro}, \ruleref{SP-recv-elim} and \ruleref{SP-recv-intro}),
monotonicity (\ruleref{SP-send-mono} and \ruleref{SP-recv-mono}),
and swapping (\ruleref{SP-swap}) follow
almost immediately from the definition, whereas
those for reflexivity (\ruleref{SP-refl}), transitivity (\ruleref{SP-trans}),
and the dual and append operator (\ruleref{SP-dual} and \ruleref{SP-append})
are proven using \ruleref{Loeb} induction.

\subsection{Protocol consistency}
\label{sec:dual_consistency}

\newcommand{\vsl}{\vec\val_1}
\newcommand{\vsr}{\vec\val_2}

To connect \pname to the semantics of channels in \Cref{sec:chan_own_model}, we define
the \emph{protocol consistency relation} $\proteval \vsl \vsr {\prot_1} {\prot_2}$,
which expresses that protocols $\prot_1$ and $\prot_2$ are \emph{consistent}
w.r.t.\ channel buffers containing values $\vsl$ and $\vsr$.
The consistency relation is defined as:
\begin{align*}
\proteval \vsl \vsr {\prot_1} {\prot_2}
  \eqdef{}& \Exists \prot. \\
  & \hspace{-13em}
  (\subprot {\recvprotT {} {\vec\val_{2.1}} {\dots
    \recvprotT {} {\vec\val_{2.\listlength{\vsr}}} \prot}} {\prot_1}) *
  (\subprot {\recvprotT {} {\vec\val_{1.1}} {\dots
    \recvprotT {} {\vec\val_{1.\listlength{\vsl}}} {\dual\prot}}} {\prot_2})
\end{align*}
Intuitively, $\proteval \vsl \vsr {\prot_1} {\prot_2}$
ensures that for all messages
$\vsl = \vec\val_{1.1} \dotsc \vec\val_{1.\listlength{\vsl}}$ in transit from the endpoint described
by $\prot_1$ to the endpoint described by $\prot_2$, the protocol $\prot_2$ is
expecting to receive these message in order (and \viceversa for $\vsr$),
after which the remaining protocols $\prot$ and $\dual\prot$ are dual.
To account for weakening we close the consistency relation under subprotocols
(by using $\subprotop$ instead of equality).
Closure under the subprotocol relation additionally implicitly captures ownership of the
quantifiers and resources associated with the messages $\vsl$ and $\vsr$.
That is, since the subprotocol relations relate the protocol arguments
$\prot_1$ and $\prot_2$ with protocols that specify no quantifiers or resources.
More precisely, by the definition of the subprotocol relation
(shown in \Cref{sec:subprotocol_model}), a relation such as
$\recvprotT{}{\val}{}{\prot_1} \subprotop \recvprot{(\xdots)}{\val}{\iprop}{\prot_2}$
is equivalent to a separation implication of the form
$\TRUE \wand \Exists \xdots. \iprop \ast \later \prot_1 \subprotop \prot_2$,
where the obligation $\TRUE$ is trivial, meaning that it implicitly asserts
ownership of $\iprop$.

Finally, closure under the subprotocol relation gives that
$\proteval \vsl \vsr {\prot_1} {\prot_2}$ and $\subprot {\prot_1} {\prot_1'}$
implies $\proteval \vsl \vsr {\prot_1'} {\prot_2}$,
and ensures that the consistency relation enjoys the following rules
corresponding to creating a channel, sending a message, and receiving a message:
\begin{align*}
& \proteval \nil \nil {\prot} {\dual \prot}
\\[0.5em]
& \begin{array}{@{} l @{}}
  (\proteval \vsl \vsr {(\sendprot \xdots \val \iprop {\prot_1})} {\prot_2}
    \ast \subst \iprop {\vec\var} {\vec\term}) \wand \\
  \qquad
  \lateropt {\listlength \vsr} (
    \proteval {(\snoc {\subst \val {\vec\var} {\vec\term}} \vsl)} \vsr {\prot_1} {\prot_2})
  \end{array} \\[0.5em]
& \begin{array}{@{} l @{}}
  \proteval \vsl {(\cons \valB \vsr)} {(\recvprot \xdots \val \iprop {\prot_1})} {\prot_2}
  \wand \\
  \qquad
  \Exists \vec\varB.
    (\valB = \subst \val {\vec\var} {\vec\varB}) *
    \subst \iprop {\vec\var} {\vec\varB} \ast
    \later (\proteval \vsl \vsr {\prot_1} {\prot_2})
  \end{array}
\end{align*}
The first rule states that dual protocols are consistent w.r.t.\ a pair of
empty buffers.
The second rule states that a protocol $\sendprot \xdots \val \iprop {\prot_1}$
can be advanced to $\prot_1$ by giving up ownership of
$\subst \iprop {\vec\var} {\vec\term}$ and enqueueing the value
$\subst \val {\vec\var} {\vec\term}$ in the buffer $\vsl$.
Dually, the third rule states that given a protocol
$\recvprot \xdots \val \iprop {\prot_1}$ and a buffer that contains value
$\valB$ as its head, we learn that $\valB$ is equal to
$\subst \val {\vec\var} {\vec\varB}$, and that we can obtain ownership of
$\subst \iprop {\vec\var} {\vec\varB}$ by advancing the protocol to $\prot_1$
and dequeuing the value $\valB$ from the buffer.
Since the relation is symmetric, \ie if $\proteval \vsl \vsr {\prot_1} {\prot_2}$
then $\proteval \vsr \vsl {\prot_2} {\prot_1}$, we obtain similar rules for
the protocol $\prot_2$ on the right-hand side.

The last two rules are proven by case analysis on the subprotocol relation
($\subprotop$) in the assumption.
Since the subprotocol relation ($\subprotop$) is defined using guarded recursion,
we obtain a later modality ($\later$) for each case analysis.
To prove the first of the rules, we need to perform a number of case analyses
equal to the size of the buffer $\vsr$,
whereas for the second rule we need to perform just a single case analysis.
These later modalities are eliminated through the $\SkipN$ operation in
the $\sendname$ operation, see
\Cref{sec:chan_own_model} for further discussion.

\subsection{The \lname ghost theory}
\label{sec:ghost_theory}

\newcommand{\fighoghost}{
  \begin{figure}
  \begin{align}
  \TRUE \vs{}& \Exists \gname.
    (\gname \mapsto_{\authfull} \prot) * (\gname \mapsto_{\authfrag} \prot)
    \tagH{ho-ghost-alloc}\\
  (\gname \mapsto_{\authfull} \prot) * (\gname \mapsto_{\authfrag} \prot')
    \vs{}&
    (\gname \mapsto_{\authfull} \prot'') * (\gname \mapsto_{\authfrag} \prot'')
    \tagH{ho-ghost-update}\\
  (\gname \mapsto_{\authfull} \prot) * (\gname \mapsto_{\authfrag} \prot')
    \wand{}&
    \later (\prot = \prot') \tagH{ho-ghost-agree}
  \end{align}
  \caption{Ghost theory for higher-order ghost variables in Iris.}
  \label{fig:hoghost}
  \end{figure}}

\newcommand{\protoghosttheoryfig}{
  \begin{figure}
  \begin{align*}
  & \begin{array}{@{} l @{}}
    \TRUE \vs
    \Exists \gpname.
    \protoctx \gpname \nil \nil \ast
    \protofragleft \gpname \prot \ast
    \protofragright \gpname {\dual\prot}
  \end{array}
  \tagH{proto-alloc} \\[0.5em]
  & \begin{array}{@{} l @{}}
    \protoctx \gpname \vsl \vsr \ast
    \protofragleft \gpname {(\sendprot \xdots \val \iprop \prot)} \ast
    \subst \iprop {\vec\var} {\vec\term}
  \vs\\
  \qquad
    \left(\lateropt {\listlength\vsr}
      \protoctx \gpname {(\snoc {\subst \val {\vec\var} {\vec\term}} \vsl)} \vsr\right) \ast
    \protofragleft \gpname {(\subst \prot {\vec\var} {\vec\term})}
  \end{array}
  \tagH{proto-send-l} \\[0.5em]
  & \begin{array}{@{} l @{}}
    \protoctx \gpname \vsl \vsr \ast
    \protofragright \gpname {(\sendprot \xdots \val \iprop \prot)} \ast
    \subst \iprop {\vec\var} {\vec\term}
    \vs\\
    \qquad
    \left(\lateropt {\listlength\vsl}
      \protoctx \gpname \vsl {(\snoc {\subst \val {\vec\var} {\vec\term}} \vsr)}\right) \ast
    \protofragright \gpname {(\subst \prot {\vec\var} {\vec\term})}
  \end{array}
  \tagH{proto-send-r} \\[0.5em]
  & \begin{array}{@{} l @{}}
    \protoctx \gpname \vsl {(\cons \valB \vsr)} \ast
    \protofragleft \gpname {(\recvprot \xdots \val \iprop \prot)}
    \vs \\
    \qquad
    \later \Exists \vec\varB.
      \valB = \subst \val {\vec\var} {\vec\varB} \ast
      \subst \iprop {\vec\var} {\vec\varB} \ast
      \protoctx \gpname \vsl \vsr \ast
      \protofragleft \gpname {(\subst \prot {\vec\var} {\vec\varB})}
  \end{array}
  \tagH{proto-recv-l} \\[0.5em]
  & \begin{array}{@{} l @{}}
    \protoctx \gpname {(\cons \valB \vsl)} \vsr \ast
    \protofragright \gpname {(\recvprot \xdots \val \iprop \prot)}
    \vs \\
    \qquad
    \later \Exists \vec\varB.
      \valB = \subst \val {\vec\var} {\vec\varB} \ast
      \subst \iprop {\vec\var} {\vec\varB} \ast
      \protoctx \gpname \vsl \vsr \ast
      \protofragright \gpname {(\subst \prot {\vec\var} {\vec\varB})}
  \end{array}
  \tagH{proto-recv-r} \\
  & \protofragleft \gpname \prot * \subprot \prot {\prot'} \wand
    \protofragleft \gpname {\prot'}
  \taghref{proto-$\subprotop$-l}{proto-weaken-l} \\
  & \protofragright \gpname \prot * \subprot \prot {\prot'} \wand
    \protofragright \gpname {\prot'}
  \taghref{proto-$\subprotop$-r}{proto-weaken-r}
  \end{align*}
  \caption{The \lname ghost theory.}
  \label{fig:proto_ghost_theory}
  \end{figure}}

\newcommand{\protoctx}[3]{\mathsf{prot\_ctx}\ #1\ #2\ #3}
\newcommand{\protofragleft}[2]{\mathsf{prot\_own_l}\ #1\ #2}
\newcommand{\protofragright}[2]{\mathsf{prot\_own_r}\ #1\ #2}
\newcommand{\gnamel}{\gname_1}
\newcommand{\gnamer}{\gname_2}
\newcommand{\gpname}{\chi}

To provide a general interface for making \lname's reasoning principles
independent of \heaplang,
we employ a standard ghost theory approach in Iris to compartmentalise channel
ownership.
In \Cref{sec:chan_own_model} we define the connective $\interp \chan \prot$
for channel endpoint ownership that links the ghost theory to the buffers of
our implementation of channels in \heaplang.

\fighoghost
\protoghosttheoryfig

The \lname ghost theory is similar in its interface to the ghost theory for
contributions that we used in \Cref{sec:integration}.
We define three new logical connectives---an authority
$\protoctx{\gpname}{\vsl}{\vsr}$, and tokens
$\protofragleft{\gpname}{\prot_l}$ and
$\protofragright{\gpname}{\prot_r}$---and prove rules about how they can be
allocated, updated, and used.
Similar to prior ghost theories, the identifier
$\gpname$ associates the connectives to each other.
The $\protoctx{\gpname}{\vsl}{\vsr}$ connective can be thought of as
an authority that governs the global state of the buffers $\vsl$ and $\vsr$.
The tokens $\protofragleft{\gpname}{\prot_l}$ and
$\protofragright{\gpname}{\prot_r}$ provide local views of the buffers state
in terms of the protocols $\prot_l$ and $\prot_r$.
As we will see in \Cref{sec:chan_own_model}, the authority can be shared
using a lock, while the tokens provide unique ownership of each endpoint.

To define the connectives of the \lname ghost theory
we use Iris's existing ghost theory for higher-order ghost variables,
revolving around the two connectives
$\gname \mapsto_{\authfull} \prot$ and $\gname \mapsto_{\authfrag} \prot'$,
which we call the inner and outer fragments, respectively.
As before, the $\gname$ is the ghost identifier that associates the connectives.
The fragments can be thought of as two pieces of a single variable,
which can only be updated in the presence of both fragments.
As a result, we know that inner and outer fragment with the same ghost identifier
$\gname$ always point to the same protocol $\prot$.
This is made precise by the rules as shown in \Cref{fig:hoghost}.
In particular, higher-order ghost variables are allocated in pairs
$\gname \mapsto_{\authfull} \prot$ and
$\gname \mapsto_{\authfrag} \prot$ for an identical protocol $\prot$
(\ruleref{ho-ghost-alloc}),
and they can only be updated together (\ruleref{ho-ghost-update}).
This means that they will always hold the same protocol
(\ruleref{ho-ghost-agree}).
The subtle part of the higher-order ghost variables is that they involve ownership
of a protocol of type $\iProto$, which is defined in terms of Iris propositions
$\iProp$.
Due to the dependency on $\iProp$ (which is covered in detail in
\Cref{sec:protocol_model,sec:equation_solution}) the rule \ruleref{ho-ghost-agree} only gives the
equality between the protocols under a later modality ($\later$).

With Iris's higher-order ghost variables at hand, we can define the \lname ghost theory
connectives as:
\begin{align*}
\protoctx{(\gnamel, \gnamer)}{\vsl}{\vsr} \eqdef{}&
  \Exists \prot_1, \prot_2.
  \begin{array}[t]{@{} l @{}}
  \gnamel \mapsto_{\authfull} \prot_1 \ast
  \gnamer \mapsto_{\authfull} \prot_2\ \ast\\
  \later \proteval \vsl \vsr {\prot_1} {\prot_2}
  \end{array}\\
\protofragleft{(\gnamel, \gnamer)}{\prot_l} \eqdef{}&
  \Exists \prot_l'.
  \gnamel \mapsto_{\authfrag} \prot_l' \ast
  \later (\subprot{\prot_l'}{\prot_l})
  \\
\protofragright{(\gnamel, \gnamer)}{\prot_r} \eqdef\ &
  \Exists \prot_r'.
  \gnamer \mapsto_{\authfrag} \prot_r' \ast
  \later (\subprot{\prot_r'}{\prot_r})
\end{align*}
Since we use two higher-order ghost variables,
our identifiers $\gpname \bnfdef (\gnamel, \gnamer)$
are pairs of Iris ghost identifiers.
The authority $\protoctx{(\gnamel, \gnamer)}{\vsl}{\vsr}$
asserts ownership of the inner fragments of the higher-order ghost
variables
$\gnamel \mapsto_{\authfull} \prot_1$ and $\gnamer \mapsto_{\authfull} \prot_2$
for some protocols $\prot_1$ and $\prot_2$.
It then asserts that the buffers $\vsl$ and $\vsr$ are consistent with respect to
those protocols $\prot_1$ and $\prot_2$
(via $\proteval{\vsl}{\vsr}{\prot_1}{\prot_2}$).
The tokens
$\protofragleft{(\gnamel, \gnamer)}{\prot_l}$ and
$\protofragright{(\gnamel, \gnamer)}{\prot_r}$ respectively
assert ownership of the outer higher-order ghost variable fragments
$\gnamel \mapsto_{\authfrag} \prot_l'$ and $\gnamer \mapsto_{\authfrag} \prot_r'$.
Here $\prot_l'$ and $\prot_r'$ are protocols that are weaker than the protocol
arguments $\prot_l$ and $\prot_r$ (via $\subprot{\prot'_l}{\prot_l}$ and
$\subprot{\prot'_r}{\prot_r}$).
The explicit weakening under the subprotocol relation may seem redundant,
as weakening is already accounted for in $\protevalname$.
However, it allows us to weaken the
protocols of the tokens without the presence of the authority as
shown momentarily.
The later modality ($\later$) makes sure that $\protofragleft{(\gnamel, \gnamer)} \prot$
and $\protofragright{(\gnamel, \gnamer)} \prot$ are contractive in $\prot$.

With the definitions of the ghost theory connectives at hand,
we prove the rules of the ghost theory presented in
\Cref{fig:proto_ghost_theory}.
The rule \ruleref{proto-alloc} corresponds to allocation of a buffer pair, the rules
\ruleref{proto-send-l} and \ruleref{proto-send-r} correspond to sending a
message, and the rules \ruleref{proto-recv-l} and \ruleref{proto-recv-r}
correspond to receiving a message.
Finally, the rules \ruleref{proto-weaken-l} and \ruleref{proto-weaken-r} captures
that we can weaken the protocols of the tokens without the presence of the
authority.
The rules of \Cref{fig:proto_ghost_theory} are proven through a combination of
the rules for higher-order ghost
state from \Cref{fig:hoghost}, and the rules for the protocol consistency
relation $\protevalname$ from \Cref{sec:dual_consistency}.

\subsection{The model of channel ownership}
\label{sec:chan_own_model}

To link the physical contents of the bidirectional channel $\chan$ to the
\lname ghost theory we define the channel ownership connective as follows:
\begin{align*}
\interp \chan \prot \eqdef{}&
  \Exists \gpname,l,r,\lockvar.
  \begin{array}[t]{@{} l}
  \left(\begin{array}{@{} l @{}}
     (\chan = (l,r,\lockvar) * \protofragleft{\gpname}{\prot})\ \lor\\
     (\chan = (r,l,\lockvar) * \protofragright{\gpname}{\prot})
   \end{array}\right) \ast\\
  \islock \lockvar
    {(\Exists \vec\val_1\,\vec\val_2.
    \llistref{l}{\vec\val_1} \ast
    \llistref{r}{\vec\val_2} \ast
    \protoctx{\gpname}{\vec\val_1}{\vec\val_2})}\
  \end{array}
\end{align*}
The predicate states that the referenced channel endpoint $\chan$
is either the left $(l,r,\lockvar)$ or the right $(r,l,\lockvar)$ side of a channel,
and that we have exclusive ownership of the ghost token
$\protofragleft{\gpname}{\prot}$ or $\protofragright{\gpname}{\prot}$ for the
corresponding side.
Iris's lock representation predicate $\logdefemph{is\_lock}$
(previously presented in \Cref{sec:integration}) is used to make sharing of the
buffers possible.
The lock invariant is governed by lock $\lockvar$, and carries
the ownership $\llistref{l}{\vec\val_1}$
and $\llistref{r}{\vec\val_2}$ of the mutable linked lists containing the channel
buffers, as well as $\protoctx{\gpname}{\vec\val_1}{\vec\val_2}$, which
asserts protocol consistency of the buffers with respect to the protocols.

With the definition of the channel endpoint ownership along with the ghost theory and
lock rules we then prove the channel rules \ruleref{Ht-new},
\ruleref{Ht-send} and \ruleref{Ht-recv} from \Cref{fig:actris_logic}.
The proofs are carried out through symbolic execution to the point
where the critical section is entered, after which the rules of the \lname ghost theory
(\Cref{fig:proto_ghost_theory}) are used to allocate or update the ghost state appropriately so that it matches
the physical channel buffers.

\paragraph{\textbf{The need for skip instructions.}}

The rules \ruleref{proto-send-l} and \ruleref{proto-send-r} from
\Cref{fig:proto_ghost_theory} contain a number of later modalities ($\later$)
proportional to the other endpoint's buffer.
As explained in \Cref{sec:dual_consistency} these later modalities are the consequence
of having to perform a number of case analyses on the subprotocol relation,
which is defined using guarded recursion, and thus contains a later modality
for each recursive unfolding.

To eliminate these later modalities, we instrument the code of the \sendname
function with the $\SkipN\ (\llistlength\ r)$ instruction, which performs a number
of skips equal to the size of the other endpoint's buffer $r$.
The $\SkipN$ instruction has the following specification:
\[
  \textstyle\hoare{\later^{n}\iprop}{\SkipN\ n}{\iprop}
\]
Instrumentation with skip instructions is used often in work on step-indexing,
see \eg~\cite{DBLP:conf/esop/SvendsenSB16,giarrusso-ICFP2020}.
Instrumentation is needed because current step-indexed logics like Iris
unify physical/program steps and logical steps, \ie for
each physical/program step at most one later can be eliminated from the hypotheses.
In recent work by
Svendsen \etal~\cite{DBLP:conf/esop/SvendsenSB16},
Matsushita and Jourdan~\cite{flexible_later}, and Spies \etal~\cite{spies_pldi2021}
more liberal versions of step-indexing have been proposed, but none of these
versions of step-indexing have been integrated into the main Coq development of Iris and \heaplang.

\subsection{Adequacy of \lname}
\label{sec:adequacy}

Having constructed the model of \lname in Iris,
we obtain the following main result,
as first presented in \Cref{sec:iris:adequacy}:

\begin{thm}[Adequacy of \lname]
\label{thm:adequacy}
Let $\fpred \in \Val \to \mProp$ be a meta-level (\ie Coq) predicate over values and
suppose $\hoare \TRUE \expr {\Ret \val. \fpred\ \val}$ is
derivable in Iris, then
$\safe{\expr}$ and
$\postvalid{\expr}{\fpred}$.
\end{thm}

Since \lname is an internal logic embedded in Iris, the proof is an immediate
consequence of Iris's adequacy theorem
(\Cref{thm:adequacy_pre}).

\subsection{Solving the recursive domain equation for protocols}
\label{sec:equation_solution}

\newcommand{\Ppol}[1]{X^{#1}}
\newcommand{\Pneg}{\Ppol{-}}
\newcommand{\Ppos}{\Ppol{+}}

Recall the recursive domain equation for \pname from \Cref{sec:protocol_model}:
\[
\iProto \cong{}
  1 + (\newaction \times (\Val \to \latert \iProto \to \iProp))
\]
This recursive domain equation shows that $\iProto$ depends on the type $\iProp$
of Iris propositions.
To use types that depend on $\iProp$ as part of higher-order ghost state in Iris,
such types need to be bi-functorial in $\iProp$.
Hence, this means that to construct $\iProto$, in a way that it can be used
in combination with the higher-order ghost variables in \Cref{fig:hoghost},
we need to solve the following recursive domain equation:
\[
  \iProto(\Pneg,\Ppos) \cong{}
  1 + (\newaction \times (\Val \to \latert \iProto(\Ppos,\Pneg) \to \Ppos))
\]
Since the recursive occurrence of $\iProto$ appears in negative position, the
polarity needs to be inverted for $\iProto$ to be bi-functorial.

The version of Iris's recursive domain equation solver based
on~\cite{america-JCSS1989,birkedal-TCS2010} as mechanised in Iris's Coq
development is not readily able to construct a solution of $\iProto(\Pneg,\Ppos)$.
Concretely, the solver can only construct solutions of non-parameterised
recursive domain equations.
While a general construction for solving such recursive domain
equations exists~\cite[\S~7]{birkedal-LMCS2012}, that construction has not been
mechanised in Coq.
We circumvent this shortcoming by solving the following recursive domain
equation instead, in which we unfold the recursion once by hand:
\[\begin{array}{@{}l@{}}
  \iProto_2(\Pneg,\Ppos) \cong{}\\
  \quad
    \begin{array}{@{}l@{}}
      1 + \Big(\newaction \times \big(\Val \to
        \latert (1 + (\newaction \times
        (\Val \to \latert \iProto_2(\Pneg,\Ppos)\to \Pneg))) \to
        \Ppos\big)\Big)
    \end{array}
\end{array}\]
Here, the polarity in the recursive occurrence is fixed, allowing us to solve
$\iProto_2(\Pneg,\Ppos)$ using Iris's existing recursive domain equation solver.
This is sufficient because a solution of $\iProto_2(\Pneg,\Ppos)$ is isomorphic
to a solution of $\iProto(\Pneg,\Ppos)$.

\section{Coq mechanisation}
\label{sec:coq}

\newcommand{\coqfig}{
\begin{figure}[t!]
\begin{tabular}{l|l|l}
\textbf{Component} & \textbf{Sections} & \textbf{$\sim$LOC} \\ \hline\hline
The \lname model & \Cref{sec:protocol_model}--\Cref{sec:ghost_theory} & 1500 \\
Channel implementation and proof rules & \Cref{sec:channel_implementation,sec:chan_own_model} & 350 \\
Tactics for symbolic execution & \Cref{sec:coq_prog_proof} & 500 \\
Utilities (linked lists, permutations, \etc) & n.a. &  450\\
Authoritative contribution ghost theory & \Cref{sec:integration} & 150 \\
Recursive domain equation theory solver & \Cref{sec:equation_solution} & 100 \\
Examples: & & \\
$\bullet$ Basic examples & \Cref{sec:intro,sec:locks} & 400 \\
$\bullet$ Coarse-grained channel-based merge sort & \Cref{sec:basics}--\Cref{sec:delegation} & 250 \\
$\bullet$ Fine-grained channel-based merge sort & \Cref{sec:dependent} & 300 \\
$\bullet$ Mapper with swapping & \Cref{sec:subprotocol_swapping} & 400 \\
$\bullet$ List reversal & \Cref{sec:subprotocol_reuse} & 100 \\
$\bullet$ Channel-based load-balancing mapper & \Cref{sec:mapper} & 200 \\
$\bullet$ Channel-based map-reduce & \Cref{sec:map_reduce} & 300 \\ \hline
\textbf{Total} & & 5000 \\
\end{tabular}
\caption{Overview of components of the \lname Coq mechanisation.}
\label{fig:coq}
\end{figure}}

The definition of the \lname logic, its model, and the proofs of all examples
in this paper have been fully mechanised using the Coq proof
assistant~\cite{coq}.
In this section we will elaborate on the mechanisation effort
(\Cref{sec:coq_effort}), and go through the full proof of a message-passing program
(\Cref{sec:coq_prog_proof}) and a subprotocol relation (\Cref{sec:coq_subprot_proof})
showcasing the tactics for \lname.
We display proofs and proof states taken directly from the Coq
mechanisation, which differ in notation from the paper as shown in \Cref{fig:coq_notation}.

\begin{figure}
  \begin{tabular}{l | l @{\quad} | l}
    & \textbf{Notation on paper} & \textbf{Notation in Coq} \\\hline\hline
    Send &
    $\sendprot{\var_1 \ldots \var_n}{\val}{\iprop}\prot$ &
    \lstinline|<! x_1 .. x_n> MSG v {{ P }}; prot| \\
    Receive &
    $\recvprot{\var_1 \ldots \var_n}{\val}{\iprop}\prot$ &
    \lstinline|<? x_1 .. x_n> MSG v {{ P }}; prot| \\
    End &
    $\protend$ &
    \lstinline|END| \\
    Dual &
    $\dual{\prot}$&
    \lstinline|iProto_dual prot| \\
    Literals &
    $\TT$, $5$, $\True$&
    \lstinline|#()|, \lstinline|#5|, \lstinline|#true| \\
    \Binders &
    $\var$, $\varB$, $\varC, \_$&
    \lstinline|"x"|, \lstinline|"y"|, \lstinline|"z"|, \lstinline|<>| \\
    Types &
    $1$, $\nat$, $\integer$&
    \lstinline|()|, \lstinline|nat|, \lstinline|Z|
  \end{tabular}
  \caption{Overview of notations in the \lname Coq mechanisation.}
  \label{fig:coq_notation}
\end{figure}

\subsection{Mechanisation effort}
\label{sec:coq_effort}

The mechanisation of \lname is built on top of the mechanisation of Iris
\cite{krebbers-ESOP2017,jung-ICFP06,jung-JFP2018}.
To carry out proofs in separation logic, we use the MoSeL Proof Mode
(formerly Iris Proof Mode)~\cite{krebbers-POPL2017,krebbers-PACMPL2018}, which
provides an embedded proof assistant for separation logic in Coq.
Building \lname on top of the Iris and MoSeL framework in Coq has a number of tangible
advantages:
\begin{itemize}
\item By defining channels on top of \heaplang,
  we do not have to define a full programming language
  semantics, and can reuse all of the program libraries and Coq machinery,
  including the tactics for symbolic execution of non message-passing programs.
\item Since \lname is mechanised as an Iris library
  we get all of the features of Iris for free,
  such as the ghost state mechanisms for reasoning about concurrency.
\item When proving the \lname proof rules, we can make use of the MoSeL Proof
  Mode to carry out proofs directly using separation logic, thus reasoning at
  a high level of abstraction.
\item We can make use of the extendable nature of the MoSeL Proof Mode to
  define custom tactics for symbolic execution of message-passing programs.
\end{itemize}
These advantages made it possible to mechanise \lname, along with the examples of
the paper, with a small Coq development of a
total size of about 5000 lines of code (comments and whitespace included).
The line count of the different components are shown in \Cref{fig:coq}.

\coqfig

\subsection{Tactic support for session type-based reasoning}
\label{sec:coq_prog_proof}

To carry out interactive \lname proofs using symbolic execution, we follow the
methodology described in the original Iris Proof Mode paper \cite{krebbers-POPL2017}.
In particular, this means that the logic in Coq is presented in weakest precondition
style rather than using Hoare triples.
For handling \sendname or \recvname we define the following tactics:
\begin{center}
\lstinline|wp_send (t1 .. tn)$\! $ with "[H1 .. Hn]"| \quad and \quad
\lstinline|wp_recv (y1 .. yn)$\! $ as "H"|.
\end{center}
These tactics roughly perform the following actions:
\begin{itemize}
\item Find a \sendname or \recvname in evaluation position of the program
  under consideration.
\item Find a corresponding $\interp \chan \prot$ hypothesis in the separation
  logic context.
\item Normalise the protocol $\prot$ using the rules for duals, composition,
  recursion, and swapping so it has a
  $\sendprot {\xdots} \val \iprop \prot$ or
  $\recvprot {\xdots} \val \iprop \prot$ construct in its head position.
\item In case of \lstinline|wp_send|, instantiate the variables $\xdots$ using
  the terms \lstinline|(t1 .. tn)|, and create a goal for the proposition
  $\iprop$ with the hypotheses \lstinline|[H1 .. Hn]|.
  Hypotheses prefixed with $\mathtt{\$}$ will automatically be consumed to resolve a
  subgoal of $\iprop$ if possible.
  In case the terms \lstinline|(t1 .. tn)| are omitted, an attempt is made to
  determine these using unification.
\item In case of \lstinline|wp_recv|, introduce the variables $\xdots$ into the
  context by naming them \lstinline|(y1 .. yn)|, and create a hypothesis
  \lstinline|H| for $\iprop$.
\end{itemize}
The implementation of these tactics follows the approach by Krebbers \etal~\cite{krebbers-POPL2017}.
The protocol normalisation is implemented via logic programming with type
classes.

As an example we will go through a proof of the following program:
\[
\begin{array}[t]{@{} l @{\ } l @{}}
\mathtt{prog\_ref\_swap\_loop} \langdef{}
\Lam \_. &
\LetNoIn {\chan} = \start
  (\Rec {\mathit{go}} \chan' =
    \begin{array}[t]{@{} l @{}}
      \Let l = {\recv \chan'} in \\
        l \leftarrow \deref l + 2; \\
      \send {\chan'} \TT;\ \mathit{go}\ \chan')\ \In \\
    \end{array}\\
& \Let {l_1} = \newref 18 in \Let {l_2} = \newref 20 in\\
& \send \chan l_1;\ \send \chan l_2; \\
& \recv \chan;\ \recv \chan;\\
& \deref l_1 + \deref l_2
\end{array}
\]
Here, the forked-off thread acts as a service that recursively receives locations,
adds 2 to their stored number, and then sends back a flag indicating that the
location has been updated.
The main thread, acting like a client, first allocates two new references,
to $18$ and $20$, respectively, which are both sent to the service
after which the update flags are received.
It finally dereferences the updated locations, and adds their values together,
thus returning $42$.
To verify this program, we use the following recursive protocol:
\[
  \mathtt{prot\_ref\_loop} \eqdef \MU (\recvar:\iProto).
  \sendprot{(\loc : \Loc) (\var : \integer)}{\loc}{\loc \mapsto \var}
  \recvprot{}{\TT}{\loc \mapsto \var+2}
  \recvar
\]
The (forked-off) service follows the (dual of) the protocol exactly,
while the main thread follows a weakened version.
The recursion is unfolded
twice, after which the second send has been swapped ahead of the first receive,
allowing it to first send both values before receiving:
\[
  \begin{array}{@{} l @{\ } l}
  \mathtt{prot\_ref\_loop}
  \subprotop
  &\begin{array}[t]{@{} l}
  \sendprot{(\loc_1:\Loc)(\var_1:\integer)}{\loc_1}{\loc_1 \mapsto \var_1} \\
  \sendprot{(\loc_2:\Loc)(\var_2:\integer)}{\loc_2}{\loc_2 \mapsto \var_2} \\
  \recvprot{}{\TT}{\loc_1 \mapsto (\var_1+2)} \\
  \recvprot{}{\TT}{\loc_2 \mapsto (\var_2+2)}
  \mathtt{prot\_ref\_loop}
  \end{array}
  \end{array}
\]
The full Coq proof of the program is shown in \Cref{fig:program_proof}.
\begin{figure}
\begin{lstlisting}[numbers=left,xleftmargin=1cm,escapechar=|]
Lemma prog_ref_swap_loop_spec : ∀ Φ, Φ$\ $#42 -∗ WP prog_ref_swap_loop #() {{ Φ$\ $}}.
Proof.
  iIntros (Φ) "HΦ". wp_lam. |\label{line:proofsetup}|
  wp_apply (start_chan_spec prot_ref_loop); iIntros (c) "Hc". |\label{line:start}|
  - iLöb as "IH". wp_lam. |\label{line:forksetup}|
     wp_recv (l x) as "Hl". wp_load. wp_store. wp_send with "[$\$$Hl]". |\label{line:forkbody}|
     do 2 wp_pure _. by iApply "IH". |\label{line:forkclose}|
  - wp_alloc l1 as "Hl1". wp_alloc l2 as "Hl2". |\label{line:mainalloc}|
     wp_send with "[$\$$Hl1]". wp_send with "[$\$$Hl2]". |\label{line:mainsend}|
     wp_recv as "Hl1". wp_recv as "Hl2". |\label{line:mainrecv}|
     wp_load. wp_load. |\label{line:mainload}|
     wp_pures. by iApply "HΦ". |\label{line:mainclose}|
Qed.
\end{lstlisting}
\caption{Proof of message-passing program}
\label{fig:program_proof}
\end{figure}
The proven lemma is logically equivalent to the specification
$\hoare{\TRUE}{\mathtt{prog\_ref\_swap\_loop}\ ()}{ \Ret \val. \val = 42 }$,
but is presented in weakest precondition style as is common in Iris in Coq.
The initial proof state is:
\begin{lstlisting}
--------------------------------------∗
∀ Φ, Φ$\ $#42 -∗ WP prog_ref_swap_loop #() {{ v, Φ$\ $v }}
\end{lstlisting}
We start the proof on \lineref{line:proofsetup} by introducing the
postcondition \lstinline{Φ}, and the hypothesis
\lstinline{HΦ : Φ$\ $#42},
and then continue by evaluating the lambda expression with \lstinline{wp_lam}.
On \lineref{line:start} we apply the specification
\lstinline{start_chan_spec}, which is the weakest precondition variant of
\ruleref{Ht-start} for
$\startname$ by picking the expected protocol \lstinline{prot_ref_loop}.
This leaves us with two subgoals, separated by bullets ``\lstinline{-}'':
one for the forked-off thread, and one for the main thread.

\subsubsection*{Proof of the forked-off thread}
In the proof of the recursively-defined forked-off thread
we use \lstinline{iLöb as "IH"} for \ruleref{Loeb} induction on \lineref{line:forksetup}.
This leaves us with the proof state:
\begin{lstlisting}
"IH" : ▷ (c ↣$\ $iProto_dual prot_ref_loop -∗
              WP (rec: "go" "c'" :=
                      let: "l" := recv "c'" in
                      "l" <- ! "l" + #2;;
                      send "c'" #();; "go" "c'") c {{ _, True }})
--------------------------------------□
"Hc" : c ↣$\ $iProto_dual prot_ref_loop
--------------------------------------∗
WP (rec: "go" "c'" :=
        let: "l" := recv "c'" in
        "l" <- ! "l" + #2;;
        send "c'" #();; "go" "c'") c {{ _, True }}
\end{lstlisting}
We now resolve the application of \lstinline{c} to the recursive function
with \lstinline{wp_lam}.
This lets us strip the later from the \ruleref{Loeb} induction hypothesis, as the program
has taken a step.
The proof state is then as follows:
\begin{lstlisting}
"IH" : c ↣$\ $iProto_dual prot_ref_loop -∗ WP prog_rec c {{ _, True }}
--------------------------------------□
"Hc" : c ↣$\ $iProto_dual prot_ref_loop
--------------------------------------∗
WP let: "l" := recv c in
    "l" <- ! "l" + #2;;
    send c #();; prog_rec c {{ _, True }}
\end{lstlisting}
For brevity's sake we abbreviate the recursive code in \lstinline{"IH"} as
\lstinline{prog_rec c}.

On \lineref{line:forkbody} we resolve the proof of the body of the recursive function.
So far, the proof only used Iris's standard tactics, we now use the
\lname tactic for receive
\lstinline{wp_recv (l x)$\! $ as "Hl"}, to resolve the
receive in evaluation position, introducing the received \binders
\lstinline{l} and \lstinline{x},
along with the predicate of the protocol \lstinline{l ↦$\ $#x} naming it
\lstinline{Hl}.
To do so, the protocol is normalised, unfolding the recursive definition once,
as well as resolving the dualisation of the head,
turning it into a receive as expected.
This leads to the following proof state:
\begin{lstlisting}
"IH" : c ↣$\ $iProto_dual prot_ref_loop -∗ WP prog_rec c {{ _, True }}
--------------------------------------□
"Hl" : l ↦$\ $#x
"Hc" : c ↣$\ $iProto_dual (<?> MSG #() {{ l ↦$\ $#(x + 2) }}; prot_ref_loop)
--------------------------------------∗
WP let: "l" := #l in
    "l" <- ! "l" + #2;;
    send c #();; prog_rec c {{ _, True }}
\end{lstlisting}
We then use the \heaplang tactics \lstinline{wp_load} and \lstinline{wp_store} to resolve the
dereferencing and updating of the location:
\begin{lstlisting}
"IH" : c ↣$\ $iProto_dual prot_ref_loop -∗ WP prog_rec c {{ _, True }}
--------------------------------------□
"Hl" : l ↦$\ $#(x + 2)
"Hc" : c ↣$\ $iProto_dual (<?> MSG #() {{ l ↦$\ $#(x + 2) }}; prot_ref_loop)
--------------------------------------∗
WP send c #();; prog_rec c  {{ _, True }}
\end{lstlisting}
We then use the \lname tactic \lstinline{wp_send with "[$\$$Hl]"} to resolve the
send operation in evaluation
position, by giving up the ownership of \lstinline{"Hl"}.
Again, the protocol is automatically normalised by resolving the dualisation of the receive
($\RECV$) to obtain the send ($\SEND$) as expected.

We finally close the proof of the forked-off thread on \lineref{line:forkclose}.
We first take two pure evaluation steps revolving the sequencing of operations with
\lstinline{do 2 wp_pure _} to reach the recursive call.
This results in the proof state:
\begin{lstlisting}
"IH" : c ↣$\ $iProto_dual prot_ref_loop -∗ WP prog_rec c {{ _, True }}
--------------------------------------□
"Hc" : c ↣$\ $iProto_dual prot_ref_loop
--------------------------------------∗
WP prog_rec c {{ _, True }}
\end{lstlisting}
We then use \lstinline{by iApply "IH"} to close the proof by using the
\ruleref{Loeb} induction hypothesis.

\subsubsection*{Proof of the main thread}
The proof of the main thread follows similarly.
On \lineref{line:mainalloc} we use \lstinline{wp_alloc l1 as "Hl1"} and
\lstinline{wp_alloc l2 as "Hl2"}, to resolve the allocations of the new locations,
binding the logical variables of the locations to \lstinline{l1} and \lstinline{l2},
and adding hypotheses \lstinline{"Hl1"} and \lstinline{"Hl2"} for ownership
of these locations to the separation logic proof context.
The proof state is then:
\begin{lstlisting}
"HΦ" : Φ$\ $#42
"Hc" : c ↣$\ $prot_ref_loop
"Hl1" : l1 ↦$\ $#18
"Hl2" : l2 ↦$\ $#20
--------------------------------------∗
WP send c #l1;; send c #l2;; recv c;; recv c;; ! #l1 + ! #l2 {{ v, Φ$\ $v }}
\end{lstlisting}
On \lineref{line:mainsend},
we resolve the first send operation with the \lname tactic
\lstinline{wp_send with "[$\$$Hl1]"}, by giving up
ownership of the location \lstinline|l1|.
Here, the protocol is normalised by unfolding the recursive definition, after which the
head symbol is a send ($\SEND$) as expected.
The resulting proof state is as follows:
\begin{lstlisting}
"HΦ" : Φ$\ $#42
"Hl2" : l2 ↦$\ $#20
"Hc" : c ↣$\ $(<?> MSG #() {{ l1 ↦$\ $#(18 + 2) }}; prot_ref_loop)
--------------------------------------∗
WP send c #l2;; recv c;; recv c;; ! #l1 + ! #l2 {{ v, Φ$\ $v }}
\end{lstlisting}
To resolve the second send operation, we need to weaken the protocol using swapping
(rule \ruleref{SP-swap'}), which is taken care of automatically by the \lname tactic
\lstinline{wp_send with "[$\$$Hl2]"}.
The normalisation detects that the protocol has a
receive ($\RECV$) as a head symbol, and therefore attempts swapping.
To do so it steps ahead of the receive ($\RECV$), and unfolds the recursive definition,
which results in a send ($\SEND$) as the first symbol after the head.
It then detects that there are no dependencies between the two, and can thus apply
the swapping rule \ruleref{SP-swap'}, moving the send ($\SEND$) ahead of the
receive ($\RECV$).
With the head symbol now being a send ($\SEND$), the symbolic execution continues as normal,
resulting in the proof state:
\begin{lstlisting}
"HΦ" : Φ$\ $#42
"Hc" : c ↣$\ $(<?> MSG #() {{ l1 ↦$\ $#(18 + 2) }};
                  <?> MSG #() {{ l2 ↦$\ $#(20 + 2) }}; prot_ref_loop)
--------------------------------------∗
WP recv c;; recv c;; ! #l1 + ! #l2 {{ v, Φ$\ $v }}
\end{lstlisting}
On \lineref{line:mainrecv} we then proceed as expected with \lstinline{wp_recv as "Hl1"}
and \lstinline{wp_recv as "Hl2"}, to resolve the receive operations, giving us back the updated
point-to resources:
\begin{lstlisting}
"HΦ" : Φ$\ $#42
"Hl1" : l1 ↦$\ $#(18 + 2)
"Hl2" : l2 ↦$\ $#(20 + 2)
"Hc" : c ↣$\ $prot_ref_loop
--------------------------------------∗
WP ! #l1 + ! #l2 {{ v, Φ$\ $v }}
\end{lstlisting}
At \lineref{line:mainload} we then continue by using \lstinline{wp_load} twice to
dereference the reacquired and updated locations, and then use trivial symbolic execution
using \lstinline|wp_pures|
to resolve the remaining computations.
On \lineref{line:mainclose} we finally close the proof by applying the
hypothesis \lstinline{"HΦ"} about the postcondition.

\subsection{Tactic support for subprotocols}
\label{sec:coq_subprot_proof}

While the \lname tactics automatically apply the subprotocol rules during
symbolic execution, as shown in \Cref{sec:coq_prog_proof},
we sometimes want to prove subprotocol relations as explicit lemmas.
We have tactic support for such proofs as well.
We extend the existing MoSeL tactics
\lstinline{iIntros}, \lstinline{iExists}, \lstinline{iFrame}, \lstinline{iModIntro},
and \lstinline{iSplitL/iSplitR} to automatically use the subprotocol rules to
turn the goal into an equivalent goal where the regular Iris tactics apply.
\begin{itemize}
\item \lstinline{iIntros (x1 .. xn)$\! $ "H1 .. Hm"}
  transforms the subprotocol goal to
  begin with \lstinline{n} universal quantification and \lstinline{m} implications, using the rules
  \ruleref{SP-send-elim} and \ruleref{SP-recv-elim},
  and then introduces the quantifiers (naming them \lstinline{x1 .. xn}) into the
  Coq context, and the hypotheses (naming them \lstinline{H1 .. Hm}) into the
  separation logic context.
\item \lstinline{iExists (t1 .. tn)} transforms the subprotocol goal to start with
  \lstinline{n} existential quantifiers, using the
  \ruleref{SP-send-intro}, \ruleref{SP-recv-intro} and \ruleref{SP-trans} rules,
  and then instantiates these quantifiers with the terms \lstinline{t1 .. tn} specified by the pattern.
\item \lstinline{iFrame "H"} transforms the subprotocol goal into a separating
  conjunction between the payload predicates of the head symbols of either
  protocol, using the rules
  \ruleref{SP-send-intro} and \ruleref{SP-recv-intro},
  and then tries to solve the payload predicate subgoal using \lstinline|"H"|.
\item \lstinline{iModIntro} transforms the subprotocol goal into a
  goal starting with a later modality ($\later$), using the rules
  \ruleref{SP-send-mono} and \ruleref{SP-recv-mono},
  and then introduces that later by stripping off a later from any hypothesis
  in the separation logic context.
\item \lstinline{iSplitL/iSplitR "H1 .. Hn"} transforms the subprotocol goal into a
  separating conjunction between the payload predicates of the head symbols of either
  protocol, using the
  \ruleref{SP-send-intro}, \ruleref{SP-recv-intro} and \ruleref{SP-trans} rules,
  and then creates two subgoals.
  For \lstinline{iSplitL} the left subgoal is given the hypotheses
  \lstinline{H1 .. Hn} from the separation logic context,
  while the right subgoal is given any remaining hypotheses, and \viceversa for
  \lstinline{iSplitR}.
\end{itemize}
The extensions of these tactics are implemented by defining custom type class
instances that hook into the existing MoSeL tactics as described by Krebbers \etal~\cite{krebbers-POPL2017}.

\begin{figure}
\begin{lstlisting}[numbers=left,xleftmargin=1cm,escapechar=|]
Lemma list_rev_subprot :
  ⊢$\ $(<! (l : loc) (vs : list val)> MSG #l {{ llist l vs }};
      <?> MSG #() {{ llist internal_eq l (reverse vs) }}; END) ⊑
     (<! (l : loc) (xs : list T)> MSG #l {{ llistI IT l xs }};
      <?> MSG #() {{ llistI IT l (reverse xs) }}; END).
Proof.
  iIntros (l xs) "Hl". |\label{line:sp_send_intro}|
  iDestruct (Hlr with "Hl") as (vs) "[Hl HIT]". |\label{line:sp_lr_split}|
  iExists l, vs. iFrame "Hl". |\label{line:sp_send_elim}|
  iModIntro.|\label{line:sp_send_mono}| iIntros "Hl". |\label{line:sp_recv_intro}|
  iSplitL. |\label{line:sp_recv_elim}|
  { rewrite big_sepL2_reverse_2. iApply Hlr. |\label{line:sp_sg_11}|
    iExists (reverse vs). iFrame "Hl HIT". } |\label{line:sp_sg_12}|
  done. |\label{line:sp_sg_2}|
Qed.
\end{lstlisting}
\caption{Proof of subprotocol relation}
\label{fig:subprot_proof}
\end{figure}

To demonstrate these tactics, we will go through a proof of the subprotocol relation
for the list reversing service presented in \Cref{sec:subprotocol_reuse}:
\[
  \begin{array}{@{} l @{} l @{}}
  &\sendprot{(\loc:\Loc)(\vec{\val}:\List\ \Val)}{\loc}{\llistref{\loc}{\vec{\val}}}
  \recvprot{}{\TT}{\llistref{\loc}{\listrev{\vec{\val}}}}
  \protend\\
  \subprotop{} &
  \sendprot{(\loc:\Loc)(\vec{\var}:\List\ \tvar)}{\loc}
  {\llistrefI{\interpvar_\tvar}{\loc}{\vec{\var}}}
  \recvprot{}{\TT}
  {\llistrefI{\interpvar_\tvar}{\loc}{\listrev{\vec{\var}}}}
  \protend
  \end{array}
\]
Recall that the following conversion between
the list representation predicate with payload $\llistrefI{\interpvar_\tvar}{\loc}{\vec{\var}}$
and one without payload $\llistref{\loc}{\vec{\val}}$ holds:
\[
  \texttt{Hlr} \quad : \quad
  \textstyle
  \llistrefI{\interpvar_\tvar}{\loc}{\vec{\var}} \ \wandIff \
  (\Exists \vec{\val}.
  \llistref{\loc}{\vec{\val}} \ast
  \Sep_{(\var,\val) \in (\vec{\var},\vec{\val})}.
  \interpvar_\tvar\ \var\ \val)
\]
The full Coq proof of the subprotocol relation is shown in
\Cref{fig:subprot_proof}.
The initial proof state is identical to the lemma statement.
On \lineref{line:sp_send_intro} we start by introducing the
\binders \lstinline{l}, \lstinline{xs} and the payload
\lstinline{llistI IT l xs} of the weaker protocol with the tactic
\lstinline{iIntros (l xs)$\! $ "Hl"}.
This tactic will implicitly apply the rule \ruleref{SP-send-elim}, so the goal
starts with a universal quantification
\lstinline{∀ (l : loc) (xs : list T). llistI IT l xs -∗ ...},
which is then introduced based on the regular Iris introduction pattern.
This gives us:
\begin{lstlisting}
"Hl" : llistI IT l vs
--------------------------------------∗
(<! (l : loc) (vs : list val)> MSG #l {{ llist l vs }};
 <?> MSG #() {{ llist l (reverse vs) }}; END) ⊑
(<!> MSG #l; <?> MSG #() {{ llistI IT l (reverse xs) }}; END)
\end{lstlisting}
To obtain the payload predicate expected by the stronger protocol,
we use the lemma $\mathtt{Hlr}$,
to derive \lstinline{llist l vs} and
\lstinline{[∗ list] x;v ∈$\ $xs;vs, IT x v} from
\lstinline{llistI l xs} with the tactic
\lstinline{iDestruct (Hlr with "Hl")$\! $ as (vs) "[Hl HIT]"} on
\lineref{line:sp_lr_split}.
The resulting proof state is:
\begin{lstlisting}
"Hl" : llist l vs
"HIT" : [∗ list] x;v ∈$\ $xs;vs, IT x v
--------------------------------------∗
(<! (l : loc) (vs : list val)> MSG #l {{ llist l vs }};
 <?> MSG #() {{ llist l (reverse vs) }}; END) ⊑
(<!> MSG #l; <?> MSG #() {{ llistI IT l (reverse xs) }}; END)
\end{lstlisting}
At \lineref{line:sp_send_elim} we
instantiate the \binders of the stronger protocol with the
\binders \lstinline{l} and \lstinline{vs} using \lstinline{iExists l, vs}.
This will implicitly apply the rules \ruleref{SP-send-intro} and \ruleref{SP-trans},
which makes the goal
start with \lstinline{∃ (l : loc)$\! $ (vs : list val)}, so the existentials
can be instantiated.
To resolve the payload predicate obligation \lstinline{llist l vs}, we use
\lstinline{iFrame "Hl"}.
This uses the rules \ruleref{SP-send-intro} and \ruleref{SP-trans} to turn the goal into
\lstinline{llist l vs ∗$\ $...}, where the left subgoal is resolved using
\lstinline{"Hl"}.
We then have the following remaining proof state:
\begin{lstlisting}
"HIT" : [∗ list] x;v ∈$\ $xs;vs, IT x v
--------------------------------------∗
(<!> MSG #l; <?> MSG #() {{ llist l (reverse vs) }}; END) ⊑
(<!> MSG #l; <?> MSG #() {{ llistI IT l (reverse xs) }}; END)
\end{lstlisting}
As the head symbols of both protocols are sends ($\SEND$) with no \binders or
payload predicates, we use \lstinline{iModIntro}
on \lineref{line:sp_send_mono}, which first applies
\ruleref{SP-send-mono} to step over the sends, and then
introduces the later modality ($\later$).
This gives us the proof state:
\begin{lstlisting}
"HIT" : [∗ list] x;v ∈$\ $xs;vs, IT x v
--------------------------------------∗
(<?> MSG #() {{ llist l (reverse vs) }}; END) ⊑
(<?> MSG #() {{ llistI IT l (reverse xs) }}; END)
\end{lstlisting}
On \lineref{line:sp_recv_intro}, similarly to before, we use
\lstinline{iIntros "Hl"}, to introduce the payload predicate,
but this time we do it for the stronger protocol,
as dictated by \ruleref{SP-recv-elim}:
\begin{lstlisting}
"HIT" : [∗ list] x;v ∈$\ $xs;vs, IT x v
"Hl" : llist l (reverse vs)
--------------------------------------∗
(<?> MSG #() ; END) ⊑
(<?> MSG #() {{ llistI IT l (reverse xs) }}; END)
\end{lstlisting}
To resolve the payload predicate of the weaker protocol, we use
\lstinline{iSplitL "Hl HIT"} on \lineref{line:sp_recv_elim}, that first use
\ruleref{SP-recv-intro} and \ruleref{SP-trans}, to turn the goal into
\lstinline{llistI IT l (reverse xs)$\ $∗$\ $...}, and then use the goal splitting pattern
of Iris, to give us two subgoals, where we use the hypotheses \lstinline{"Hl"} and
\lstinline{"HIT"} in the left subgoal.
The first subgoal is then:
\begin{lstlisting}
"HIT" : [∗ list] x;v ∈$\ $xs;vs, IT x v
"Hl" : llist l (reverse vs)
--------------------------------------∗
llistI IT l (reverse xs)
\end{lstlisting}
On \lineref{line:sp_sg_11}, we first use the lemma \lstinline{Hlr}
in the right-to-left
direction, and then rewrite the hypothesis \lstinline{"HIT"} using a
lemma from the Iris library with \lstinline{rewrite big_sepL2_reverse_2}.
We do this to obtain \lstinline{[∗ list] x;v ∈$\ $reverse xs;reverse vs, IT x v},
in order to match the proof goal.
This gives the proof obligation:
\begin{lstlisting}
"HIT" : [∗ list] x;v ∈$\ $reverse xs;reverse vs, IT x v
"Hl" : llist l (reverse vs)
--------------------------------------∗
∃$\ $vs : list val, llist l vs ∗$\ $([∗ list] x;v ∈ reverse xs;vs, IT x v)
\end{lstlisting}
We finally close the proof on \lineref{line:sp_sg_12}
with \lstinline{iExists (reverse vs)}, followed by
\lstinline{iFrame "Hl HIT"}, as the goal matches the hypotheses exactly,
when picking \lstinline{reverse vs} as the existential quantification.
We then move on to the second subgoal:
\begin{lstlisting}
--------------------------------------∗
(<?> MSG #(); END) ⊑$\ $(<?> MSG #(); END)
\end{lstlisting}
We resolve this subgoal, on \lineref{line:sp_sg_2},
with the tactic \lstinline{done},
which tries to close the proof, by automatically applying
\ruleref{SP-refl}.

\section{Related work}
\label{sec:related_work}

This section elaborates on the relation to
message passing in separation logic (\Cref{sec:related_sep}) and
process calculi (\Cref{sec:related_process}),
session types (\Cref{sec:related_session}),
session subtyping (\Cref{sec:related_session_subtyping}),
endpoint sharing (\Cref{sec:related_manifest_sharing}),
and verification of map-reduce (\Cref{sec:related_map_reduce}).

\subsection{Message passing and separation logic}
\label{sec:related_sep}

Lozes and Villard~\cite{villard-APLAS2009,villard-ICE2012} present a logic for contract-based
reasoning about programs in a small imperative language with
bi-directional asynchronous channels.
Contracts are represented by finite-state automata with labelled send or receive
transitions, equipped with separation logic predicates.
Similar to session types (and \lname), contracts have a notion of duality, but
unlike \lname they do not support dependencies between messages.
Their logic
supports ownership transfer (including ownership transfer of channels, akin to
delegation), session-type like choice, and a form of recursive contracts.
Their language has a close operation for channel deallocation
instead of being garbage collected.
A restriction to structured concurrency
(\ie par instead of fork-based), structured channel deallocation (\ie must close both
endpoints together) and linear (instead of affine) logic ensures memory-leak freedom.
A form of channel sharing is supported, which we further discuss in \Cref{sec:related_manifest_sharing}.

Craciun \etal~\cite{craciun-ICECCS2015} introduced \emph{session logic}, a variant of
separation logic that includes
predicates for protocol specifications similar to ours.
This work includes support for mutable state,
ownership transfer (including ownership transfer of channels, akin to delegation),
session-type like choice using a special type of disjunction operator on
the protocol level, and a sketch of an approach to verify deadlock
freedom of programs.
Combined, these features allow them to verify
interesting and non-trivial
message-passing programs.
Their logic as a whole is not
higher-order, which means that sending functions over channels is not
possible.
Moreover, their logic does not support protocol-level \binders that can connect the
transferred message with the tail protocol.
It is therefore not possible to model dependent protocols like we do in \lname.
Their work includes a notion of subtyping as weakening and strengthening of the
payload predicates, however they do not consider swapping, and do not allow
manipulation of resources as a part of their subtyping relation.
There also exists
no support for other concurrency primitives such as locks, which by
extension means that manifest sharing is not possible.
In \lname we get this for free by building on top of Iris, and reusing its ghost state mechanism.
Their work has not been
mechanised in a proof assistant, but example programs can
be checked using the HIP/SLEEK verifier.

The original Iris paper~\cite{jung-POPL2015} includes a small
message-passing language with
channels that do not preserve message order.
It was included to demonstrate that Iris is flexible enough to handle other
concurrency models than standard shared-memory concurrency.
Since the Hoare triples for send and receive reason about
the entire channel buffer, protocol reasoning
must be done via STSs or other forms of ghost state.

Hamin and Jacobs~\cite{hamin-ECOOP2019} take an orthogonal direction and use separation
logic to prove deadlock freedom of programs that communicate via
message passing using a custom logic tailored to this purpose.
They do not provide abstractions akin to our session-type based protocols.
Instead one has to reason using invariants and ghost state explicitly.

Mansky \etal~\cite{mansky-OOPSLA2017} verify the functional
correctness of a message-passing system written in C using the VST framework
in Coq~\cite{appel2014vst}.
While they do not verify message-passing programs
like we do, they do verify that the implementation of their message-passing
system is resilient to faulty behaviour in the presence of malicious senders
and receivers.

Tassarotti \etal~\cite{tassarotti-ESOP2017} prove correctness
and termination preservation of a compiler from a simple language with session
types to a functional language with mutable state, where channels are
implemented using references on the heap.
This work is also done in Iris in Coq.
The session types they consider are more like standard session types, which
cannot express functional properties of messages, but only their types.

The Disel logic by Sergey \etal~\cite{sergey-POPL2018} and the Aneris logic
by Krogh-Jespersen \etal~\cite{krogh-jespersen} can be used to reason about message-passing programs
that work on network sockets.
Channels can only be used to send strings, are not order preserving, and
messages can be dropped but not duplicated.
Since only strings are sent over channels complex data (such as functions) must
be marshalled and unmarshalled in order to be sent over the network.
Both Disel and Aneris therefore address a different problem than we do.

SteelCore \cite{DBLP:journals/pacmpl/SwamyRFMAM20} is a framework for
concurrent separation logic embedded in the F\textsuperscript{$\star$} language.
SteelCore has been used to encode unidirectional synchronous channels that
can be typed with protocols akin to session types.
Their protocols are defined as a dependent sequence of value obligations with
associated separation logic predicates, dictating what can be sent over the channel,
including the transfer of ownership.
Channels are first-class and can also be transferred (akin to delegation),
but their protocols do not include higher-order protocol-level \binders,
or subtyping.
They postulated that their approach scales to bidirectional asynchronous
communication, but left that for future work.

\subsection{Separation logic and process calculi}
\label{sec:related_process}

Another approach to verify message-passing programs is to combine
separation logic and process calculus.
Neither of the approaches below support delegation or concurrency paradigms
other than message passing.

Francalanza \etal~\cite{francalanza-LMCS2011} use separation logic to verify
programs written in a CCS-like language.
Channels model memory location, which has the effect that their input-actions
behave a lot like our updates of mutable state with variable
substitutions updating the state. As a proof of concept they prove the
correctness of an in-place quick-sort algorithm.

Oortwijn \etal~\cite{oortwijn-PLACES2016} use separation logic and the mCRL2 process
calculus to model communication protocols.
The logic itself operates on a high level of abstraction and deals exclusively
with intraprocess communication where a fractional separation logic is
used to distribute channel resources to concurrent threads.
Protocols are extracted from code, but there is no formal connection
between the specification logic and the underlying language.

\subsection{Session types}
\label{sec:related_session}

Seminal work on linear type systems for the $\pi$-calculus by
Kobayashi \etal~\cite{kobayashi-POPL1996} led to the creation of binary session types
by Honda \etal~\cite{honda-ESOP1998}, and consequentially multiparty session types
by Honda \etal~\cite{honda-POPL2008}.

Later work by Dardha \etal~\cite{dardha-PPDP2012} helped merge the linear type
systems of Kobayashi with Honda's session types, which facilitated the
incorporation of session types in mainstream programming
languages like Go~\cite{lange-ICSE2018},
OCaml~\cite{padovani-JFP2017, imai-soCP2019}, and
Java~\cite{hu-ECOOP2010}.
These works focus on adding session-typed support for
message passing in existing languages, but do not
target functional correctness.

Bocchi \etal~\cite{bocchi-CONCUR2010} pushed the boundaries of
what can be verified with (multiparty) session types while staying within a
decidable fragment of first-order logic.
They use first-order predicates to describe
properties of values being sent and received.
Decidability is maintained by imposing restrictions on these predicates, such
as ensuring that nothing is sent that will be invalidated down the line.
The constraints on the logic do, however, limit what programs can be
verified.
The work includes standard subtyping on communicated values and on choices,
but no notion of swapping sends ahead of receives.

Caires and Pfenning~\cite{caires-CONCUR2010} discovered a correspondence between intuitionistic
linear logic and $\pi$-calculus with session types, which was extended with
quantifiers and dependent types by Toninho \etal~\cite{toninho_PPDP11}.
These quantifiers range over both terms and propositions of an LF-based
logic~\cite{cervesato_lics96}, and can be used to specify basic properties of
the exchanged values.
Toninho and Yoshida~\cite{DBLP:conf/fossacs/ToninhoY18} extended this work by allowing the
structure of the protocol to depend on the quantifiers.
This notion of dependency allows for protocols where the length of the (tail)
protocol depends on the values that were previously exchanged, similar to what
we do in \Cref{sec:dependent}.
Finally, Das and Pfenning~\cite{das_concur2020, das_ppdp2020} developed a
dependent session-type system with domain-specific logic for verifying
arithmetic properties of programs with message passing.

Another approach to dependent session types was carried out by
Thiemann and Vasconcelos~\cite{DBLP:journals/pacmpl/ThiemannV20} who introduced label-dependent
session types.
They unify universal and existential quantifiers with the
send and receive primitives of conventional session types.
Hence, similar to \lname, the choice connectives ($\branchop$ and $\selectop$)
can be derived.

Toninho \etal~\cite{toninho_tgc14} and Lindley and
Morris~\cite{lindley_icfp2016} developed session-type systems with
termination guarantees in the presence of recursive (session) types.
This is achieved by imposing a discipline similar to (co)inductive definitions
in Coq and Agda.
In contrast, \lname poses no usage discipline on recursive \pname,
and hence guarantees partial correctness.

\subsection{Session subtyping}
\label{sec:related_session_subtyping}

\lname's subprotocol relation is inspired by the notion of session subtyping,
for which seminal work was carried out by Gay and Hole~\cite{gay-AI2005}.
Mostrous \etal~\cite{mostrous-ESOP2009} extended session subtyping to
multiparty asynchronous session types, and as part of that, introduced the notion of
swapping sends ahead of receives for independent channels.
Mostrous \etal~\cite{mostrous-IaC2015} later considered swapping over the same channel in
the context of binary session types.
Our subprotocol relation is most closely related to the work of
Mostrous \etal~\cite{mostrous-IaC2015},
although they define subtyping as a simulation on infinite trees,
using so-called asynchronous contexts, whereas we define it using Iris's
support for guarded recursion.
It should be noted that the work by Gay and Hole~\cite{gay-AI2005} differs
from the work by Mostrous \etal~\cite{mostrous-ESOP2009} and
Mostrous \etal~\cite{mostrous-IaC2015} in the orientation of the subtyping relation,
as discussed by Gay~\cite{DBLP:conf/birthday/Gay16}.
Our subprotocol relation uses the orientation of
Gay and Hole~\cite{gay-AI2005}.

Session subtyping for recursive type systems is universally carried out as a type
simulation on infinite trees
\cite{gay-AI2005, mostrous-ESOP2009,mostrous-IaC2015},
which complicates subtyping under the recursion operator.
Bernardi \etal~\cite{bernardi_tgc2014} and Gay \etal~\cite{gay-PLACES2020} provide further insights on
this problem, although they primarily investigate duality rather than subtyping.

To reason about recursive subtyping, Brand and Henglein~\cite{brandt-1998FI} present
a coinductive formulation of subtyping
(which they apply to regular type systems, rather than session types).
We use a similar coinductive formulation, but instead of ordinary coinduction,
we use Iris's support for guarded recursion, which lets us
prove subtyping relations of recursive protocols using \ruleref{Loeb} induction.

\subsection{Endpoint sharing}
\label{sec:related_manifest_sharing}

One of the key features of conventional session types is that endpoints are owned
by a single thread.
While endpoints can be delegated (\ie transferred from one thread to
another), they typically cannot be shared (\ie be accessed by multiple threads
concurrently).
However, as demonstrated in \Cref{sec:integration}, sharing channels
endpoints is often desirable, and possible in \lname.

As a simple way to relax this limitation of sharing in conventional
session types,
Vasconcelos~\cite{vasconcelos-IC2012} allows session types of the form
$(\MU \recname. \sendtype{\tvar}{\recname})$ or
$(\MU \recname. \recvtype{\tvar}{\recname})$ to be shared.
Lozes and Villard~\cite{villard-ICE2012} present a similar idea in the context of their
contract-based separation logic (see also \Cref{sec:related_sep}) by equipping
the connective for channel endpoint ownership with a fractional permission.
If the fraction is smaller than 1, then the endpoint can be shared, but at the
cost of only permitting transitions to the same contract state.
Using fractional permissions they prove a lock specification \`a la
Gotsman \etal~\cite{gotsman_locks} of an implementation of locks in terms of channels.
This approach to locks is dual to ours in \lname, where we implement
channels in terms of locks.
Unlike Iris (and \lname), their logic does not support ghost state, so
it cannot express complex protocols like the ones from \Cref{sec:integration}.

In the $\pi$-calculus community there has been prior work
on endpoint sharing, \eg by
Atkey \etal~\cite{atkey2016}, Kobayashi~\cite{kobayashi-CONCUR2016}, and
Padovani~\cite{padovani-CSL2014}.
The latest contribution in this line of work is by
Balzer \etal~\cite{balzer-ESOP2019}, who
developed a type system based on session types with
support for manifest sharing.
Manifest sharing is the notion of sharing a channel endpoint
between multiple processes using a lock-like structure to ensure mutual
exclusion.
Their key idea to ensure mutual exclusion using a type system is to use
adjoint modalities to connect two classes of types: types
that are linear, and thus denote unique channel ownership, and types that are
unrestricted, and thus can be shared.
The approach to endpoint sharing in \lname is different:
\pname do not include a built-in notion for endpoint sharing, but can be combined
with Iris's general-purpose mechanisms for sharing, like locks.

\subsection{Verification of map-reduce}
\label{sec:related_map_reduce}

To our knowledge the only verification related to the map-reduce model
\cite{dean-OSDI2004} is by
Ono \etal~\cite{ono-SEFM2011}, who made two mechanisations in Coq. The first took
a functional model of map-reduce and verified a few specific mappers
and reducers, extracted these to Haskell, and ran them using Hadoop
Streaming. The second did the same by annotating Java mappers and
reducers using JML
and proving them correct using the Krakatoa tool \cite{marche-JLAP2004}, using
a combination of
SAT-solvers and the Coq proof assistant. While they worked on
verifying specific mappers and reducers, our case study focuses on
verifying the communication of a map-reduce model that can later be parameterised with
concrete mappers and reducers.

\section{Conclusion and future work}
\label{sec:future_work}

In this paper, we have given a comprehensive account of the \lname concurrent
separation logic for proving functional correctness of programs that
combine message-passing with other programming and concurrency paradigms.
The core feature of \lname its its mechanism of \pname, which is inspired by session types.
Considering the rich literature on session types and concurrent separation
logic, we expect there to be many promising directions for future work.

\subsubsection*{Multi-party}

The formalism of multi-party session types \cite{honda-POPL2008}
applies to message-passing communication between more
than two parties (threads or processes).
The key ingredient of multi-party session types is the notion of a
\emph{global protocol}, which specifies the permitted communication for
multiple parties of a system.
From the global protocol one can then generate \emph{local protocols} for the
individual parties.
It would be interesting to explore a multi-party version of \pname.
Prior work by Costea \etal~\cite{costea-APLAS2018} on multi-party session logic and
Zhoud \etal~\cite{multiparty_refinements} on refined multiparty session types could serve
as a starting point.

\subsubsection*{Deadlock freedom}

As discussed in \Cref{sec:adequacy_pre_actris}, deadlocks are valid behaviours
according to the notion of safety used in Iris (and thus \lname).
Many conventional session type systems do not consider deadlocks to be valid
behaviours, but achieve that at the
expense of prohibiting valid (deadlock free) programs that can be verified
in \lname.

A direction for future work is to develop a variant of \lname that incorporates
the usual restrictions of session-type systems like
linearity and a $\startname$ primitive for combined channel and thread creation.
To prove an adequacy theorem that ensures that this variant of \lname indeed
prohibits deadlocks, one needs to change the
model of \lname to ensure acyclicity of the dependency structure among the threads
and channels.
This could be achieved by building upon recent work
by Bizjak \etal~\cite{bizjak-PACMPL2019} on linearity in Iris and by Jacobs \etal~\cite{connectivity_graphs}
on a separation-logic based proof method for deadlock freedom of session types.
Additionally, one could consider a version
of \lname without garbage collection but with a $\derivedkw{close}$
instruction for channel
deallocation, and prove that it indeed guarantees memory-leak freedom.

Another direction for future work is to develop a separation logic
that combines session-type based deadlock freedom with lock-order based
deadlock freedom to prove deadlock freedom of programs that combine
message passing with other concurrency mechanisms like locks.
The work by Hamin and Jacobs~\cite{hamin-ECOOP2019} on reasoning about lock orders in separation
logic, and the work by Balzer \etal~\cite{balzer-ESOP2019} on deadlock freedom for manifest
sharing might provide valuable insights, but figuring out how to combine these
two approaches with
Iris and \lname is a challenging open problem.

\section*{Acknowledgments}

We thank the anonymous reviewers of both this paper and the
POPL'20 conference version for their helpful feedback.
We are grateful to Andreea Costea, Daniel Gratzer, Dani\"el Louwrink, Fabrizio Montesi,
Marco Carbone,
and the participants of the Iris workshop 2019 for discussions.
The third author (Robbert Krebbers) was supported by the Dutch Research Council (NWO), project 016.Veni.192.259.

\bibliographystyle{alphaurl}
\bibliography{main}

\end{document}